\documentclass[aps,prx,preprint,onecolumn,citeautoscript,nofootinbib,eqsecnum,superscriptaddress]{revtex4-2}
\usepackage[T1]{fontenc}
\usepackage{comment}
\usepackage{subcaption}
\usepackage{amsmath}
\usepackage{amssymb}
\usepackage{amsthm}
\usepackage{amsfonts}
\usepackage{bm,bbm}
\usepackage{braket}
\usepackage{xcolor}
\usepackage{pifont}
\usepackage{slashed}
\usepackage[mathscr]{euscript}
\usepackage[shortlabels]{enumitem}
\usepackage{tikz-feynman}
\usetikzlibrary{arrows}

\usepackage[papersize={8.5in,11in}]{geometry}

\usepackage{float}
\allowdisplaybreaks

\usepackage{graphicx}
\usepackage[colorlinks=true]{hyperref}
\renewcommand{\approx}{\simeq}
\renewcommand{\Re}{\text{Re}}
\renewcommand{\Im}{\text{Im}}

\definecolor{wrongultramarine}{rgb}{1,0.5,0}

\newcommand{\rd}{{\rm d}}
\newcommand{\sgn}{{\rm sgn\,}}
\newcommand{\Tr}{{\rm \, Tr\,}}
\newcommand{\calN}{{\mathcal N}}
\newcommand{\calG}{{\mathcal G}}
\newcommand{\bk}{{\bm k}}
\newcommand{\bq}{{\bm q}}
\newcommand{\br}{{\bm r}}
\newcommand{\nn}{\nonumber \\}
\newcommand{\beq}{\begin{equation}}
\newcommand{\eeq}{\end{equation}}
\def\bea{\begin{eqnarray}}
\def\eea{\end{eqnarray}}

\hypersetup{
    bookmarks=true,         % show bookmarks bar?
    unicode=false,          % non-Latin characters
    pdftoolbar=true,        % show Acrobat
    pdfmenubar=true,        % show Acrobat
    pdffitwindow=false,     % window fit to page when opened
    pdfstartview={FitH},    % fits the width of the page to the window
    pdftitle={Large N theory of critical Fermi surfaces},    % title
    pdfauthor={Ilya Esterlis, Haoyu Guo, Aavishkar A. Patel, Subir Sachdev},     % author
    pdfsubject={},   % subject of the document
    pdfcreator={},   % creator of the document
    pdfproducer={}, % producer of the document
    pdfkeywords={} {} {}, % list of keywords
    pdfnewwindow=true,      % links in new window
    colorlinks=true,       % false: boxed links; true: colored links
    linkcolor=magenta, %red,          % color of internal links (change box color with linkbordercolor)
    citecolor=blue,        % color of links to bibliography
    filecolor=magenta,      % color of file links
    urlcolor=blue           % color of external links
}

\tikzset{
  mid arrow/.style={postaction={decorate,decoration={
        markings,
        mark=at position .575 with {\arrow[#1]{stealth}}
      }}},
  near arrow/.style={postaction={decorate,decoration={
        markings,
        mark=at position .275 with {\arrow[#1]{stealth}}
      }}},
   far arrow/.style={postaction={decorate,decoration={
        markings,
        mark=at position .800 with {\arrow[#1]{stealth}}
      }}},
   boson/.style={decorate, draw=black,
    decoration={snake,amplitude=1pt, segment length=5pt},
      },
   mid triangle/.style={postaction={decorate,decoration={
        markings,
        mark=at position .575 with {\arrow[#1]{triangle 45}}
      }}}
}

\begin{document}

\title{Large $N$ theory of critical Fermi surfaces II: conductivity}

\author{Haoyu Guo}
\affiliation{Department of Physics, Harvard University, Cambridge MA 02138, USA}

\author{Aavishkar A. Patel}
\affiliation{Center for Computational Quantum Physics, Flatiron Institute, New York,
New York, 10010, USA}

\author{Ilya Esterlis}
\affiliation{Department of Physics, Harvard University, Cambridge MA 02138, USA}

\author{Subir Sachdev}
\affiliation{Department of Physics, Harvard University, Cambridge MA 02138, USA}

\date{\today}% It is always \today, today,

\begin{abstract}
A Fermi surface coupled to a scalar field can be described in a $1/N$ expansion by choosing the fermion-scalar Yukawa coupling to be random in the $N$-dimensional flavor space, but invariant under translations. We compute the conductivity of such a theory in two spatial dimensions for a critical scalar. We find a Drude contribution, and verify that the proposed $1/\omega^{2/3}$ contribution to the optical conductivity at frequency $\omega$ has vanishing co-efficient for a convex Fermi surface. We also describe the influence of impurity scattering of the fermions, and find that while the self energy resembles a marginal Fermi liquid, the resistivity and optical conductivity behave like a Fermi liquid.
\end{abstract}

\maketitle

\newpage

\tableofcontents

\section{Introduction}
%Cite \cite{Esterlis:2021eth,Patel:2022gdh}
One of the cornerstones of modern condensed matter theory is the Fermi liquid (FL) theory. The central assumption of FL is the existence of well-defined quasiparticles as elementary excitations of the system. Due to these quasiparticles, at low temperatures $(T)$ the resistivity scales as $\rho=\rho_0+AT^2$ \cite{coleman2015}. However, in the study of strongly correlated systems such as half-filled Landau level, metallic quantum critical points and gapless quantum spin liquids \cite{PALee89,Polchinski:1993ii,HLR,Kim94,Altshuler94,Nayak:1994ng,sungsik1,metlitski1,mross,sungsik3,metlitski5,Hartnoll:2014gba,Patel_viscosity,HolderMetzner1,HolderMetzner2,Raghu1,Raghu2,Torroba1,Torroba2,Hooley15,Patel2018mag,Chowdhury:2018sho,Moon2010,Chubukov1,Chubukov2,Chubukov3,Berg1,Berg2,Berg3,DCBerg,DebanjanAPS,Patel:2018eak,Ilya1,Ilya2,Wang:2019bpd,Wang:2020dtj,Altman1,Patel:2016wdy,Patel:2019qce,oganesyan2001},
the strong interaction destroys the quasiparticles and the resulting system is often dubbed a non-Fermi liquid (NFL).

The universal low-energy physics of non-Fermi liquids can be captured by the model of critical Fermi surface (FS) \cite{Lee_ARCMP}, in which a Fermi surface of free fermions is coupled to critically fluctuating bosons.
Inspired by Sachdev-Ye-Kitaev (SYK) models \cite{SY92,kitaev2015talk,SS15,JMDS16},
a previous work by the same authors \cite{Esterlis:2021eth} (hereafter referred to as I) has proposed a controlled theory to perform $1/N$ expansion of the problem: the main idea is to make the Yukawa coupling between fermions and bosons a random variable in the $N$-dimensional flavor indices, but uniform in space.

The SYK model has an emergent time reparameterization symmetry, and consequently fluctuations at a frequency scale $\omega \sim 1/N$ are very strong, and change the critical behavior at $\omega \ll 1/N$ \cite{Bagrets:2016cdf,JMDS16,Kitaev:2017awl}. In contrast, it was shown in I that the critical Fermi surface does {\it not\/} have any emergent time reparameterization symmetry, and so the criticality of the corresponding $1/N$ expansion is expected to be more stable than that of the SYK model. This is also reflected in the fact that the large $N$ entropy density of the critical Fermi surface vanishes as $T \rightarrow 0$, while that of the SYK model has a non-zero limit as $T \rightarrow 0$.

In the study of strongly interacting metals such as cuprate high-$T_c$ superconductors, near critical doping the normal state shows `strange' metallic behavior \cite{ProustTaillefer2019,bruin,Zaanen,Gael21,Paschen22,Sankar2022}. This includes $T\ln(1/T)$ specific heat and a linear-in-temperature resistivity much smaller than the quantum resistivity unit (in 2D, $h/e^2$), which is related to a `Planckian' dissipation time $\hbar/(k_B T)$. These experimental observations can be encapsulated into a phenomenological theory called the marginal Fermi liquid \cite{MFL89}. It is conjectured that a marginal Fermi liquid could emerge from a non-Fermi liquid.

In this work, we study the transport properties of the critical Fermi surface from I, focusing on the electrical conductivity of the fermions. We will study the translationally invariant (clean) model of I, and also consider the effect of spatial potential disorder on the fermions. We demonstrate that neither theory gives rise to linear-in-temperature resistivity due to various cancellations, even though the fermions in the latter do acquire a marginal Fermi liquid self energy.

Building on the lessons learned in this paper, a mechanism for strange metal behavior with linear-in-temperature resistivity is proposed in a companion paper \cite{Patel:2022gdh}. We argue in Ref.~\cite{Patel:2022gdh} that the ingredient missing in the present paper is spatial randomness in the interaction term between the fermions and bosons.

In Sec.~\ref{sec:summary}, we review the previous work of I and summarize the main results of the present paper. In Sec.~\ref{sec:uniform} we present detailed derivation of results related to the clean model, and in Sec.~\ref{sec:potential} we discuss adding spatial potential disorder to the clean theory.
Appendix~\ref{app:Kim} reviews an earlier computation of the optical conductivity in the clean case in Ref.~\cite{Kim94}, and shows that they overlooked a cancellation between diagrams.

   % We found that in the translational invariant case, the DC resistivity diverges in agreement with memory matrix theory, and the NFL scaling cancels in optical conductivity due to strong boson drag, in contrast to previous literature \cite{PALee89,Kim94}.

   % The results above show that a strange metal cannot emerge from a translational invariant (clean) NFL, and motivate us to consider effects of disorder. We found that the potential disorder alters the critical point from having boson dynamical exponent $z=3$ in the clean case to $z=2$ in the disordered (dirty) case. At the dirty critical point, the fermion acquires a marginal Fermi liquid self energy.

\section{Summary of Main Results}\label{sec:summary}
\subsection{Translational Invariant (Clean) Model}

We start by reviewing the SYK-inspired large $N$ theory of the two-dimensional quantum-critical metal \cite{Esterlis:2021eth,Altman1}. The imaginary time ($\tau$) action for the fermion field $\psi_i$ and scalar field $\phi_i$ (with $i=1 \ldots N$ a flavor index) is \cite{Esterlis:2021eth}
\begin{align}
&\mathcal{S}_g = \int d\tau\sum_{\bk}\sum_{i=1}^N\psi^\dagger_{i\bk}(\tau)\left[\partial_\tau +\varepsilon (\bk) \right]\psi_{i\bk}(\tau) \nn
&+\frac{1}{2}\int d\tau \sum_{\bq}\sum_{i=1}^N \phi_{i\bq}(\tau)\left[-\partial_\tau^2 + K \bq^2 +m_b^2\right]\phi_{i,-\bq}(\tau) \nn
&+\frac{g_{ijl}}{N} \int d\tau d^2 r \sum_{i,j,l=1}^N \psi^\dagger_{i}(\br,\tau)\psi_{j}(\br,\tau)\phi_{l}(\br,\tau)\,,
\label{eq:latticeaction}
\end{align}
where the fermion dispersion $\varepsilon(\bk)$ determines the Fermi surface, the scalar mass $m_b$ has to be tuned to criticality and is needed for infrared regularization but does not appear in final results, and $g_{ijl}$ is space independent but random in flavor space with
\beq
\overline{g_{ijl}} = 0 \,, \quad  \overline{{g}^\ast_{ijl}g_{abc}} = {g}^2\,\delta_{ia}\delta_{jb}\delta_{lc}\,,
\eeq
where the overline represents average over flavor space.
The hypothesis is that a large domain of flavor couplings all flow to the same universal low energy theory (as in the SYK model), so we can safely examine the average of an ensemble of theories. Momentum is conserved in each member of the ensemble, and the flavor-space randomness does not lead to any essential difference from non-random theories. This is in contrast to position-space randomness which we consider later, which does relax momentum and modify physical properties.

The flavor-space average of the partition function of $\mathcal{S}_g$ leads to a `$G$-$\Sigma$' theory, whose large $N$ saddle point of (\ref{eq:latticeaction}) has singular fermion ($\Sigma$) and boson ($\Pi$) self energies at $T=0$  \cite{Esterlis:2021eth}
\begin{align}
    \Pi(i\omega, \bq) =
    -c_b \frac{|\omega|}{|\bq|}\,,& \quad \Sigma (i\omega, \bk) = - i c_f \mbox{sgn}(\omega) |\omega|^{2/3}\,,  \nonumber \\
      c_b= \frac{g^2}{2 \pi \kappa v_F}\,,& \quad c_f = \frac{g^2}{2 \pi v_F \sqrt{3}} \left( \frac{2 \pi v_F \kappa}{K^2 g^2} \right)^{1/3}\,.
\label{eq:sigmag}
\end{align}
These results are obtained on a circular Fermi surface with curvature $\kappa=1/m$ where $m$ is the effective mass of the fermions. In this work, we re-derived the above results using the full Fermi surface, while in the previous work in I, we have only considered the theory of two antipodal patches around $\pm \bk_0$ on the Fermi surface to which $\bq$ is tangent, with axes chosen so that $\bq = (0,q)$ and fermionic
dispersion $\varepsilon(\pm \bk_0 + \bk) = \pm v_F k_x + \kappa k_y^2/2$. This is because transport computation requires including momenta beyond patch theories.

The large $N$ computation of the optical conductivity at zero temperature $(T=0)$ yields only the clean Drude result $\mathrm{Re}[\sigma(\omega)]/N=\pi\mathcal{N}v_F^2\delta(\omega)/2$, where $\mathcal{N}=m/(2\pi)$ is the fermion density of states at the Fermi level. This is obtained for a circular Fermi surface when only states on the Fermi surface are considered (it is shown in  Ref.~\cite{Rech06} that Fermi surface curvature is important for current vertices, and our approach implicitly includes these effects). By coincidence, this result agrees with the patch theory, but we will show that the patch theory fails to fully capture transport properties. The absence of a $\omega \neq 0$ contribution is tied to an exact cancellation between self-energy and vertex diagrams arising from momentum conservation. Previous literature which obtained a $|\omega|^{-2/3}$ optical conductivity \cite{PALee89,Kim94} didn't fully account for this cancellation; but other works \cite{ChubukovMaslov,Maslov12,Maslov17} did find the cancellation, and argued that it was present only for convex Fermi surfaces. In the appendix we reproduce the calculations of \cite{Kim94} and show that the cancellation indeed happens after obtaining a numerical coefficient undetermined in \cite{Kim94}.
Furthermore, this cancellation can be recast into a kinematical constraint for all odd harmonics of the Fermi surface \cite{LedwithAoP,LedwithArxiv}: all odd harmonic modes relax slowly even for a general Fermi surface, and the leading order contribution to relaxation is due to states not exactly on the Fermi surface. When these additional relaxation are included, we expect the optical conductivity to scale as $\sigma(\omega)\sim 1/(-i\omega+\# \omega^2)\sim1/(-i\omega)+\#|\omega|^0$ (see Eq.\eqref{eq:sigmaxxD}). Note that this $\omega^2$ scattering rate is still more singular than a scattering rate in a \emph{translational invariant} Fermi liquid \cite{LedwithAoP,LedwithArxiv}.  We have ignored umklapp processes as they are not universal.  We also note our large-$N$ result agrees with recent arguments based on anomalies  of the $N=1$ theory \cite{ShiElse2022}.

\subsection{Model with Potential Disorder}

    The results above show that a clean non-Fermi liquid cannot demonstrate linear-in-temperature resistivity, and this motivates us to consider effects of spatial disorder. As a first attempt we consider adding potential disorder:
    \begin{align}
& \mathcal{S}_v =   \frac{1}{\sqrt{N}} \int d^2 r d \tau \, v_{ij} (\br)  \psi_i^{\dagger} (\br, \tau) \psi_j(\br, \tau) \nonumber \\
& \overline{v_{ij} (\br)} = 0 \,, \quad \overline{v^\ast_{ij} (\br) v_{lm} (\br')} = v^2 \, \delta(\br-\br') \delta_{il} \delta_{jm}
\end{align}
and here the overline is an now average over spatial co-ordinates and flavor space. The large $N$ limit of the $G$-$\Sigma$ theory of $\mathcal{S}_g+\mathcal{S}_v$
is described in Sec.~\ref{sec:potential}, and
yields results similar to earlier studies \cite{HLR,ChubukovMaslov,Berg1}. The low frequency boson propagator now has the diffusive form $\sim (q^2 + c_d |\omega|)^{-1}$ with $z=2$ (in contrast to $z=3$ of the clean theory), while the fermion self energy has an elastic scattering term, along with a marginal Fermi liquid \cite{MFL89} inelastic term at low frequencies
\begin{align}
\Pi (i \omega, {\bm q}) & =-\frac{\mathcal{N} g^2|\omega|}{\Gamma}, \quad \Gamma = 2\pi v^2 \mathcal{N}, \\
\Sigma(i\omega, {\bm k} = k_F \hat{k}) & = - i \frac{\Gamma}{2} \mbox{sgn}(\omega) - \frac{ig^2 \omega}{2\pi^2 \Gamma}\ln\left(\frac{e \Gamma^3}{\mathcal{N} g^2 v_F^2 |\omega|}\right), \nonumber
\end{align}
at $T = 0$.
However, the marginal Fermi liquid self energy, while leading to a $T\ln(1/T)$ specific heat, does {\it not\/} lead to the claimed \cite{MFL89} linear-$T$ term in the DC resistivity, as it arises from forward scattering of electrons off the $\mathbf{q}\sim0$ bosons (this has also been noted in the recent work of Ref.~\cite{Foster22}). These forward scattering processes are unable to relax either current or momentum due to the small wavevector of the bosons involved and the momentum conservation of the $g$ interactions. As a result, even a perturbative computation of the conductivity at $\mathcal{O}(g^2)$  shows a cancellation  between the interaction-induced self energy contributions and the interaction-induced vertex correction, leading to a DC conductivity that is just a constant, set by the elastic potential disorder scattering rate $\Gamma$. A full summation of all diagrams at large $N$ shows that the $g$ interactions only renormalize the frequency term in the Drude formula:
\begin{equation}\label{}
  \frac{1}{N}\Re[\sigma(\omega\gg T)]=\frac{1}{2}\frac{\mathcal{N}v_F^2\Gamma}{\tilde{\omega}^2+\Gamma^2}\,,
\end{equation} where
\begin{equation}\label{}
  \tilde{\omega}=\omega\left(1-\frac{g^2}{2\pi^2\mathcal{N}^2v_F^4}\left[\frac{v_F \Lambda}{4}+\frac{\Gamma}{4\pi}\ln\left(\frac{\Gamma}{e \Lambda v_F}\right)\right]\right),
\end{equation}
and $\Lambda\sim k_F$ is a UV momentum cutoff. In the limit of large Fermi energy (and hence large $\mathcal{N}v_F^2$), this renormalization is negligible and $\tilde{\omega}\approx\omega$. In addition, the boson drag only corrects $\Gamma$ by order $\omega^2$. The leading frequency dependence of the optical conductivity at frequencies $\omega\ll \Gamma$ is therefore just a constant, and there is no linear in frequency correction. Correspondingly, in the DC limit, there is no linear in $T$ correction, and a conventional $T^2$ correction is expected.

\section{Spatially uniform quantum-critical metal}
\label{sec:uniform}

\subsection{The model and notations}

    In this section, we review some properties of the clean model studied in the previous work of I, and recapitulate some useful notations. For simplicity, we will work with $K=1$ (boson velocity set to one) which can be restored by dimensional analysis.
\subsubsection{Lagrangian and $G$-$\Sigma$ action}
    We write the action in Eq.\eqref{eq:latticeaction} as a Lagrangian below
\begin{equation}\label{eq:L_clean}
  \mathcal{L}=\sum_{i}\psi_i^\dagger (\partial_\tau+\varepsilon_k-\mu)\psi_i+\frac{1}{2}\sum_{i}\phi_i(-\partial_\tau^2+\omega_q^2+m_b^2)\phi_i+\sum_{ijl}\frac{g_{ijl}}{N}\psi_i^\dagger \psi_j \phi_l\,.
\end{equation}Here $\varepsilon_k=k^2/(2m)$ and $\omega_q^2=q^2$ which physically describe the dispersions of fermions and bosons respectively, should be understood as differential operators that act on the fields. The Yukawa couplings $g_{ijl}=g_{jkl}^{*}$ are Gaussian random variables with zero mean and variance $\overline{|g_{ijl}|^2}=g^2$. Throughout the paper we work in $2+1$ dimensions.

    Assuming the system self averages, we perform disorder average over $g_{ijl}$ with a simple replica, and next we introduce bilocal variables
\begin{equation}\label{eq:bilocal_def}
  \begin{split}
     G(x_1,x_2) & = -\frac{1}{N}\sum_{i}\psi_i(x_1)\psi_i^\dagger(x_2)\,, \\
     D(x_1,x_2)  & = \frac{1}{N}\sum_i \phi_i(x_1)\phi_i(x_2)\,,
  \end{split}
\end{equation} as well $\Sigma(x_1,x_2)$ and $\Pi(x_1,x_2)$ as Lagrangian multipliers to enforce the above definitions, to obtain the $G$-$\Sigma$ action
\begin{equation}\label{eq:S_Gsigma_clean}
\begin{split}
   \frac{1}{N}S[G,\Sigma,D,\Pi] =& -\ln\det\left(\left(\partial_\tau+\varepsilon_k-\mu\right)\delta(x-x')+\Sigma\right) +\frac{1}{2} \ln\det\left(\left(-\partial_\tau^2+\omega_q^2+m_b^2\right)\delta(x-x')-\Pi\right)  \\
   &-\Tr\left(\Sigma\cdot G\right)+\frac{1}{2}\Tr\left(\Pi\cdot D\right)+\frac{g^2}{2}\Tr\left((GD)\cdot G\right)\,.
\end{split}
\end{equation} Here $\delta(x-x')$ denotes a spacetime delta function.

We pause briefly the explain our notation, which is the same as in Ref.~\cite{Gu:2019jub}. For two bilocal functions $f,g$, we define their inner product as
\begin{equation}\label{}
  \Tr(f\cdot g)\equiv f^T g\equiv\int \rd x_1 \rd x_2 f(x_2,x_1)g(x_1,x_2)\,.
\end{equation}The action of a linear functional $A$ is defined as:
\begin{equation}\label{}
  A[f](x_1,x_2)\equiv \int \rd x_3 \rd x_4 A(x_1,x_2;x_3,x_4)f(x_3,x_4)\,.
\end{equation} The transpose acts both on functions and on functionals:
\begin{equation}\label{}
  f^T(x_1,x_2)\equiv f(x_2,x_1)\,,
\end{equation}
\begin{equation}\label{}
  A^T(x_1,x_2;x_3,x_4)\equiv A(x_4,x_3;x_2,x_1)\,.
\end{equation}
\subsubsection{Saddle point}
   Going back to the action \eqref{eq:S_Gsigma_clean} and differentiating it, we obtain
\begin{equation}\label{eq:deltaSstar}
  \frac{\delta S}{N}=\Tr\left(\delta\Sigma\cdot(G_*[\Sigma]-G)+\delta G\cdot(\Sigma_*[G]-\Sigma)+\frac{1}{2}\delta\Pi\cdot(D-D_*[\Pi])+\frac{1}{2}\delta D\cdot (\Pi-\Pi_*[D])\right)\,,
\end{equation} where
    \begin{eqnarray}
      G_*[\Sigma](x_1,x_2) &=& (-\partial_\tau+\mu-\varepsilon_k-\Sigma)^{-1}(x_1,x_2)\,,\label{eq:Gstar} \\
      \Sigma_*[G](x_1,x_2) &=& \frac{g^2}{2}G(x_1,x_2)\left(D(x_1,x_2)+D(x_2,x_1)\right)\,, \\
      D_*[\Pi](x_1,x_2) &=& (-\partial_\tau^2+\omega_q^2+m_b^2-\Pi)^{-1}(x_1,x_2)\,, \\
      \Pi_*[D](x_1,x_2) &=& -g^2 G(x_1,x_2)G(x_2,x_1)\,.\label{eq:Sigmastar}
    \end{eqnarray} In the first and the third line the inverse is in the functional sense. Therefore the saddle point equations are simply
\begin{equation}\label{}
  G=G_*[\Sigma]\,,\quad \Sigma=\Sigma_*[G]\,,\quad D=D_*[\Pi]\,,\quad \Pi=\Pi_*[D]\,.
\end{equation}

\subsubsection{Fluctuations about the saddle point}
  We can further expand \eqref{eq:S_Gsigma_clean} to second order around the saddle point to obtain the fluctuations around the saddle point. Define the collective notation $\mathcal{G}_a=(D,G)$ and $\Xi_a=(\Pi,\Sigma)$, where $a=b,f$ denotes boson/fermion. The gaussian fluctuations around the saddle point is described by
  \begin{equation}\label{eq:deltaS1}
  \frac{1}{N}\delta^2 S=\frac{1}{2}\begin{pmatrix}
                        \delta \Xi^T & \delta \calG^T
                      \end{pmatrix}\Lambda
                      \begin{pmatrix}
                        W_\Sigma & -1 \\
                        -1 & W_G
                      \end{pmatrix}
                      \begin{pmatrix}
                        \delta \Xi \\
                        \delta \calG
                      \end{pmatrix}\,,
\end{equation} where $\Lambda=\text{diag}(-1/2,1)$ acts on the $b,f$ indices, and $W_\Sigma$ and $W_G$ are defined by
\begin{equation}\label{}
  W_\Sigma(x_1,x_2;x_3,x_4)_{a\tilde{a}}=\frac{\delta \calG_*[\Xi]_a(x_1,x_2)}{\delta \Xi_{\tilde{a}}(x_3,x_4)},\qquad W_G(x_1,x_2;x_3,x_4)_{a\tilde{a}}=\frac{\delta \Xi_*[\calG]_a(x_1,x_2)}{\delta\calG_{\tilde{a}}(x_3,x_4)}\,.
\end{equation}

  Later for the evaluation of the conductivity, we will be using fluctuation of self energies, which is given by
\begin{equation}\label{}
  \braket{\delta\Xi_a(x_1,x_2)\delta\Xi_{\tilde{a}}(x_4,x_3)}=\left[W_G\frac{1}{W_\Sigma W_G-1}\Lambda^{-1}\right]_{a\tilde{a}}(x_1,x_2;x_3,x_4)\,.
\end{equation}

    For the $G$-$\Sigma$ action \eqref{eq:S_Gsigma_clean}, $W_\Sigma$ and $W_G$ are given by Feynman diagrams
    \begin{equation}\label{eq:Wsigmagraph}
  W_\Sigma(x_1,x_2;x_3,x_4)=\begin{pmatrix}
                      \begin{tikzpicture}[baseline={([yshift=-4pt]current bounding box.center)}]
                     \draw[thick, boson] (40pt,12pt)--(0pt,12pt);
                     \path[mid triangle] (40pt,12pt)--(0pt,12pt);
                     \draw[thick, boson] (0pt,-12pt)--(40pt,-12pt);
                     \path[ mid triangle] (0pt,-12pt)--(40pt,-12pt);
                     \node at (-5pt,12pt) {\scriptsize $1$};
                     \node at (-5pt,-12pt) {\scriptsize $2$};
                     \node at (48pt,12pt) {\scriptsize $3$};
                     \node at (48pt,-12pt) {\scriptsize $4$};
                     \end{tikzpicture}
                        & 0 \\
                      0 &
                      \begin{tikzpicture}[baseline={([yshift=-4pt]current bounding box.center)}]
                     \draw[thick, mid arrow] (40pt,12pt)--(0pt,12pt);
                     \draw[thick, mid arrow] (0pt,-12pt)--(40pt,-12pt);
                     \node at (-5pt,12pt) {\scriptsize $1$};
                     \node at (-5pt,-12pt) {\scriptsize $2$};
                     \node at (48pt,12pt) {\scriptsize $3$};
                     \node at (48pt,-12pt) {\scriptsize $4$};
                     \end{tikzpicture}
                    \end{pmatrix},
\end{equation}
\begin{equation}\label{eq:WGgraph}
  W_G(x_1,x_2;x_3,x_4)=\begin{pmatrix}
                 0 & -g^2\left(\begin{tikzpicture}[baseline={([yshift=-4pt]current bounding box.center)}]
                                  \draw[thick, dashed] (20pt,12pt)--(0pt,12pt);
                                  \draw[thick, dashed] (0pt,-12pt)--(20pt,-12pt);
                                  \draw[thick, mid arrow] (20pt,12pt)--(20pt,-12pt);
                                  \node at (-5pt,12pt) {\scriptsize $1$};
                                  \node at (-5pt,-12pt) {\scriptsize $2$};
                                  \node at (24pt,12pt) {\scriptsize $3$};
                                  \node at (24pt,-12pt) {\scriptsize $4$};
                               \end{tikzpicture}
                               +
                               \begin{tikzpicture}[baseline={([yshift=-4pt]current bounding box.center)}]
                                  \draw[thick, dashed] (20pt,-12pt)--(0pt,12pt);
                                  \draw[thick, dashed] (0pt,-12pt)--(20pt,12pt);
                                  \draw[thick, mid arrow] (20pt,12pt)--(20pt,-12pt);
                                  \node at (-5pt,12pt) {\scriptsize $1$};
                                  \node at (-5pt,-12pt) {\scriptsize $2$};
                                  \node at (24pt,12pt) {\scriptsize $3$};
                                  \node at (24pt,-12pt) {\scriptsize $4$};
                               \end{tikzpicture}
                               \right) \\
                 \frac{g^2}{2}\left(\begin{tikzpicture}[baseline={([yshift=-4pt]current bounding box.center)}]
                                  \draw[thick, dashed] (20pt,12pt)--(0pt,12pt);
                                  \draw[thick, dashed] (0pt,-12pt)--(20pt,-12pt);
                                  \draw[thick, mid arrow] (0pt,-12pt)--(0pt,12pt);
                                  \node at (-5pt,12pt) {\scriptsize $1$};
                                  \node at (-5pt,-12pt) {\scriptsize $2$};
                                  \node at (24pt,12pt) {\scriptsize $3$};
                                  \node at (24pt,-12pt) {\scriptsize $4$};
                               \end{tikzpicture}+
                                \begin{tikzpicture}[baseline={([yshift=-4pt]current bounding box.center)}]
                                  \draw[thick, dashed] (20pt,-12pt)--(0pt,12pt);
                                  \draw[thick, dashed] (0pt,-12pt)--(20pt,12pt);
                                  \draw[thick, mid arrow] (0pt,-12pt)--(0pt,12pt);
                                  \node at (-5pt,12pt) {\scriptsize $1$};
                                  \node at (-5pt,-12pt) {\scriptsize $2$};
                                  \node at (24pt,12pt) {\scriptsize $3$};
                                  \node at (24pt,-12pt) {\scriptsize $4$};
                               \end{tikzpicture}

                 \right) & g^2\begin{tikzpicture}[baseline={([yshift=-4pt]current bounding box.center)}]
                                  \draw[thick, dashed] (20pt,12pt)--(0pt,12pt);
                                  \draw[thick, dashed] (0pt,-12pt)--(20pt,-12pt);
                                  \draw[thick, boson] (20pt,-12pt)--(20pt,12pt);
                                  \path[mid triangle] (20pt,-12pt)--(20pt,12pt);
                                  \node at (-5pt,12pt) {\scriptsize $1$};
                                  \node at (-5pt,-12pt) {\scriptsize $2$};
                                  \node at (24pt,12pt) {\scriptsize $3$};
                                  \node at (24pt,-12pt) {\scriptsize $4$};
                               \end{tikzpicture}
               \end{pmatrix}\,,
\end{equation}
where a black arrowed line denotes fermion propagator, a wavy arrowed line denotes boson propagator (the arrow denotes momentum), and a dashed line denotes spacetime $\delta$-function. The first entry is boson and the second entry is fermion. Recalling $\Lambda=\mathrm{diag}(-1/2,1)$, we see that $\Lambda W_\Sigma$ and $\Lambda W_G$ are explicitly symmetric as required by quadratic expansion.

In momentum space, we can explicitly write down the action of $W_\Sigma$ and $W_G$:
\begin{equation}\label{}
  W_\Sigma\begin{pmatrix}
            B(k,p) \\
            F(k,p)
          \end{pmatrix}=\begin{pmatrix}
                          G(k+p/2)G(k-p/2) & 0 \\
                          0 & D(k+p/2)D(k-p/2)
                        \end{pmatrix}
                        \begin{pmatrix}
                          B(k,p) \\
                          F(k,p)
                        \end{pmatrix}\,.
\end{equation}
\begin{equation}\label{}
  W_G \begin{pmatrix}
        B(k,p) \\
        F(k,p)
      \end{pmatrix}=\begin{pmatrix}
                      \tilde{B}(k,p) \\
                      \tilde{F}(k,p)
                    \end{pmatrix}\,,
\end{equation}where
\begin{equation}\label{eq:tB}
\begin{split}
  &\tilde{B}(k_1,p)=-g^2\int\frac{\rd^3 k_2}{(2\pi)^{3}}\left[G(k_2-k_1)F(k_2,p)+G(k_1-k_2)F(-k_2,p)\right],
\end{split}
\end{equation}
\begin{equation}\label{eq:tF}
\begin{split}
   \tilde{F}(k_1,p) & = g^2\int\frac{\rd^3 k_2}{(2\pi)^{3}}\left[\frac{1}{2}G(k_1-k_2)\left(B(k_2,p)+B(-k_2,p)\right)+D(k_1-k_2)F(k_2,p)\right]\,.
\end{split}
\end{equation} Here $p$ denotes the CoM 3-momentum and $k$ denotes the relative 3-momentum. Unless stated explicitly, we will be using $\int \rd\omega/(2\pi)$ and $T\sum_{\omega_n}$ interchangeably.

\subsubsection{Relation to patch theories}

  In the previous paper I, we have studied the same theory within patch approximations $\varepsilon_k=\pm k_x+k_y^2$. In this paper, we will take a different route by working with the full Fermi surface and taking a patch-like approximation at a later stage. While patch theories produce the correct solution to the saddle point equations, they are inadequate for transport computations. In particular, within the patch theory the vector nature of the current operator is neglected, and it behaves very similar to the density operator. For example, in the single patch theory they are exactly proportional and in the two patch theory with two antipodal patches, they differ by a $\mp$ sign on the left/right patch. Due to this similarity, current-current correlation function can be inferred from the density-density correlation function, and this results in zero conductivity at non-zero frequency.

  As we will see later in the theory of the full Fermi surface, the current operator as a vector, is susceptible to additional scattering events than the density operator, which is a scalar. These scattering events are due to bosons carrying momentum tangential to the Fermi surface. Because the current operator contains $l=1$ angular harmonics, there is a phase shift $e^{-i l \theta_{kk'}}$ associated with the scattering event $k\to k'$, which is absent for scalar operators. This effect has the same origin as the $(1-\cos\theta)$ factor in the transport scattering rate of Boltzmann equations, and this factor is set to zero in the patch theory.

\subsection{Expression for Conductivity}

\subsubsection{Polarization Bubble}
In this section we derive an expression for the conductivity from the $G$-$\Sigma$ action. To define the electric current, we use the minimal coupling scheme $\partial_\mu\to \partial_\mu+iA_\mu$, i.e. $k_\mu\to k_\mu+A_\mu$, so the only relevant term is the fermion determinant term as the following:
\begin{equation}\label{eq:SGSigma_conductivity}
  S[G,\Sigma,D,\Pi;A]=-\ln\det((\partial_\tau+\varepsilon_{k+A}-\mu)\delta(x-x')+\Sigma)-\Tr(\Sigma\cdot G)+S_b[D,\Pi]+S_{int}[G,D]\,,
\end{equation} where $S_b[D,\Pi]$ denotes the kinetic terms for the boson and $S_{int}[G,D]$ describes the interactions.

The conductivity is given by Kubo formula
\begin{equation}\label{}
  \sigma^{\mu\nu}(\omega)=i\frac{\Pi^{\mu\nu}(i\omega_n\to \omega+i0,k=0)}{\omega}\,,
\end{equation} and here the polarization $\Pi^{\mu\nu}$ is defined in real space by
\begin{equation}\label{eq:PiA}
  \Pi_A^{\mu\nu}(x,x')=-\left.\frac{\delta^2 \ln Z[A]}{\delta A_\mu(x) \delta A_\nu(x')}\right\vert_{A=0}\,,
\end{equation}
where $Z[A]=\int\mathcal{D}G\mathcal{D}\Sigma\mathcal{D}D\mathcal{D}\Pi e^{-S}$ is the partition function.
We can alternatively write the above expression as
\begin{equation}\label{eq:PiA2}
  \Pi_A^{\mu\nu}(x,x')=\left.\left\langle\frac{\delta^2 S}{\delta A_\mu(x)\delta A_\nu(x')}-\frac{\delta S}{\delta A_\mu(x)}\frac{\delta S}{\delta A_\nu(x')}\right\rangle_c\right\vert_{A=0}\,,
\end{equation} where the average only includes connected diagrams, and it is performed over bilocal fields. In the leading large-$N$ order, we can take $S$ to be the saddle-point action. The expression in fourier space is given by
\begin{equation}\label{}
  \Pi_A^{\mu\nu}(p)=-\frac{(2\pi)^{3}}{\delta(0)}\left.\frac{\delta^2 \ln Z[A]}{\delta A_\mu(-p) \delta A_\nu(p)}\right\vert_{A=0}\,,
\end{equation} where $A_\mu(x)=\int \frac{\rd^3 p}{(2\pi)^3}A_\mu(p)e^{i\vec{p}\cdot\vec{x}-ip_0 x_0}$.

Let's now compute the functional derivatives in \eqref{eq:PiA2}. Expanding \eqref{eq:SGSigma_conductivity} in $A$ by
\begin{equation}\label{}
  S[A]=S_0+\delta_A S+\delta_A^2 S\,,
\end{equation} where for the first order term we have
\begin{equation}\label{eq:deltaAS}
  \delta_A S=N\int_{x,x'} G_*[\Sigma](x,x') \delta_A\varepsilon_{k+A}(x',x).
\end{equation} Here $G_*[\Sigma]$ is a functional of $\Sigma$ which defines the RHS of SD equations:
\begin{equation}\label{}
  G_*[\Sigma]=\frac{1}{-\partial_\tau+\mu-\varepsilon_{k+A}-\Sigma}\,,
\end{equation} and $\partial_\tau,\mu,\varepsilon_{k+A},\Sigma$ should be understood as bilocal fields or functionals on local fields.

We can proceed to second order in the expansion, which yields
\begin{equation}\label{eq:deltaA2S}
  \delta_A^2 S=\frac{N}{2}\left[\int_{x,x',y,y'}G_*[\Sigma](x,y)\delta_A \varepsilon_{k+A}(y,y')G_*[\Sigma](y',x')\delta_A\varepsilon_{k+A}(x',x)+2\int_{x,x'} G_*[\Sigma](x,x')\delta_A^2 \varepsilon_{k+A}(x',x)\right]\,.
\end{equation}

We can see that the first term of \eqref{eq:PiA2} comes from \eqref{eq:deltaA2S}, which can be evaluated directly at the saddle point. The first term in \eqref{eq:deltaA2S} is a current-current correlator and the second term is a contact term. The second term of \eqref{eq:PiA2}, however, is zero at the saddle point (since they are disconnected) and must be evaluated using fluctuations of the bilocal fields, i.e. summing the ladder diagrams.

\subsubsection{Vertex functions}

To write down explicit expressions for the functional derivatives, we need to calculate the vertex functions $\delta_A \varepsilon_{k+A}$. For simplicity, we shall assume that we only turn on gauge field in the $x$-direction, and it is independent of $y$: $A_x(\tau,x,y)=A_x(\tau,x)$. Under this assumption, the kinetic term $\varepsilon_{k+A}$ is
\begin{equation}\label{}
  \varepsilon_{k+A}=\varepsilon_k(k_x+A_x,k_y)\,,
\end{equation} where $\varepsilon_k$ is a (smooth) function describing the dispersion, but the arguments $k_x+A_x$ and $k_y$ are operators. Our above assumptions of $A_x$ means that $A_x$ commutes with $k_y$, and therefore we can unambiguously write down a Taylor expansion for $\varepsilon_k$:
\begin{equation}\label{}
  \varepsilon_k(k_x+A_x,k_y)=\sum_{n=0}^\infty \frac{1 }{n!}f_x^{(n)}(0)(k_x+A_x)^n\,,
\end{equation} where $f_x(k_x)=\varepsilon_k(k_x,k_y)$.

\begin{comment}
For simplicity, we shall assume that the dispersion $\varepsilon_k$ depends on momentum in the following fashion:
\begin{equation}\label{eq:veksep}
  \varepsilon_{(k_x,k_y)}=f_x(k_x)+f_y(k_y),
\end{equation} where $f_{x,y}$ are smooth functions (Taylor expandable around zero). This assumption is already sufficient for quadratic band or the square-lattice tight-binding model. When we implement minimal coupling $k_\mu\to k_\mu+A_\mu$, $k_\mu$ and $A_\nu$ are in general not commuting, so we need to promote $\varepsilon_{k}$ to operator. The usual strategy is to use the Taylor series of $\varepsilon_k$.  However, given that $k_x+A_x$ not commuting with $k_y+A_y$,  the Taylor series recipe doesn't work for a general function  $\varepsilon_k(k_x+A_x,k_y+A_y)$, but it still works for functions satisfying \eqref{eq:veksep}.
\end{comment}
Let's first calculate $\delta_{A_x}\varepsilon_{k+A}$, we can expand $\varepsilon_{k+A}$ to first order in $A_x$:
\begin{equation}\label{}
  \delta_{A_x}\varepsilon_{k+A}=\sum_{n=0}^{\infty}\frac{1}{n!}f_x^{(n)}(0)\left(k_x^{n-1}A_x+k_x^{n-2}A_xk_x+\dots+A_x k_x^{n-1}\right)\,.
\end{equation} This is an operator equation, where the matrix elements are
\begin{equation}\label{}
  k_x(x,x')=-i\partial_x \delta(x-x'),\qquad A_x(x,x')=A_x(x)\delta(x-x')\,.
\end{equation} Insert these matrix elements into $\delta_{A_x}\varepsilon_{k+A}$, and we obtain
\begin{equation}\label{}
\begin{split}
  \frac{\delta \varepsilon_{k+A}}{\delta A_x(x_0)}(x_1,x_2)=& \sum_{n=0}^{\infty} \frac{1}{n!}f_x^{(n)}(0) \sum_{m=0}^{n-1} (k_x^{n-1-m})(x_1,x_0)(k_x^{m})(x_0,x_2)\\
  =&\sum_{n=0}^{\infty} \frac{1}{n!}f_x^{(n)}(0) \sum_{m=0}^{n-1} (-i\partial_{x_1})^{n-1-m}(i\partial_{x_2})^{m}\delta(x_1-x_0)\delta(x_2-x_0)\\
  =&\sum_{n=0}^{\infty} \frac{1}{n!}f_x^{(n)}(0) \frac{(-i\partial_{x_1})^{n}-(i\partial_{x_2})^{n}}{(-i\partial_{x_1})-(i\partial_{x_2})}\delta(x_1-x_0)\delta(x_2-x_0)\\
  =& \frac{f_x(-i\partial_{x_1})-f_x(i\partial_{x_2})}{(-i\partial_{x_1})-(i\partial_{x_2})}\delta(x_1-x_0)\delta(x_2-x_0)\,.
\end{split}
\end{equation} Here $\partial_x$ only acts on the $x$-component, but the delta functions are over the spacetime.
We can also write it in momentum space as
\begin{equation}\label{}
  \frac{\delta \varepsilon_{k+A}}{\delta A_x(r)}(p,q)=\Gamma^x(p,q)\delta(r+q-p)\,,\quad \Gamma^x(p,q)=\frac{f_x(p_x)-f_x(q_x)}{p_x-q_x}\,.
\end{equation} We remind the reader that here  the external momentum $r$ has no $y$ component $r=(r_0,r_x,0)$.

To obtain the second derivative, we write
\begin{equation}\label{}
  \delta_{A_x}^2 \varepsilon_{k+A}=\sum_{n=0}^{\infty}\frac{1}{n!}f_x^{(n)}(0)\sum_{a=0,b=0,a+b\leq n-2} k_x^{a} A_x k_x^{b}A_x k_x^{n-2-a-b}\,.
\end{equation} The expression for the functional derivative is complicated for general external momentum, but we only need it for the case where the two $A_x$'s carry opposite momenta, and the functional derivative simplifies to
\begin{equation}\label{}
  \frac{\delta^2 \varepsilon_{k+A}}{\delta A_x(-r)\delta A_x(r)}(p,r+p)=\frac{ \delta(0)}{(2\pi)^{3}} \Delta^x(p,r)\,,\quad \Delta^x(p,r)=2\left.\left(\frac{\rd}{\rd p_x}\frac{f_x(p_x)-f_x(q_x)}{p_x-q_x}\right)\right\vert_{q=r+p}\,.
\end{equation}

The expression for $\Gamma^x$ and $\Delta^x$ can be further simplified by noticing that in conductivity calculations we only need the homogeneous limit $r_x=0$, and we obtain
\begin{equation}\label{eq:vertex_zerop}
  \Gamma^x(p,p)=\frac{\partial \varepsilon_k(p)}{\partial p_x}\,,\quad \Delta^x(p,0)=\frac{\partial^2 \varepsilon_k(p)}{\partial p_x^2}\,.
\end{equation}

Therefore we can write down the contribution to $\Pi_A$ from the first term of \eqref{eq:PiA2}, which originates from \eqref{eq:deltaA2S}:
\begin{equation}\label{eq:PiAxx1}
  \Pi_{A1}^{xx}(r)=N \int \frac{\rd^3 p}{(2\pi)^{3}}G_*[\Sigma](p)\Gamma^x(p,r+p)G_*[\Sigma](r+p)\Gamma^x(r+p,p)\,,
\end{equation}
\begin{equation}\label{eq:PiAxx2}
  \Pi_{A2}^{xx}(r)=N\int \frac{\rd ^3 p}{(2\pi)^3}G_*[\Sigma](p)\Delta^x(p,r)\,.
\end{equation}
These two terms are the same as the conventional current-current correlator term and the diamagnetic term. This can be seen from the example
\begin{equation}\label{}
  \varepsilon_k=\frac{k_x^2+k_y^2}{2m}\,\quad \Gamma^x(p,p)=\frac{p_x}{m},\quad \Delta^x(p,0)=\frac{1}{m}\,,
\end{equation}which agrees with well-known results.

\begin{comment}
For square lattice we have
\begin{equation}\label{}
  \varepsilon_{k}=-2t(\cos k_x+\cos k_y),\quad \Gamma^x(p,q)=-2t\frac{\cos p_x-\cos q_x}{p_x-q_x}=4t\sin\frac{p_x+q_x}{2}\frac{\sin\frac{p_x-q_x}{2}}{p_x-q_x},\quad
\end{equation}
\begin{equation}\label{}
  \Delta_x(p,r)=-4t\frac{\cos(p_x+r_x)+r_x\sin p_x-\cos p_x}{r_x^2}\,.
\end{equation}
For the calculation of conductivity, we usually take $\vec{r}=0$, therefore have the simplified case
\begin{equation}\label{}
  \Gamma^{x}(p,p)=2t \sin p_x\,,\qquad \Delta_x(p,0)=2t \cos p_x\,.
\end{equation}
\end{comment}

Finally, we look at the second term of \eqref{eq:PiA2}. At leading $N$ order we can expand $G_*[\Sigma]$ and obtain
\begin{equation}\label{eq:PiAxx3}
\begin{split}
  \Pi_{A3}^{xx}(r)=&-\frac{N^2}{(2\pi)^{3}\delta(0)}\int \frac{\rd ^3 p\rd^3 q}{(2\pi)^{6}}G_*[\Sigma](p)G_*[\Sigma](p+r)G_*[\Sigma](q)G_*[\Sigma](q+r)\Gamma^x(p+r,p)\Gamma^x(q,q+r)\\
  &\times\left\langle\delta \Sigma(p,p+r) \delta \Sigma(q+r,q)\right\rangle\,.
\end{split}
\end{equation} Here the $\delta\Sigma(p,q)$ is the fourier transform of the fluctuating bilocal field
\begin{equation}\label{}
  \delta\Sigma(p,q)=\int \rd^3 x \rd^3 y \delta\Sigma(x,y)e^{-i(\vec{p}\cdot\vec{x}-p_0x_0)}e^{i(\vec{q}\cdot\vec{y}-q_0y_0)}.
\end{equation} The correlator $\braket{\delta\Sigma \delta\Sigma}\propto N^{-1}$ is calculated in the previous paper I, where we have derived the expression:
\begin{equation}\label{}
  \left\langle\delta \Sigma(p,p+r) \delta \Sigma(q+r,q)\right\rangle=N^{-1}(2\pi)^{3}\delta(0)\left[W_G\frac{1}{K_G-1}\Lambda^{-1}\right](p+r/2,q+r/2;r)\,,
\end{equation} where the delta-function comes from energy-momentum conservation. The first two arguments on the RHS label the relative momenta and the third argument denotes the CoM momentum. Since we are looking at fermionic components, the matrix $\Lambda$ can be replaced by identity.

Also notice that the $GG$ factors in \eqref{eq:PiAxx1} and \eqref{eq:PiAxx3} are nothing but $W_\Sigma$, we can therefore write $\Pi_{A1}+\Pi_{A3}$ as
\begin{equation}\label{eq:PiAxx13}
  \Pi^{xx}_{A13}(r)\equiv\Pi^{xx}_{A1}(r)+\Pi^{xx}_{A3}(r)=N (\Gamma^x)^T\frac{1}{W_\Sigma^{-1}-W_G}\Gamma^x\,,
\end{equation} where the vertex function $\Gamma^x$ is viewed as a two-point function by ignoring the leg with external momentum $r$, and thus can be acted by $W_\Sigma$.

The total polarization is therefore
\begin{equation}\label{eq:PiAxxtot}
  \Pi_A^{xx}=\Pi_{A13}^{xx}+\Pi_{A2}^{xx}\,.
\end{equation}

  The above formalism can also be used to derive the charge-charge polarization function. Using the minimal coupling scheme $\partial_\tau\to \partial_\tau+i A_\tau$, we obtain the vertex function
  \begin{equation}\label{}
    \Gamma^\tau(p,q)=i\,.
  \end{equation} There is no diamagnetic term for charge, so the charge-charge (density-density) polarization function is
\begin{equation}\label{}
  \Pi^{\tau\tau}_{A}=N (\Gamma^\tau)^T\frac{1}{W_\Sigma^{-1}-W_G}\Gamma^\tau\,.
\end{equation}

\subsubsection{Polarization bubble at the DC limit}

  In this section we show that at the DC limit $p_x=0,p_0\to 0$, the polarization bubble vanishes in the presence of U(1) symmetry:
\begin{equation}\label{}
  \Pi^{xx}_A(p_x=0,p_0\to 0)=0\,.
\end{equation} Here, we use $p_n$ to denote the discrete Matsubara frequency and $p_0$ to denote the frequency continued to real time, i.e. $ip_n\to p_0+i\eta$.

  We introduce a renormalized vertex function $V^\mu$:
  \begin{equation}\label{}
    V^\mu=W_\Sigma^{-1}\frac{1}{W_\Sigma^{-1}-W_G}\Gamma^\mu\,.
  \end{equation}

  Therefore the current-current (paramagnetic) contribution to the polarization is
  \begin{equation}\label{}
    \Pi_{A13}^{xx}(p_n,\vec{p}=0)=NT\sum_{q_n}\int \frac{\rd^2 \vec{q}}{(2\pi)^2}\Gamma^x(q,q)G(q+p)G(q)V^x(p+q,q)\,.
  \end{equation} Here we have used the fact that the bare vertex $\Gamma^x(q,q+p)=\Gamma^x(q,q)$ because $\vec{p}=0$.

  The diamagnetic term is
  \begin{equation}\label{}
    \Pi_{A2}^{xx}(p_n,\vec{p}=0)=NT\sum_{q_n}\int \frac{\rd^2 \vec{q}}{(2\pi)^2}\Delta^x(q,0)G(q)\,.
  \end{equation}Using \eqref{eq:vertex_zerop}, we can integrate by parts in $q_x$ to obtain
  \begin{equation}\label{eq:PiA2_IBP}
    \Pi_{A2}^{xx}(p_n,\vec{q}=0)=-NT\sum_{q_n}\int \frac{\rd^2 \vec{q}}{(2\pi)^2}\Gamma^x(q,q)G(q)^2\left(\Gamma^x(q,q)+\frac{\partial \Sigma(q)}{\partial q_x}\right)\,.
  \end{equation}

  We therefore needs to show that the renormalized vertex $V^x(q,q+p)$ cancels the terms in the parenthesis in \eqref{eq:PiA2_IBP} when $p_0\to 0$.

  Using the U(1) Ward identity \eqref{eq:U1Ward2} in the next section, we have
  \begin{equation}\label{}
    p_\mu V^\mu(p+q,q)=G^{-1}(q)-G^{-1}(q+p)\,.
  \end{equation} Plugging in $p_\mu=(-p_n,p_x,0)$ and expanding the Green's functions, we get
  \begin{equation}\label{}
    -p_n V^\tau(p+q,q)+p_xV_x(p+q,q)=-ip_n+(\varepsilon_{p+q}-\varepsilon_q)+\left(\Sigma(p+q)-\Sigma(q)\right)\,.
  \end{equation} Taking the limit $p_x\to 0$ on both sides, and matching to linear order in $p_x$, we obtain
  \begin{equation}\label{eq:Vx}
    V^x=\Gamma^x+p_n \frac{\partial V^\tau }{\partial p_x}+\frac{\partial \Sigma(p_n+q_n,\vec{p})}{\partial q_x}\,.
  \end{equation} Here both $V^x$ and $\Gamma^x$ are evaluated at $(p+q,q)$ with $\vec{p}=0$, and the derivative of $V^\tau$ is
  \begin{equation}\label{}
    \frac{\partial V^\tau}{\partial p_x}\equiv\left.\frac{\partial V^\tau(k,q)}{\partial k_x}\right\vert_{k=(p_n+q_n,\vec{q})}\,.
  \end{equation}
  Now, the function $V^x$ given by \eqref{eq:Vx}, viewed as a function of $p_n$ can be analytically continued to the complex $p_n$ plane and it has a branch cut at $p_n=-q_n$. There is no ambiguity in taking the limit $p_n\to \eta$, and because $\partial V^\tau/\partial p_x$ is finite, we have
  \begin{equation}\label{}
    V^x=\Gamma^x+\frac{\partial \Sigma(q)}{\partial q_x}\,,
  \end{equation} and therefore $\Pi^{xx}(p_n\to 0,\vec{p}=0)=0$.

\subsection{Ward Identities} \label{sec:Ward}

    For the clean model, Ward identities are an important tool that makes the evaluation of conductivities possible. The main idea is the following: We will apply arguments similar to Prange and Kadanoff \cite{PrangeKadanoff} to integrate out momentum dependence in electron Green's functions and reduce the kernel $W_\Sigma^{-1}-W_G$ in \eqref{eq:PiAxx13} to act only in frequency and angular harmonic space. Since the current vertex function is a first angular harmonics proportional $\cos\theta_k$ (but frequency-independent), the conductivity can be schematically written as an inner product
    \begin{equation}\label{eq:sigmaxxschematic}
        \sigma^{xx}(i\Omega)\sim \braket{\cos\theta|\frac{1}{W_\Sigma^{-1}-W_G}|\cos\theta}\sim \left(\int \rd\theta \cos^2\theta \right) \braket{1|\frac{1}{W_\Sigma^{-1}-W_G^{(1)}}|1}
    \end{equation} Here due to rotation symmetry $W_\Sigma$ and $W_G$ can be decomposed into blocks acting on angular harmonics (each block is a functional in frequency space) which we have factored out. $W_\Sigma$ is the same for all angular harmonics, while $W_G$ acts as $W_G^{(l)}$ in the $l$-th angular harmonic sector.

    The U(1) Ward identity yields an eigenvector equation satisfying
    \begin{equation}\label{}
      (W_\Sigma^{-1}-W_G^{(0)})\ket{1}=\Omega\ket{1}\,.
    \end{equation} Physically, the difference between $W_G^{(0)}$ and $W_G^{(1)}$ is small in $1/k_F$ due to small angle scattering and can be calculated by gradient expansion.  Therefore, the conductivity can be calculated using first order perturbation as
    \begin{equation}\label{}
      \sigma^{xx}\sim \frac{1}{\Omega+\delta\lambda}\,,
    \end{equation} where
    \begin{equation}\label{}
      \delta\lambda\propto \braket{1|W_G^{(0)}-W_G^{(1)}|1}\,.
    \end{equation}
    A conventional $|\omega|^{-2/3}$ conductivity \cite{Kim94,ChubukovMaslov0} corresponds to $\delta\lambda\propto \Omega^{4/3}$, but our more careful computation show that $\delta\lambda=0$ due to momentum conservation. In section \ref{sec:conductivity}, we will formalize the above discussions.

   % Under this reduction, the Ward identities become a statement of eigenvector of $W_\Sigma^{-1}-W_G$ , and it turns out that the vertex function $\Gamma^x$ is very close to this eigenvector: The U(1) Ward identity yields an eigenvector which is just a zeroth angular harmonics $1$, and the vertex function is a first angular harmonics $\cos\theta_k$. Under the action of $W_\Sigma^{-1}-W_G$, the difference between the images of these two functions is small in the $k_F\to\infty$ limit, and can be calculated using gradient expansion.

   % which makes a resummed perturbation theory possible.

\subsubsection{Master Ward identity}

  We first present a master Ward identity which includes both U(1) symmetry and diffeomorphism invariance. We write the $G$-$\Sigma$ action in the form
  \begin{equation}\label{eq:S_Gsigma2}
\begin{split}
   \frac{S}{N} =& -\ln\det\left(\sigma_f+\Sigma\right) +\frac{1}{2} \ln\det\left(-\sigma_b-\Pi\right)  -\Tr\left(\Sigma\cdot G\right)+\frac{1}{2}\Tr\left(\Pi\cdot D\right)+\frac{g^2}{2}\Tr\left((GD)\cdot G\right)\,,
\end{split}
\end{equation}where
\begin{equation}\label{}
  \sigma_f(x,x')=(\partial_\tau+\varepsilon_k-\mu)\delta(x-x')\,,
\end{equation}and
\begin{equation}\label{}
  \sigma_b(x,x')=(\partial_\tau^2-\omega_q^2)\delta(x-x')
\end{equation}are the UV sources.

  Consider the following change of variables $(G,\Sigma,D,\Pi,\sigma_f,\sigma_b)\to(\tilde{G},\tilde{\Sigma},\tilde{D},\tilde{\Pi},\tilde{\sigma}_f,\tilde{\sigma}_b)$ which makes the action invariant:
\begin{equation}\label{eq:dylG}
  G(x_1,x_2)=\left|\frac{\partial y_1}{\partial x_1}\right|^{\Delta}\left|\frac{\partial y_2}{\partial x_2}\right|^{\Delta}\tilde{G}(y_1,y_2)e^{i\left(\lambda(y_1)-\lambda(y_2)\right)}\,,
\end{equation}
\begin{equation}\label{eq:dylSigma}
  \Sigma(x_1,x_2)=\left|\frac{\partial y_1}{\partial x_1}\right|^{1-\Delta}\left|\frac{\partial y_2}{\partial x_2}\right|^{1-\Delta}\tilde{\Sigma}(y_1,y_2)e^{i\left(\lambda(y_1)-\lambda(y_2)\right)}\,,
\end{equation}
\begin{equation}\label{}
  D(x_1,x_2)=\left|\frac{\partial y_1}{\partial x_1}\right|^{1-2\Delta}\left|\frac{\partial y_2}{\partial x_2}\right|^{1-2\Delta}\tilde{D}(y_1,y_2)\,,
\end{equation}
\begin{equation}\label{}
  \Pi(x_1,x_2)=\left|\frac{\partial y_1}{\partial x_1}\right|^{2\Delta}\left|\frac{\partial y_2}{\partial x_2}\right|^{2\Delta}\tilde{\Pi}(y_1,y_2)\,,
\end{equation}
\begin{equation}\label{eq:dylsigmaf}
  \sigma_f(x_1,x_2)=\left|\frac{\partial y_1}{\partial x_1}\right|^{1-\Delta}\left|\frac{\partial y_2}{\partial x_2}\right|^{1-\Delta}\tilde{\sigma}_f(y_1,y_2)e^{i\left(\lambda(y_1)-\lambda(y_2)\right)}\,,
\end{equation}
\begin{equation}\label{eq:dylsigmab}
  \sigma_b(x_1,x_2)=\left|\frac{\partial y_1}{\partial x_1}\right|^{2\Delta}\left|\frac{\partial y_2}{\partial x_2}\right|^{2\Delta}\tilde{\sigma}_b(y_1,y_2)\,.
\end{equation} Here $|\partial y/\partial x|$ is the Jacobian of $y=y(x)$ , and $\Delta$ is an arbitrary real number.

Define $\delta_{\lambda,y} G=\tilde{G}(x_1,x_2)-G(x_1,x_2)$ and similarly for other variables, we can write down a master Ward identity
\begin{equation}\label{eq:MasterWard1}
  \Tr(\frac{\delta S}{\delta G}\delta_{\lambda,y} G+\frac{\delta S}{\delta \Sigma}\delta_{\lambda,y}\Sigma+\frac{\delta S}{\delta D}\delta_{\lambda,y}D+\frac{\delta S}{\delta \Pi}\delta_{\lambda,y}\Pi)=-\Tr(\frac{\delta S}{\delta \sigma_f}\delta_{\lambda,y}\sigma_f+\frac{\delta S}{\delta \sigma_b}\delta_{\lambda,y}\delta \sigma_b)\,.
\end{equation}

Taking functional derivatives of the master Ward identity \eqref{eq:MasterWard1} at the saddle point and using \eqref{eq:deltaSstar}, we obtain
\begin{equation}\label{}
  \int \rd x_1\rd x_2 \left(\frac{\delta \Sigma_*(x_2,x_1)}{\delta G(x_3,x_4)}\delta_{y,\lambda} G(x_1,x_2)-\frac{1}{2}\frac{\delta \Pi_*(x_2,x_1)}{\delta G(x_3,x_4)}\delta_{y,\lambda} D(x_1,x_2)\right)=\delta_{y,\lambda} \Sigma(x_4,x_3)\,,
\end{equation}
\begin{equation}\label{}
  \int \rd x_1\rd x_2 \left(\frac{\delta \Sigma_*(x_2,x_1)}{\delta D(x_3,x_4)}\delta_{y,\lambda} G(x_1,x_2)-\frac{1}{2}\frac{\delta \Pi_*(x_2,x_1)}{\delta D(x_3,x_4)}\delta_{y,\lambda} D(x_1,x_2)\right)=-\frac{1}{2}\delta_{y,\lambda} \Pi(x_4,x_3)\,,
\end{equation}
\begin{equation}\label{}
  -\delta_{y,\lambda} G(x_4,x_3)+\int \rd x_1\rd x_2 \frac{\delta G_*(x_2,x_1)}{\delta \Sigma(x_3,x_4)}\delta_{y,\lambda}\Sigma(x_1,x_2)=-\int \rd x_1\rd x_2\frac{\delta G_*(x_2,x_1)}{\delta \Sigma(x_3,x_4)}\delta_{y,\lambda} \sigma_f(x_1,x_2)\,,
\end{equation}
\begin{equation}\label{}
  \frac{1}{2}\delta_{y,\lambda} D(x_4,x_3)-\frac{1}{2}\int \rd x_1\rd x_2 \frac{\delta D_*(x_2,x_1)}{\delta \Pi(x_3,x_4)}\delta_{y,\lambda}\Pi(x_1,x_2)=\frac{1}{2}\int \rd x_1\rd x_2\frac{\delta D_*(x_2,x_1)}{\delta \Pi(x_3,x_4)}\delta_{y,\lambda} \sigma_b(x_1,x_2)\,.
\end{equation}
Matching the above functional derivatives with the definitions of $W_\Sigma$ and $W_G$, and using the property that  $\Lambda W_\Sigma$ and $\Lambda W_G$ are symmetric, we can bring the above four equations into a compact form
\begin{equation}\label{eq:Ward1}
  (\delta_{y,\lambda}\Pi,\delta_{y,\lambda}\Sigma)^T=W_G(\delta_{y,\lambda}D,\delta_{y,\lambda}G)^T\,,
\end{equation}
\begin{equation}\label{eq:Ward2}
  (W_\Sigma^{-1}-W_G)(\delta_{y,\lambda} D,\delta_{y,\lambda} G)^T=(\delta_{y,\lambda} \sigma_b,\delta_{y,\lambda} \sigma_f)^T\,,
\end{equation} and here the transpose only acts on $b,f$ indices and doesn't act on functions.
\subsubsection{U(1) Ward identity}
Setting $y(x)=x$, we obtain the U(1) Ward identity:
\begin{equation}\label{}
  \delta_\lambda \Sigma=W_G \delta_\lambda G\,,
\end{equation}
\begin{equation}\label{}
  (W_\Sigma^{-1}-W_G)\delta_\lambda G= \delta_\lambda \sigma_f\,.
\end{equation} Here the bosons are not charged under U(1) and therefore dropped.

Using the transformations \eqref{eq:dylG}, \eqref{eq:dylSigma} and \eqref{eq:dylsigmaf}, we can explicitly write down $\delta_\lambda \Sigma$ and $\delta_\lambda G$ in momentum space:
\begin{equation}\label{}
  \delta_\lambda G(k,p)=i\left[G\left(k-\frac{p}{2}\right)-G\left(k+\frac{p}{2}\right)\right]\lambda(p)\,,
\end{equation}
\begin{equation}\label{}
  \delta_\lambda \Sigma(k,p)=i\left[\Sigma\left(k-\frac{p}{2}\right)-\Sigma\left(k+\frac{p}{2}\right)\right]\lambda(p)\,,
\end{equation}
\begin{equation}\label{}
  \delta_\lambda \sigma_f(k,p)=i\left[\sigma_f\left(k-\frac{p}{2}\right)-\sigma_f\left(k+\frac{p}{2}\right)\right]\lambda(p)\,.
\end{equation} Here $\lambda(p)=\int \rd^3 x \lambda(x)e^{-ip\cdot x}$, and $p\cdot x=\vec{p}\cdot\vec{x}-p_0 x_0$. Using $\sigma_f(k)=-ik_0+\varepsilon_k-\mu$, and the vertex functions, we can rewrite $\delta_\lambda \sigma_f$ as
\begin{equation}\label{}
  \delta_\lambda \sigma_f(k,p)=-i\lambda(p) p_\mu \Gamma^\mu(k+p/2,k-p/2)\,,
\end{equation} where $p_\mu=(-p_n,\vec{p})$.

Factoring out $i\lambda(p)$, the U(1) Ward identity then reduces to the statements
\begin{equation}\label{eq:U1Ward1}
  \Sigma\left(k-\frac{p}{2}\right)-\Sigma\left(k+\frac{p}{2}\right)=W_G\left[G\left(k-\frac{p}{2}\right)-G\left(k+\frac{p}{2}\right)\right]\,,
\end{equation}and
\begin{equation}\label{eq:U1Ward2}
  G\left(k-\frac{p}{2}\right)-G\left(k+\frac{p}{2}\right)=\frac{1}{W_\Sigma^{-1}-W_G}\left[-p_\mu \Gamma^\mu\right](k,p)\,.
\end{equation}

The above two Ward identities are easy to check using the saddle point equations. The first identity \eqref{eq:U1Ward1} follows from the fact that $W_G=\delta \Sigma/\delta G$ and that $\Sigma$ is linear in $G$. By using explicit forms of $W_\Sigma$ and $W_G$, the second identity \eqref{eq:U1Ward2} is equivalent to
\begin{equation}\label{}
  \left[\Sigma(k-p/2)+G^{-1}(k-p/2)-\Sigma(k+p/2)-G^{-1}(k+p/2)\right]=p_\mu \Gamma^\mu(k+p/2,k-p/2)\,,
\end{equation}which is trivially satisfied by the vertex functions.

\subsubsection{Density-Density Correlation Function}

  We can use the Ward identity to compute the density-density correlation function at the limit $\vec{p}=0$. Setting $p_\mu=(\Omega_n,0)$, and using $\Gamma^\tau=i$, the Ward identity \eqref{eq:U1Ward2} yields
  \begin{equation}\label{eq:U1Ward_p=0}
    \frac{1}{W_\Sigma^{-1}-W_G}[1](r,p)=\frac{1}{i\Omega_n}\left[G\left(ir_n-i\Omega_n/2,\vec{r}\right)-G\left(ir_n+i\Omega_n/2,\vec{r}\right)\right]\,,
  \end{equation} therefore
  \begin{equation}\label{eq:Pi00}
    \Pi^{00}(i\Omega_n,\vec{p}=0)=T\sum_{r_n}\int \frac{\rd^2\vec{r}}{(2\pi)^2}\frac{1}{i\Omega_n}\left[G\left(ir_n-i\Omega_n/2,\vec{r}\right)-G\left(ir_n+i\Omega_n/2,\vec{r}\right)\right]=0\,,
  \end{equation}which agrees with \cite{Kim94}. A corollary of this result is that in a patch theory, the conductivity vanishes. This is because $\varepsilon_k=\pm k_x+k_y^2$ implies $\Gamma^x=\pm 1$ and $\Delta^x=0$, and therefore $\Pi^{xx}$ is proportional to $\Pi^{00}$.

\subsubsection{Diffeomorphism Ward identity}

Now we want to derive the Ward identity for translation symmetry, by setting $\lambda=0$.
Let $y^{\mu}=x^\mu+\epsilon^\mu$, we can compute:
\begin{equation}\label{eq:deltay1}
  \delta_{y,\lambda=0} G=-\left(\Delta \partial_\mu \epsilon^\mu(x_1)+\Delta \partial_\mu \epsilon^\mu(x_2)+\epsilon^\mu(x_1)\partial_{x_1^\mu}+\epsilon^\mu(x_2)\partial_{x_2^\mu}\right)G(x_1,x_2)\,,
\end{equation}
\begin{equation}\label{}
  \delta_{y,\lambda=0} \Sigma=-\left((1-\Delta) \partial_\mu \epsilon^\mu(x_1)+(1-\Delta) \partial_\mu \epsilon^\mu(x_2)+\epsilon^\mu(x_1)\partial_{x_1^\mu}+\epsilon^\mu(x_2)\partial_{x_2^\mu}\right)\Sigma(x_1,x_2)\,,
\end{equation}
\begin{equation}\label{}
  \delta_{y,\lambda=0} D=-\left((1-2\Delta) \partial_\mu \epsilon^\mu(x_1)+(1-2\Delta) \partial_\mu \epsilon^\mu(x_2)+\epsilon^\mu(x_1)\partial_{x_1^\mu}+\epsilon^\mu(x_2)\partial_{x_2^\mu}\right)D(x_1,x_2)\,,
\end{equation}
\begin{equation}\label{}
  \delta_{y,\lambda=0} \Pi=-\left(2\Delta \partial_\mu \epsilon^\mu(x_1)+2\Delta \partial_\mu \epsilon^\mu(x_2)+\epsilon^\mu(x_1)\partial_{x_1^\mu}+\epsilon^\mu(x_2)\partial_{x_2^\mu}\right)\Pi(x_1,x_2)\,,
\end{equation}
\begin{equation}\label{}
  \delta_{y,\lambda=0} \sigma_f=-\left(\left(1-\Delta\right)\partial_\mu \epsilon^\mu(x_1)+(1-\Delta)\partial_\mu \epsilon^\mu(x_2)+\epsilon^\mu(x_1)\partial_{x_1^\mu}+\epsilon^\mu(x_2)\partial_{x_2^\mu}\right)\sigma_f(x_1,x_2)\,,
\end{equation}
\begin{equation}\label{eq:deltay6}
  \delta_{y,\lambda=0} \sigma_b=-\left(2\Delta \partial_\mu \epsilon^\mu(x_1)+2\Delta \partial_\mu \epsilon^\mu(x_2)+\epsilon^\mu(x_1)\partial_{x_1^\mu}+\epsilon^\mu(x_2)\partial_{x_2^\mu}\right)\sigma_b(x_1,x_2)\,.
\end{equation}

 Since the choice of $\Delta$ is arbitrary, we expect all terms proportional to $\Delta$ to cancel identically in the master Ward identity \eqref{eq:MasterWard1}. This cancellation involves an extra ingredient, which is the UV regularization of the determinant terms \cite{Gu:2019jub}: $\det(\sigma_f+\Sigma)\to \det(\sigma_f+\Sigma)/\det(\sigma_f)$, $\det(\sigma_b+\Pi)\to \det(\sigma_b+\Pi)/\det(\sigma_b)$. After using this regularization, the cancellation of $\Delta$ terms becomes manifest. This regularization term is unimportant for the derived Ward identities \eqref{eq:Ward1} and \eqref{eq:Ward2} because they are obtained from functional derivatives of \eqref{eq:MasterWard1} with respect to bi-local fields, but the regularization term is independent of the fields.

We can rewrite the above infinitesimal transformations in fourier space as
\begin{equation}\label{eq:dyA}
\begin{split}
  &\delta_{y,\lambda=0} A(k,p)=-ip_\mu \epsilon^\mu(p)\left(\Delta_{A}-\frac{1}{2}\right)\left[A\left(k-\frac{p}{2}\right)+A\left(k+\frac{p}{2}\right)\right]-ik_\mu \epsilon^\mu(p)\left[A\left(k-\frac{p}{2}\right)-A\left(k+\frac{p}{2}\right)\right]\\
  &=-ip_\mu\epsilon^\mu \Delta_A\left[A\left(k-\frac{p}{2}\right)+A\left(k+\frac{p}{2}\right)\right]-i\epsilon^\mu(p)\left(k-\frac{p}{2}\right)_\mu A\left(k-\frac{p}{2}\right)+i \epsilon^\mu(p)\left(k+\frac{p}{2}\right)_\mu A\left(k+\frac{p}{2}\right)
\end{split}
\end{equation} Here $p_\mu=\eta_{\mu\nu}p^\nu$, with $\eta_{\mu\nu}=(-,+,+)$. $k$ denotes relative momentum and $p$ denotes CoM momentum. $A=G,\Sigma,D,\Pi,\sigma_f,\sigma_b$ and $\Delta_G$ denotes the corresponding value of $\Delta$ appeared above.

 The two Ward identities \eqref{eq:Ward1} and \eqref{eq:Ward2} with diffeomorphism can also be verified by using the saddle point equations.

 The Noether theorem states that
 \begin{equation}\label{}
   \delta_y S=-\int \rd^3 x T^{\mu\nu} \partial_\mu \varepsilon_\nu(x)\,,
 \end{equation} where $T^{\mu\nu}$ is the stress tensor. We are interested in the consequences of momentum conservation ($T^{0i}$) at the transport limit, therefore we set $\varepsilon^0=0$ and $p^\mu=(p_n,0)$ in \eqref{eq:dyA}. Applying this to $\delta_{y}\sigma_f$ and $\delta_y \sigma_b$, we can read out the momentum vertices:
 \begin{equation}\label{}
   \delta_{y,\lambda=0}\sigma_f(k,p)=i\Gamma^{\mu}(k+p/2,k-p/2)k_{\nu}p_\mu\varepsilon_\nu\,,
 \end{equation}
 \begin{equation}\label{}
   \delta_{y,\lambda=0}\sigma_b(k,p)=i\tilde{\Gamma}^{\mu}(k+p/2,k-p/2)k_\nu p_\mu\varepsilon_\nu\,,
 \end{equation} where $\Gamma^\mu$ is the electron current vertex and $\tilde{\Gamma}^\mu=(k_n,-\vec{k})$. The momentum vertices are therefore read out to be $\Gamma^0 k_i$ and $\tilde{\Gamma}^0 k_i$.

\subsection{Solving the saddle point}
We now solve the saddle point equations on the whole FS. We work in the units where the boson velocity $\sqrt{K}=1$. The boson self energy is
\begin{equation}\label{}
  \Pi(i\Omega_n,\vec{q})=-g^2 T \sum_{\omega_n}\int \frac{\rd^2 \vec{k}}{(2\pi)^2}\frac{1}{i\omega_n-\xi_k-\Sigma(i\omega_n)}\frac{1}{i\omega_n+i\Omega_n-\xi_{k+q}-\Sigma(i\omega_n+i\Omega_n)}\,.
\end{equation}We expand the dispersion with $\xi_{k+q}=\xi_k+v_F q\cos\theta_{kq}$\, and then we can perform the integral over $\theta_{kq}$ and $\xi_k$ to obtain
\begin{equation}\label{}
  \Pi(i\Omega_n,\vec{q})=\pi\calN g^2 T \sum_{\omega_n}\frac{\sgn \omega_n\left(\sgn(\omega_n+\Omega_n)-\sgn \omega_n\right)}{\sqrt{v_F^2 q^2-(i\Omega_n-\Sigma(i\omega_n+i\Omega_n)+\Sigma(i\omega_n))^2}}\,,
\end{equation} where $\calN=\frac{m}{2\pi}$ is the fermion DoS. In the denominator, only the $v_F q$ term is relevant, and we get
\begin{equation}\label{}
  \Pi(i\Omega_n,q)=-\gamma \frac{|\Omega_n|}{q}\,,\quad \gamma=\frac{\calN g^2}{v_F}\,.
\end{equation} As a sanity check, we compare with patch theory where $m=1/2$ and $v_F=1$, we get $\gamma=g^2/(4\pi)$ which agrees with two-patch theory in I. At zero Mastsubara frequency, we also need to include a thermal mass term in the boson propagator
\begin{equation}\label{}
  D(0,q)=\frac{1}{\vec{q}^2+\Delta(T)^2}\,,
\end{equation} where $\Delta(T)^2\sim T\ln(1/T)$ \cite{millis,Hartnoll:2014gba,Esterlis:2021eth}.

The electron self energy $\Sigma=\Sigma_Q+\Sigma_T$ can be decomposed into a quantum part $\Sigma_Q\propto |\omega|^{2/3}$ and a thermal part $\Sigma_T\propto T^{1/2}$. The quantum part is
\begin{equation}\label{}
  \Sigma_Q(i\omega_n,k)=g^2 \int \frac{\rd^2 q}{(2\pi)^2}T \sum_{\Omega_n\neq 0}\frac{1}{q^2+\gamma\frac{|\Omega_n|}{q}}\frac{1}{i\omega_n-i\Omega_n-\xi_{k-q}-\Sigma(i\omega_n-i\Omega_n)}\,.
\end{equation} We expand $\xi_{k-q}=\xi_k-q v_F \cos\theta_q$, and then integrate over $\theta_q$ to get
\begin{equation}\label{}
  \Sigma_Q(i\omega_n,k)=g^2 T \int_0^{\infty} \frac{q\rd q}{2\pi} \sum_{\Omega_n\neq 0} \frac{1}{q^2+\gamma\frac{|\Omega_n|}{q}}\frac{i\sgn(\Omega_n-\omega_n)}{\sqrt{(v_F q)^2+A(\omega_n)^2}}\,,
\end{equation} where $A(\omega_n)=\omega_n+i\Sigma(\omega_n)$. We now evaluate the $q$ integral. Due to the boson propagator, the typical value of $q$ is of order $|\Omega_n|^{1/3}$, which  is larger than $A(\omega_n)$ in the scaling sense. Therefore we can drop $A(\omega_n)$ in the second factor and obtain
\begin{comment}
Next, we expand $\xi_{k+q}=\xi_k+q_{\parallel}v_F+\frac{q_\perp^2}{2m}$. To perform the integral over $q$, we can ignore the $q_\parallel$ dependence in the first factor of the integrand. This approximation is valid in the scaling sense: The second factor of the integrand has a peak in $q_\parallel$ with width of the order $\Im \Sigma_T\sim T^{1/2}$, but in the first factor the width is of order $|\Omega|^{1/3}\sim T^{1/3}$. Therefore the $q_\parallel$ dependence in the first factor is slower and can be ignored.
 In summary, we can replace $q$ by $|q_\perp|$ in the first factor.
\end{comment}
\begin{equation}\label{eq:SigmaQ_val}
\begin{split}
  \Sigma_Q(i\omega_n,k)&=\frac{ig^2}{v_F} T\sum_{\Omega_n}\int_0^{\infty} \frac{\rd q}{2\pi}\frac{\sgn(\Omega_n-\omega_n)}{q^2+\gamma\frac{|\Omega_n|}{q}}\\
  &=\frac{i g^2}{3\sqrt{3}v_F \gamma^{1/3}}T\sum_{\Omega_n}\frac{\sgn(\Omega_n-\omega_n)}{|\Omega_n|^{1/3}}\\
  &=-\frac{i 2^{2/3} g^2 T^{2/3} \sgn(\omega_n)}{3 \sqrt{3}
   \pi^{1/3} \gamma^{1/3} v_F}H_{1/3}\left(\frac{|\omega_n|}{2\pi T}-\frac{1}{2}\right)\\
   &=-\frac{i g^2}{2\sqrt{3}\pi v_F \gamma^{1/3}}\sgn(\omega_n)|\omega_n|^{2/3}\,,\quad (T=0)\,.
\end{split}
\end{equation} The above result also agrees with two-patch theory in \cite{Esterlis:2021eth} when $\gamma=g^2/(4\pi)\,,v_F=1$. Here $H_{1/3}(x)$ is $\verb|HarmonicNumber[x,1/3]|$ in Mathematica.

The thermal part of the self-energy is
        \begin{equation}\label{}
          \Sigma_T(i\omega_n,k)=g^2 T \int \frac{\rd^2 q}{(2\pi)^2}\frac{1}{q^2+\Delta(T)^2}\frac{1}{i \omega_n-\xi_{k-q}-\Sigma_Q(i\omega_n)-\Sigma_T(i\omega_n)}.
        \end{equation} Evaluating the $q$ integral, we obtain
        \begin{equation}\label{}
          \Sigma_T(i\omega_n)=-i\sgn \omega_n \frac{g^2 T}{2\pi}\frac{\sec^{-1}\left(\frac{v_F \Delta(T)}{|A(\omega_n)|}\right)}{\sqrt{v_F^2 \Delta(T)^2-A(\omega_n)^2}}\,,
        \end{equation}where $A(\omega_n)=\omega_n+i\Sigma_Q(i\omega_n)+i\Sigma_T(i\omega_n)$.

        At the low-frequency limit $|\omega_n+i\Sigma_Q(\omega_n)|\ll \Delta(T)$, we obtain
        \begin{equation}\label{eq:SigmaT_val}
          \Sigma_T(i\omega_n)=-i\sgn \omega_n  h(T)\,,
        \end{equation}where $h(T)$ satisfies
        \begin{equation}\label{}
          h(T)=\frac{g^2 T}{2\pi}\frac{\cos^{-1}\left(\frac{h(T)}{v_F \Delta(T)}\right)}{\sqrt{v_F^2 \Delta(T)^2-h(T)^2}}\,.
        \end{equation} Since $\Delta(T)^2/T\to \infty$ as $T\to 0$, the asymptotic behavior of $h(T)$ is
        \begin{equation}\label{}
          h(T)\to \frac{g^2 T}{4 v_F \Delta(T)}\underbrace{\frac{1}{1+\frac{g^2 T}{2\pi v_F^2 \Delta(T)^2}}}_{\approx 1}\,.
        \end{equation}

\subsection{Conductivity Computation}
\label{sec:conductivity}

We will work at zero temperature $T=0$.

\subsubsection{Prange-Kadanoff Reduction}\label{sec:PK}

 The saddle point computation above is consistent with a reduction method proposed by Prange and Kadanoff \cite{PrangeKadanoff}. It assumes that the fermionic spectral function $A(\omega,\vec{k})$  has a sharp peak in $\xi_k$ at the Fermi surface, and doesn't require a well-defined quasiparticle peak in $\omega$. Therefore as an approximation, we could restrict all fermionic momenta to be exactly on the FS, and work with the angular variables. For application to our problem, there is an additional validity requirement\footnote{In the original paper of Prange and Kadanoff \cite{PrangeKadanoff}, they were considering phonons with energy comparable to Debye frequency. There the phonon propagator is controlled by the bare dispersion and can be considered as a smooth function. In our model the boson momentum is small and the Landau damping term plays an important role.}: the typical peak in the boson propagator (as a function of momentum $q$) should be much wider than the peak in the fermion propagator (as a function of $\xi_k\sim v_F q$), i.e.
 \begin{equation}\label{eq:PKcond}
 v_F|\Im \Pi_R(\omega)|^{1/2}\gg |\Im \Sigma_R(\omega)|\,.
 \end{equation} The exponent 1/2 on the LHS is due to the fact that boson momentum appears in the propagator as $(q^2+\Pi)^{-1}$, and therefore the typical width in boson momentum is order $|\Im \Pi_R|^{1/2}$.

 The condition $\eqref{eq:PKcond}$ implies a description using only fermions on the Fermi surface: For any boson carrying momentum $q_\parallel$ normal to the Fermi surface, it will excite a fermion with energy $\xi_k\sim v_F q_\parallel$. This energy is much larger than the width determined by fermion self energy and that process has a much smaller amplitude due to small fermionic spectral weight. As a consequence, we only consider bosons that connect fermions on the Fermi surface. When the two fermion momenta are close, it also implies that the boson momentum is tangent to the Fermi surface --- a feature also seen in patch theories.

 The condition \eqref{eq:PKcond} is indeed satisfied by the clean model we are considering: the fermion self energy is of order $\Im\Sigma\sim \max(|\omega|^{2/3},T^{1/2}/\ln(1/T))$, and the boson self energy is $\Im\Pi(\omega\neq 0)\sim |\omega|/q\sim |\omega|^{2/3}$ and $\Im\Pi(\omega=0)\sim \Delta(T)^2\sim T\ln(1/T)$.   However, this condition is violated when we add disorder potential to the fermions, and therefore the method only applies to the translational invariant model.

  We now apply the reduction idea to conductivity computation. We are interested in optical conductivity and we work at $T=0$.  We  compute the paramagnetic term \eqref{eq:PiAxx13} of the polarization function:
\begin{equation}\label{}
  \Pi^{xx}(i\Omega_n,\vec{p}=0)/N=(\Gamma^x)^T \frac{1}{W_\Sigma^{-1}-W_G} \Gamma^{x}\,,
\end{equation} where we have assumed zero CoM momentum and a finite CoM frequency $\Omega_n>0$. The diamagnetic term exactly cancels the contribution of the paramagnetic term at zero frequency, so the conductivity is
\begin{equation}\label{eq:sgimaxx_subtract}
  \sigma_{xx}(\omega)=\left.\frac{\Pi^{xx}(i\Omega_n)-\Pi^{xx}(0)}{\Omega_n}\right\vert_{i\Omega_n\to \omega+i0}\,.
\end{equation}

    Near the Fermi surface, we can approximate the vertex function to be $\Gamma^{x}(k,k)=v_F \cos\theta_k$, which only contains first harmonics of $\theta_k$. We can write $\Pi^{xx}$ as an inner product of the form
\begin{equation}\label{eq:Pixx_cos}
  \Pi^{xx}(i\Omega_n)/N= v_F^2\braket{\cos\theta_k|\frac{1}{W_\Sigma^{-1}-W_G}|\cos\theta_k}\,.
\end{equation} Here the inner product is defined as
\begin{equation}\label{}
  \braket{f|g}=\int\frac{\rd \omega}{2\pi} \frac{\rd ^2 \vec{k}}{(2\pi)^2}f(\vec{k},i\omega) g(\vec{k},i\omega)\,,
\end{equation} and $\ket{\cos\theta_k}$ denotes the constant function $\cos\theta_k$.

\begin{figure}
  \centering
  \begin{subfigure}[c]{0.3\columnwidth}
  \centering
  \includegraphics[width=\textwidth]{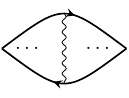}
  \caption{}
  \end{subfigure}
  \begin{subfigure}[c]{0.3\columnwidth}
  \centering
  \includegraphics[width=\textwidth]{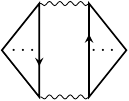}
  \caption{}
  \end{subfigure}
  \begin{subfigure}[c]{0.3\columnwidth}
  \centering
  \includegraphics[width=\textwidth]{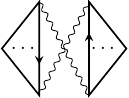}
  \caption{}
  \end{subfigure}
  \caption{\label{fig:diagrams} Feynman diagrams blocks $W_\text{MT}$ and $W_\text{AL}$ for conductivity computation in Eq.\eqref{eq:partial_inverse}. (a) is Maki-Thompson (MT). (b),(c) are Aslamazov-Larkin (AL) diagrams. %The dots mean these diagrams can be combined and repeated.
  }
\end{figure}

 Notice that $W_\Sigma$ and $W_G$ are block operators as given in \eqref{eq:Wsigmagraph} and \eqref{eq:WGgraph}, and we are only interested in the fermionic sector, we can perform a block inversion to obtain
 \begin{equation}\label{eq:partial_inverse}
      \left(\frac{1}{W_\Sigma^{-1}-W_G}\right)_{FF}=\frac{1}{W_{\Sigma,FF}^{-1}-\underbrace{W_{G,FF}}_{W_\text{MT}}-\underbrace{W_{G,FB}W_{\Sigma,BB}W_{G,BF}}_{W_\text{AL}}}\,.
    \end{equation} Here the additional subscripts refer to boson/fermion blocks of $W_\Sigma,W_G$. The two terms that emerge from the block inversion can be interpreted as Maki-Thompson (MT) diagrams and Aslamazov-Larkin (AL) diagrams. The diagrammatic representation of $W_\text{MT}$ and $W_\text{AL}$ are given in Fig.~\ref{fig:diagrams}.

\subsubsection{Maki-Thompson Diagrams}

  We apply the Prange-Kadanoff reduction to the MT diagram kernel $W_\text{MT}$, which is given by
\begin{equation}\label{eq:WMT}
  W_\text{MT}[F](\omega,\vec{k})=g^2 \int \frac{\rd \omega'}{2\pi}\frac{\rd^2 \vec{k'}}{(2\pi)^2}D(k-k')F(\omega',\vec{k}')\,.
\end{equation} We factorize the momentum integral as
\begin{equation}\label{}
  \int \frac{\rd^2\vec{k}'}{2\pi'}=\calN\int \rd\theta'\int\frac{\rd \xi_{k'}}{2\pi}\,,
\end{equation} where the density of state is $\calN=k_F/(2\pi v_F)$. Assuming the function $F$ is sharply peaked on the FS $\xi_{k'}=0$, we perform the integral over $\xi_{k'}$ first, obtaining
\begin{equation}\label{}
  \hat{F}(\omega',\theta')=\int \frac{\rd \xi_{k'}}{2\pi} F(\omega',\vec{k}')\,,
\end{equation} and other factors in \eqref{eq:WMT} are assumed to have a smooth dependence on $\xi_{k'}$, and are evaluated at $\xi_{k'}=0$. In later steps we will also integrate over $\xi_k$, and therefore we can assume $\vec{k}$ is also on the Fermi surface, we get
\begin{equation}\label{}
  W_\text{MT}[\hat{F}](\omega,\theta,\xi_k=0)=\calN g^2 \int \frac{\rd \omega'\rd \theta'}{2\pi}\frac{1}{|\vec{q}|^2+\frac{\gamma |\omega-\omega'|}{|\vec{q}|}}\hat{F}(\omega',\theta')\,,
\end{equation}where the boson momentum $\vec{q}=k_F(\hat{\theta}-\hat{\theta}')$ and $\hat{\theta}$,$\hat{\theta}'$ are unit vectors corresponding to angles $\theta,\theta'$ respectively. To carry out the $\theta'$ integral, we use a gradient expansion. Let $\theta'=\theta+\delta\theta$, and expand $\hat{F}(\omega',\theta')=\hat{F}(\omega',\theta)+\delta\theta\partial_\theta\hat{F}(\omega',\theta)+\frac{1}{2}\delta\theta^2\partial_\theta^2\hat{F}(\omega',\theta)+\dots$. The momentum $\vec{q}$ is parameterized as $|\vec{q}|=2k_F\sin(\delta\theta/2)$. The result is
\begin{equation}\label{eq:WMT2}
  W_\text{MT}[\hat{F}](\omega,\theta,\xi_k=0)=\frac{g^2}{v_F}\frac{2}{3\sqrt{3}}\int \frac{\rd \omega'}{2\pi}\left[\frac{1}{\gamma^{1/3}|\omega-\omega'|^{1/3}}\hat{F}(\omega',\theta)-\frac{\gamma^{1/3}|\omega-\omega'|^{1/3}}{2k_F^2}\partial_\theta^2\hat{F}(\omega',\theta)\right]\,.
\end{equation} Here we have only kept the leading order term in $1/k_F$ for each order of derivative in $\theta$. As we will see later, the first term in the bracket cancels the self energies. In obtaining \eqref{eq:WMT2}, we used dimensional regularization by analytically continuing the following integral
\begin{equation}\label{}
  \int_0^{\infty}\rd q \frac{q^{\eta}}{q^2+\frac{a}{q}}=\frac{\pi}{3} a^{\frac{\eta-1}{3}}\sec\left(\frac{\pi}{6}(2\eta+1)\right)\quad(a>0)\,,
\end{equation} which is only convergent for $-2<\eta<1$ but continued to all $\eta$.
\subsubsection{Aslamazov-Larkin diagram}

    Next we consider the
    \begin{equation}\label{eq:WGAL1}
    \begin{split}
      W_{\text{AL}}[F](k_1)=&-\frac{g^4}{2} \int \frac{\rd^3 q\rd^3 k_2}{(2\pi)^6}\left(G(k_1-q)+G(k_1+q)\right)\left(G(k_2-q)+G(k_2+q)\right)\\
      &\times D(q+p/2)D(q-p/2)F(k_2)\\
      &=-g^4 \int \frac{\rd^3 q\rd^3 k_2}{(2\pi)^6}G(k_1-q)\left(G(k_2-q)+G(k_2+q)\right)D(q+p/2)D(q-p/2)\\
      &\times F(k_2)\,,
    \end{split}
    \end{equation} where $p=(\Omega_n,0)$ denotes the CoM frequency. We first perform the Prange-Kadanoff reduction. We rewrite \eqref{eq:WGAL1} as ($\nu$ is the frequency component of $q$)
\begin{equation}\label{}
\begin{split}
  &W_{\text{AL}}[F](\omega_1,\vec{k}_1)=-g^4\int\frac{\rd^3 q\rd^3 k_2\rd^2\vec{k}'\rd^2\vec{k}''}{(2\pi)^6}G(\omega_1-\nu,\vec{k}')\delta(\vec{q}=\vec{k}_1-\vec{k}')\\
  &\times\left(G(\omega_2-\nu,\vec{k''})\delta(\vec{q}=\vec{k}_2-\vec{k}'')+G(\omega_2+\nu,\vec{k}'')\delta(\vec{q}=\vec{k}''-\vec{k}_2)\right)\\
 & \times D(q+p/2)D(q-p/2)F(\omega_2,\vec{k}_2)\,.
\end{split}
\end{equation}Next, we perform integrals over  $\xi_{k_2}$, $\xi_{k'}$ and $\xi_{k''}$ assuming other terms in the integrand are slow varying, and we effectively restrict all fermionic momenta to be on the FS, parameterized by angles $\theta_1,\theta_2,\theta',\theta''$. The momentum delta functions then impose the following conditions on the angles:
\begin{equation}\label{}
\begin{split}
  (\theta_2,\theta'')&=(\theta_1,\theta')~\hbox{or}~(\theta'+\pi,\theta_1+\pi)  \quad\hbox{if}\quad\vec{q}=\vec{k}_1-\vec{k}'=\vec{k}_2-\vec{k}'';\\
  (\theta'',\theta_2)&=(\theta_1,\theta')~\hbox{or}~(\theta'+\pi,\theta_1+\pi)  \quad\hbox{if}\quad\vec{q}=\vec{k}_1-\vec{k}'=\vec{k}''-\vec{k}_2\,.
\end{split}
\end{equation}We can therefore integrate out $\theta_2$ and $\theta''$, yielding
\begin{equation}\label{}
\begin{split}
  &W_{\text{AL}}[F](\omega_1,\theta_1,\xi_{k_1}=0)=\pi^2 g^4 \calN^3\int\frac{\rd \nu}{2\pi}\frac{\rd \omega_2}{2\pi}\rd\theta' \frac{1}{q k_F}D(q+p/2)D(q-p/2)\sgn(\omega_1-\nu)\\
  &\times\left[\sgn(\omega_2-\nu)\left(\hat{F}(\omega_2,\theta_1)+\hat{F}(\omega_2,\theta'+\pi)\right)+\sgn(\omega_2+\nu)\left(\hat{F}(\omega_2,\theta')+\hat{F}(\omega_2,\theta_1+\pi)\right)\right]\,.
\end{split}
\end{equation}Here the momentum $\vec{q}=k_F(\hat{\theta}_1-\hat{\theta}')$. To proceed, we should assume that the function $\hat{F}$ has a definite parity $P=\pm 1$ under inversion: $\hat{F}(\theta+\pi)=P \hat{F}(\theta)$. We obtain
\begin{equation}\label{eq:WAL2P}
\begin{split}
  &W_{\text{AL}}[F](\omega_1,\theta_1,\xi_{k_1}=0)=\frac{\pi^2}{2} g^4 \calN^3\int\frac{\rd \nu}{2\pi}\frac{\rd \omega_2}{2\pi}\rd\theta' \frac{1}{q k_F}D(q+p/2)D(q-p/2)\\
  &\times\left(\sgn(\omega_1-\nu)+P\sgn(\omega_1+\nu)\right)\left(\sgn(\omega_2-\nu)+P\sgn(\omega_2+\nu)\right)\left(\hat{F}(\omega_2,\theta_1)+P\hat{F}(\omega_2,\theta')\right)\,.
  %&\times\left[\sgn(\omega_2-\nu)\left(\hat{F}(\omega_2,\theta_1)+\hat{F}(\omega_2,\theta'+\pi)\right)+\sgn(\omega_2+\nu)\left(\hat{F}(\omega_2,\theta')+\hat{F}(\omega_2,\theta_1+\pi)\right)\right]\,.
\end{split}
\end{equation} For computation of conductivity, we are interested in odd parity modes and we set $P=-1$ from now on. Performing gradient expansion in $\theta$, we get
\begin{equation}\label{eq:WAL2}
  W_\text{AL}^{P=-1}[F](\omega_1,\theta_1,\xi_{k_1}=0)=\frac{-g^4}{6\sqrt{3}k_F v_F^3 \gamma^{2/3}}\int_{|\nu|>|\omega_1|,|\nu|>|\omega_2|}\frac{\rd \nu\rd \omega_2}{(2\pi)^2}\frac{|\nu+\Omega/2|^{1/3}-|\nu-\Omega/2|^{1/3}}{|\nu+\Omega/2|-|\nu-\Omega/2|}\partial_\theta^2 \hat{F}(\omega_2,\theta_1)\,.
\end{equation}

\subsubsection{Resummation}

  In this part we include the effects of $W_\Sigma$ in \eqref{eq:Pixx_cos} and \eqref{eq:partial_inverse}. We expand the geometric series to write
\begin{equation}\label{}
  \frac{1}{W_\Sigma^{-1}-W_{\text{MT+AL}}}=W_\Sigma+W_\Sigma W_\text{MT+AL}W_\Sigma+\dots\,.
\end{equation} In analyzing $W_\text{MT}$ and $W_\text{AL}$ in previous sections, we have assumed that they act on functions of $\xi_k$ which are sharply peaked on the Fermi surface. This assumption is justified by noting that the $W_\Sigma$ factor as a product of two fermion Green's functions which is indeed peaked on the Fermi surface. Therefore, we have
\begin{equation}\label{}
  \int \frac{\rd \xi_k}{2\pi} W_\Sigma[F](\xi_k,\omega,\theta_k)\approx L(i\omega)\theta(\Omega/2-|\omega|) F(\omega,\theta_k,\xi_k=0)\,,
\end{equation} where $\theta(\Omega/2-|\omega|)$ is the Heaviside theta function and
\begin{equation}\label{eq:def_L}
   \begin{split}
     L(i\omega)&=\int \frac{\rd \xi_k}{2\pi} W_\Sigma(\xi_k,\omega)=\int \frac{\rd \xi_k}{2\pi} G(i(\omega+\Omega/2),\xi_k) G(i(\omega-\Omega/2),\xi_k)\\
     &=\frac{i}{2} \frac{\sgn(\omega+\Omega/2)-\sgn(\omega-\Omega/2)}{i\Omega+\Sigma(i(\omega-\Omega/2))-\Sigma(i(\omega+\Omega/2))}=\frac{1}{\Omega-i\Sigma(i(\omega-\Omega/2))+i\Sigma(i(\omega+\Omega/2))}\,.
   \end{split}
   \end{equation} We see that the effect of $W_\Sigma$ is to restrict the functional space to be supported only on $[-\Omega/2,\Omega/2]$. We arrive at the following new inner product formula for $\Pi^{xx}$, which is over functions of angle $\theta$ and frequency $\omega$ ($|\omega|\leq \Omega/2$):
\begin{equation}\label{eq:Pixx_cos2}
  \Pi^{xx}(i\Omega)/N=v_F^2 \braket{\cos\theta\Vert\frac{1}{L^{-1}-W_\text{MT}-W_\text{AL}}\Vert\cos\theta}\,,
\end{equation} with a reduced inner product
\begin{equation}\label{}
  \braket{f\Vert g}=\calN\int_0^{2\pi} \rd \theta \int_{-\Omega/2}^{\Omega/2}\rd \omega f(i\omega,\theta)g(i\omega,\theta)\,.
\end{equation} The operator $L$ is defined by Eq.~\eqref{eq:def_L}, and $W_\text{MT}$ and $W_\text{AL}$ are given by Eqs.~\eqref{eq:WMT2} and \eqref{eq:WAL2} respectively, understood as functionals acting on $\hat{F}$ instead of $F$.

The vertex function $f(\theta_k)=\cos\theta_k$ appearing in \eqref{eq:Pixx_cos2} is frequency independent, allowing us to compute its image under $W_\text{MT}$ and $W_\text{AL}$ explicitly:
\begin{equation}\label{eq:WMT3}
\begin{split}
  W_\text{MT}[f](\omega,\theta)&=\frac{g^2}{v_F}\frac{2}{3\sqrt{3}}\int_{-\Omega/2}^{\Omega/2} \frac{\rd \omega'}{2\pi}\left[\frac{1}{\gamma^{1/3}|\omega-\omega'|^{1/3}}f(\theta)-\frac{\gamma^{1/3}|\omega-\omega'|^{1/3}}{2k_F^2}\partial_\theta^2f(\theta)\right]\\
  &=\left[i\Sigma(i(\omega+\Omega/2))-i\Sigma(i(\omega-\Omega/2))\right]f(\theta)\\
  &\quad-\frac{g^2 \gamma^{1/3}}{8\pi\sqrt{3}v_F k_F^2}\left[\sgn(\omega+\Omega/2)|\omega+\Omega/2|^{4/3}-\sgn(\omega-\Omega/2)|\omega-\Omega/2|^{4/3}\right]\partial_\theta^2 f(\theta)\,.
\end{split}
\end{equation}
\begin{equation}\label{eq:WAL3}
\begin{split}
  W_\text{AL}^{P=-1}[f](\omega,\theta)&=\frac{-g^4}{6\sqrt{3}k_F v_F^3 \gamma^{2/3}}\int_{|\nu|>|\omega|,|\nu|>|\omega'|,|\omega'|<\Omega/2}\frac{\rd \nu\rd \omega'}{(2\pi)^2}\frac{|\nu+\Omega/2|^{1/3}-|\nu-\Omega/2|^{1/3}}{|\nu+\Omega/2|-|\nu-\Omega/2|}\partial_\theta^2 f(\theta)\,.
  \\&=\frac{g^4}{16\pi^2\sqrt{3}k_F v_F^3 \gamma^{2/3}}\left[\sgn(\omega+\Omega/2)|\omega+\Omega/2|^{4/3}-\sgn(\omega-\Omega/2)|\omega-\Omega/2|^{4/3}\right]\partial_\theta^2 f(\theta)\,.
\end{split}
\end{equation} In obtaining \eqref{eq:WAL3}, we again used dimensional regularization on the exponents of $|\nu\pm \Omega/2|$ to drop the divergent parts at $\nu\to\pm\infty$.

Using the relation
$
\gamma=\frac{\calN g^2}{v_F}=\frac{k_F g^2}{2\pi v_F^2}\,,
$  we see that the last line of \eqref{eq:WMT3} exactly cancels \eqref{eq:WAL3}, and therefore
\begin{equation}\label{eq:eigenodd}
  \left(L^{-1}-W_\text{MT}-W^{P=-1}_\text{AL}\right)[f]=\Omega f\,.
\end{equation} That is, any odd-parity frequency-independent function $f(\theta)$ is an eigenvector of $L^{-1}-W_\text{MT}-W_\text{AL}$ with eigenvalue $\Omega$.

This implies that the conductivity  of the model is exactly Drude like
\begin{equation}\label{eq:sigmaxx_Drude}
  \sigma^{xx}(\omega)=N\frac{\calN v_F^2}{2}\frac{1}{-i\omega}\,.
\end{equation}

\subsubsection{Discussion}

    \paragraph{Change of Integration Order: }In obtaining the above results, we have exchanged the order of integration between frequency and momentum, which can potentially modify the value of the integral. However, the difference between two integration orders is due to UV divergence at large frequency and momentum. There is exactly one diagram that has this behavior, which is the one-loop bubble of fermions. By examining this diagram, it can be shown that changing the integration order just cancels the diamagnetic term \eqref{eq:PiAxx2}.

    \paragraph{Cancellation and Ward Identity}The Drude-like result \eqref{eq:sigmaxx_Drude} is due to two cancellation related symmetries: First, the cancellation between self energies and Maki-Thompson diagrams due to U(1) symmetry and charge conservation. Second, the cancellation between the Aslamazov-Larkin diagram and the remainings of Maki-Thompson diagram is due to diffeomorphism symmetry and momentum conservation. These cancellations can be related to the Ward identities derived in Sec.~\ref{sec:Ward} by Prange-Kadanoff reduction. The almost cancellation between the self energy and the MT diagram is a consequence of the U(1) Ward identity. This can be seen by integrating both sides of Eq.~\eqref{eq:U1Ward_p=0} over $\xi_r$\footnote{For the constant function 1, the AL diagrams vanishes identically as the integral \eqref{eq:WAL2P} is odd in $\omega_2$ when $P=1$.}. The cancellation between the rest of MT diagram and the AL diagram can be seen as the following: Within the Prange-Kadanoff formalism, we only consider momenta exactly on the Fermi surface. Therefore the current vertex function $v_F\cos\theta$ is proportional to the momentum vertex function $k_F\cos\theta$. Because the boson self-interaction is irrelevant at the critical point, there is no boson-boson entry in the kernel $W_G$, and from the Ward identity \eqref{eq:Ward1}, we have
    \begin{equation}\label{}
      \delta_{y}\Pi=W_{G,BF}[\delta_y G]\,.
    \end{equation} Here $\delta_y$ denotes small diffeomorhism transformation as defined in Eqs.\eqref{eq:deltay1}-\eqref{eq:deltay6}. Substitute the above into \eqref{eq:Ward2} and we obtain
    \begin{equation}\label{eq:Ward_diff}
      \left(W_\Sigma^{-1}-W_\text{MT}-W_\text{AL}\right)[\delta_y G]=\delta_y \sigma_f-W_{G,FB} W_\Sigma [\delta_y \sigma_b]\,.
    \end{equation} At the critical point, the bare boson momentum term $\sigma_b$ is also irrelevant compared to the boson self energy $\Pi$, and therefore the second term on the RHS \eqref{eq:Ward_diff} can be dropped. Multiplying $\left(W_\Sigma^{-1}-W_\text{MT}-W_\text{AL}\right)^{-1}$ on both sides and then perform Prange-Kadanoff reduction by integrating over $\xi$, we see that the momentum vertex is exactly an eigenvector of $L^{-1}-W_\text{MT}-W_\text{AL}$ with eigenvalue $\Omega$.

    \paragraph{Slow Relaxation of Odd-Parity Modes}

        We now argue that within the Prange-Kadanoff approximation,
         every odd harmonic $\cos m\theta$ satisfies the eigenvalue equation \eqref{eq:eigenodd} at any order of gradient expansion. As a corollary, \eqref{eq:sigmaxx_Drude} is valid at the critical point regardless of Fermi surface shape, as long as it has inversion symmetry and is convex. This conclusion is in disagreement with Ref.~\cite{ChubukovMaslov0} which assumed that MT and AL diagrams would not cancel.

        Eq.~\eqref{eq:eigenodd} has already been shown at second order in the gradient expansion. What happens at higher order?
         It can be seen that both in $W_\text{MT}$ and $W_\text{AL}$ associated with each $\partial_\theta$ there is a factor of $\delta\theta\approx q/k_F\sim \gamma^{1/3}|\Omega|^{1/3}/k_F$. Therefore the gradient expansion is at the same time a $1/k_F$ expansion (i.e. the series is actually in $(1/k_F)\partial_\theta$). Momentum conservation implies that the series vanishes identically for first harmonics to all orders in $1/k_F$, and therefore it must also vanish to all orders in $\partial_\theta$, given $P=-1$.

         When the Fermi surface is not exactly circular but still inversion symmetric and convex, we can decompose the current vertex into angular harmonics of the momentum angle $\theta_k$, and by inversion symmetry  it only contains odd harmonics. The convexity ensures that the number of solutions to the angular delta functions remains unchanged \cite{ChubukovMaslov} and the derivation to \eqref{eq:eigenodd} continues to hold. Since all odd-harmonics satisfy \eqref{eq:eigenodd}, the result \eqref{eq:sigmaxx_Drude} continues to hold.

         There is a more intuitive way to understand the statement in terms kinematic constraint for fermion collision. What happens in our model is a non-Fermi liquid generalization of a Fermi liquid story \cite{LedwithAoP,LedwithArxiv}. Within the Prange-Kadanoff approximation, we only consider momenta on the Fermi surface scattering onto Fermi surface. Because of momentum conservation and Pauli's exclusion principle, when two initial momenta $(\vec{k}_1,\vec{k}_2)$ are not head-on $(\vec{k}_1+\vec{k}_2\neq 0)$, the only kinematically allowed process is forward scattering or particle exchange. This process doesn't cause any relaxation. When the two initial momenta are head-on, they are allowed to scatter to any head-on pairs. However, this process only relaxes even harmonics of the Fermi surface, because a pair of head-on particles have zero overlap with odd harmonics. This intuitive picture holds for any inversion symmetric Fermi surface.
    \paragraph{Beyond Prange-Kadanoff} According to the Fermi liquid story \cite{ChubukovMaslov,LedwithAoP,LedwithArxiv}, the first correction to the eigenvalue equation \eqref{eq:eigenodd} is a superdiffusion term $\partial_\theta^4$ in the angular coordinate. The superdiffusion term can be understood as a two-particle correlated random walk on the angular coordinate which conserves center of mass coordinate due to momentum conservation. Furthermore, the superdiffusion also intertwines angular and radial relaxation, and it is therefore beyond the Prange-Kadanoff approximation. Following the analysis there, we can estimate the diffusion coefficient to be
    \begin{equation}\label{}
      D\sim\Im \Sigma_R (\delta\theta)^4\sim \frac{g^2 \gamma |\omega|^2}{k_F^4 v_F}\sim \frac{g^4|\omega|^2}{(v_Fk_F)^3}\,.
    \end{equation} This result is accurate up to logarithmic corrections of order $\ln\delta\theta$ \cite{LedwithAoP}. The optical conductivity is therefore
    \begin{equation}\label{eq:sigmaxxD}
      \sigma^{xx}(\omega)\sim \frac{1}{\omega}\braket{\cos\theta|\frac{1}{-i \omega -D \partial_\theta^4}|\cos\theta}\sim\calN v_F^2 \frac{1}{-i\omega-D}\sim\frac{1}{-i\omega}+|\omega|^0\,.
    \end{equation} This requires a non-circular Fermi surface since for first harmonics the correction term still vanishes by momentum conservation.

    At finite temperature, the quantum-critical scaling is violated by thermal fluctuations. However, we expect the angular superdiffusion picture to still hold, but with the angular step $\delta\theta\sim \Delta(T)/k_F$ where $\Delta(T)$ is the thermal mass. Therefore we have (when $\omega=0$)
    \begin{equation}\label{}
      D\sim \Im\Sigma_R (\delta\theta)^4\sim T^{5/2}\ln^{3/2}(1/T)\,.
    \end{equation}

    We should note that the scalings of the diffusion coefficients estimated above are still more singular than a \emph{translational invariant} Fermi liquid. It has been calculated in \cite{LedwithAoP,LedwithArxiv} that in a Fermi liquid where collisions conserve momentum, the diffusion coefficient at DC scales as $T^4 \ln(1/T)$. For optical conductivity, we expect a scaling of $\omega^4\ln(|\omega|)$. In contrast, in a Fermi liquid with disorder, all decay rates are expected to scale as $\omega^2$ or $T^2$.

 \paragraph{A Would-be $|\omega|^{-2/3}$ Optical Conductivity}
The MT and AL diagrams were noted in earlier work \cite{Kim94} but their cancellation was overlooked, as we review in Appendix~\ref{app:Kim}.
If we consider the MT diagram only, our calculation would reproduce the conventional $|\omega|^{-2/3}$ conductivity \cite{Kim94} (which, we maintain, is absent). This can be seen by noting that at lowest order of angular expansion, the MT diagram exactly cancels self-energy contribution (see the first line of \eqref{eq:WMT3}). This is related to the U(1) Ward identity, and can be physically interpreted as forward scattering doesn't contribute to current dissipation. We have obtained an eigenvalue statement $(L^{-1}-W^{(0)}_\text{MT})[f]=\Omega f$ which is valid only at zeroth order of gradient expansion. Effect of small angle scattering is included as a first order gradient expansion (the second line of \eqref{eq:WMT3}), which perturbs the eigenvalue equation above by a term of order $\Omega^{4/3}$, whose leading order effect is to shift the eigenvalue by an amount of order $\Omega^{4/3}$. As a result, we would obtain a Drude formula with scattering rate $\sim\Omega^{4/3}/k_F^2$, and in the $k_F\to\infty$ limit, this turns into a $|\omega|^{-2/3}$ in the conductivity:
        \begin{equation}
            \sigma^{xx}(\omega)=\frac{\calN v_F^2}{2}\frac{1}{-i\omega+\frac{\#}{k_F^2}|\omega|^{4/3}}\sim\frac{\calN v_F^2}{2}\left(\frac{1}{-i\omega}+\frac{\#}{k_F^2}|\omega|^{-2/3}\right)\,.
        \end{equation}

        The above result can also be obtained in the picture of angular diffusion on the Fermi surface. Because the constraints from momentum conservation is ignored, the leading order diffusion process is no longer a correlated diffusion but a single particle diffusion with operator $\partial_\theta^2$. From this the diffusion constant can be estimated as
\begin{equation}
    D\sim \Im\Sigma_R(\omega)\left(\delta\theta\right)^2 \sim |\omega|^{4/3}\,,
\end{equation} which agrees with results above.
\section{Potential disorder}
\label{sec:potential}

In this part we investigate the spatially disordered theory with potential ($v$) disorder, and compute its conductivity.
\subsection{Lagrangian}
The model we consider is
\begin{equation}\label{}
\mathcal{L}=\sum_{i}\psi_i^\dagger (\partial_\tau+\varepsilon_k-\mu)\psi_i+\frac{1}{2}\sum_{i}\phi_i(-\partial_\tau^2+\omega_q^2+m_b^2)\phi_i+\sum_{ijl}\frac{g_{ijl}}{N}\psi_i^\dagger \psi_j \phi_l+\sum_{ij}\frac{v_{ij} (x)}{\sqrt{N}}\psi_i^\dagger \psi_j\,.
\end{equation}Here $\varepsilon_k$ and $\omega_q^2$ should be understood as differential operators. $g_{ijl}$ is the random interaction and $v_{ij}$ is disorder. The averaging procedures of $g_{ijl}$ and $v_{ij}$ are given in Sec.~\ref{sec:summary}. The boson mass term $m_b^2$ might be replaced by a fixed length constraint as in I. We will assume that in the low-energy limit the disorder scattering rate $\Gamma=2\pi\calN v^2$ ($\calN$ is DOS) is the largest scale.
\subsubsection{Scaling Analysis}
  Assuming dynamical exponent $z=2$ for the bosons, we have $[\tau]=-2$, $[x]=[y]=-1$. At the fixed point, we assume the disorder self energy of the fermions and the boson kinetic term are invariant under scaling. We can then determine $[\psi]=2$ and $[\phi]=1$. Therefore the Yukawa coupling and the fermion-disorder coupling are irrelevant. There is also the boson mass term $\phi^2$ which is relevant and the boson self interaction $\phi^4$ which is marginal, but we assume that they have been tuned to criticality.
\subsubsection{$G$-$\Sigma$ action}
After averaging out $g_{ijl}$ and $v_{ij}$, we obtain the $G$-$\Sigma$ action
\begin{equation}\label{eq:S_Gsigma}
\begin{split}
   \frac{S}{N} =& -\ln\det\left(\left(\partial_\tau+\varepsilon_k-\mu\right)\delta(x-x')+\Sigma\right) +\frac{1}{2} \ln\det\left(\left(-\partial_\tau^2+\omega_q^2+m_b^2\right)\delta(x-x')-\Pi\right)  \\
   &-\Tr\left(\Sigma\cdot G\right)+\frac{1}{2}\Tr\left(\Pi\cdot D\right)+\frac{g^2}{2}\Tr\left((GD)\cdot G\right)+\frac{v^2}{2}\Tr((G\bar{\delta})\cdot G)\,,
\end{split}
\end{equation}where $\delta$ is a space-time delta function and $\bar{\delta}$ is a spatial delta function.

The saddle point equations are
\begin{equation}\label{}
\begin{split}
   G(i\omega_n,\vec{k}) &= \frac{1}{i\omega_n+\mu-\varepsilon_k-\Sigma(i\omega_n,\vec{k})}\,, \\
   D(i\Omega_n,\vec{q})  & =\frac{1}{\Omega_m^2+\omega_q^2+m_b^2-\Pi(i\Omega_n,\vec{k})}\,, \\
   \Sigma(x)  & = g^2 G(x)D(x)+v^2 G(x)\bar{\delta}(x)\,,  \\
   \Pi(x)  & =-g^2 G(x)G(-x)\,.
\end{split}
\end{equation}

\subsection{Solving the saddle point}

In this disordered model, the Prange-Kadanoff method does not apply. In the presence of disorder, the disorder contribution to electron self energy $\Sigma_{dis}=-i(\Gamma/2)\sgn\omega_n$ dominates at low energy. As a consequence, the peak in the electron Green's function is now wider than the peak in the boson Green's function (as we will see the boson self energy scales linearly with frequency). Therefore the Prange-Kadanoff method does not apply, and it is not legitimate in the scaling sense to neglect momentum dependence in the electron self energy. However, the momentum dependence only introduces non-dissipative corrections, and for the real part of optical conductivity we are interested in the dissipative part, so to simplify the calculation we can still set fermionic momenta to be on the Fermi surface.

\subsubsection{Boson self energy}

 Let us compute the boson self energy first, which in momentum space reads
 \begin{equation}\label{}
   \Pi(i\Omega_n,\vec{q})=-g^2 T \sum_{\omega_n}\int \frac{\rd ^2 \vec{k}}{(2\pi)^2}\frac{1}{i\omega_n-\xi_{\vec{k}}-\Sigma(i\omega_n)}\frac{1}{i\omega_n+i\Omega_n-\xi_{\vec{k}+\vec{q}}-\Sigma(i\omega_n+i\Omega_n)}\,,
 \end{equation} where we have assumed that the electron self energy takes value on the Fermi surface, and $\xi_{\vec{k}}=\varepsilon_{\vec{k}}-\mu$. Expanding in small $\vec{q}$ and around a circular Fermi surface, we have
 \begin{equation}\label{}
   \Pi(i\Omega_n,\vec{q})=-g^2 T \sum_{\omega_n}\int \frac{\rd \theta}{2\pi}\int\nu\rd \xi_{\vec{k}}\frac{1}{i\omega_n-\xi_{\vec{k}}-\Sigma(i\omega_n)}\frac{1}{i\omega_n+i\Omega_n-\xi_{\vec{k}}-\Sigma(i\omega_n+i\Omega_n)-v_F q\cos\theta}\,.
 \end{equation}
 Taking the $\xi_{\vec{k}}$ integral to be over the real line, we obtain
 \begin{equation}\label{}
 \begin{split}
   \Pi(i\Omega_n,\vec{q})&=-\pi\calN g^2 T\sum_{\omega_n}\frac{\sgn\Omega_n(\sgn(\omega_n+\Omega_n)-\sgn\omega_n)}{\sqrt{v_F^2q^2-\left(i\Omega_n-\Sigma(i\omega_n+i\Omega_n)+\Sigma(i\omega_n)\right)^2}}\\
   &\approx-\frac{\calN g^2|\Omega_n|}{\sqrt{v_F^2 \vec{q}^2+\Gamma^2}}\approx-\frac{\calN g^2|\Omega_n|}{\Gamma}\equiv-\gamma|\Omega_n|
 \end{split}
 \end{equation} Here $\calN$ is density of states.
 Here we have assumed that at low frequencies the electron self energy is dominated by disorder scattering $\Sigma\approx -i\frac{\Gamma}{2}\sgn(\omega_n)$.

 The thermal mass of this boson self energy has been calculated in the previous paper I, which is
 \begin{equation}\label{}
   \Delta(T)^2=\frac{\displaystyle -\pi \gamma T W_0\left(-\frac{1}{\pi}\ln\left(\frac{2\pi T }{\gamma e^{\gamma_E}}\right)\right)}{\ln\left( \displaystyle \frac{2\pi T }{\gamma e^{\gamma_E}}\right)}\,,\quad \gamma=\frac{\calN g^2}{\Gamma}\,.
 \end{equation} Here $\gamma_E=0.577\dots$ is Euler's constant, and $W_0$ is Lambert W-function. Also note in this section the meaning of the parameter $\gamma$ is different from Sec.~\ref{sec:uniform}.

\subsubsection{Electron self energy}

The electron self energy is given by
\begin{equation}\label{}
  \Sigma(i\omega_n,\vec{k})=g^2\int \frac{\rd^2 \vec{q}}{(2\pi)^2}T\sum_{\Omega_n}D(i\Omega_n,\vec{q})G(i\omega_n-i\Omega_n,\vec{k}-\vec{q})+v^2\int \frac{\rd^2\vec{q}}{(2\pi)^2}G(i\omega_n,\vec{q})\,.
\end{equation} The second term gives rise to the disorder contribution
\begin{equation}\label{}
  \Sigma_{dis}(i\omega_n,\vec{k})=-i\frac{\Gamma}{2}\sgn(\omega_n),\quad \Gamma=2\pi v^2 \calN\,.
\end{equation}

The first term can be split into thermal part and quantum part
\begin{equation}\label{}
  \Sigma_T(i\omega_n,\vec{k})=g^2T\int \frac{\rd^2\vec{q}}{(2\pi)^2}D(0,\vec{q})G(i\omega_n,\vec{k}-\vec{q})
\end{equation} Taking $\vec{k}$ to be on the Fermi surface, we can expand $\xi_{\vec{k}-\vec{q}}=v_Fq\cos\theta$, we obtain
\begin{equation}\label{}
  \Sigma_T(i\omega_n,\vec{k})=-\frac{g^2 T}{2\pi}\sgn(\omega_n)\frac{\sec^{-1}\left(\frac{v_F \Delta(T)}{A(\omega_n)}\right)}{\sqrt{A(\omega_n)^2-v_F^2\Delta(T)^2}}\,,
\end{equation}where $A(\omega_n)=\omega_n+i\Sigma(\omega_n)$ and $\Delta(T)$ is the thermal mass. Taking the large $\Gamma$ limit, we obtain
\begin{equation}\label{eq:SigmaT_disordered}
  \Sigma_T(i\omega_n)=-\frac{ig^2 T\sgn\omega_n}{2\pi|A(\omega_n)|}\ln\left|\frac{2A(\omega_n)}{v_F \Delta(T)}\right|= \frac{-i g^2 T \sgn\omega_n}{\pi \Gamma}\ln\left|\frac{\Gamma}{v_F \Delta(T)}\right|
\end{equation}

The quantum part is
\begin{equation}\label{}
  \Sigma_Q(i\omega_n)=g^2\int\frac{\rd^2 \vec{q}}{(2\pi)^2}T\sum_{\Omega_n\neq 0}\frac{1}{\gamma|\Omega_n|+\vec{q}^2}\frac{1}{iA(\omega_n-\Omega_n)-\xi_{\vec{k}-\vec{q}}}\,
\end{equation}   Replace $\xi_{k-q}=v_F q \cos\theta_q$ and perform the angular integral, we obtain
\begin{equation}\label{eq:SigmaQ0}
  \Sigma_Q(i\omega_n)=g^2 T\sum_{\Omega_n\neq 0} \int \frac{q\rd q}{(2\pi)} \frac{1}{q^2 +\gamma|\Omega_n|}\frac{-i\sgn (\omega_n-\Omega_n)}{\sqrt{v_F^2 q^2+A(\omega_n-\Omega_n)^2}}
\end{equation}
Using low frequency and low energy approximations, we perform the frequency sum first and get
\begin{equation}\label{}
  \Sigma_Q(i\omega_n)=\frac{-ig^2\sgn\omega_n}{2\pi^2\gamma}\int_0^\infty\frac{q\rd q}{\sqrt{v_F^2 q^2+\Gamma^2/4}}\left[\psi\left(\frac{|\omega_n|}{2\pi T}+\frac{1}{2}+\frac{q^2}{2\pi T\gamma}\right)-\psi\left(1+\frac{q^2}{2\pi T\gamma}\right)\right]\,.
\end{equation} At zero temperature, the above reduces to
\begin{equation}\label{eq:SigmaQ_disordered_T=0}
\begin{split}
  \Sigma_Q(i\omega)&=\frac{-ig^2\sgn\omega}{2\pi^2\gamma}\int_0^\infty\frac{q\rd q}{\sqrt{v_F^2 q^2+\Gamma^2/4}}\ln\left(1+\frac{|\omega| \gamma}{q^2}\right)\\
  &=\frac{-ig^2\sgn\omega}{2\pi^2\gamma}\left(\frac{- \Gamma}{2 v_F^2}\right)\left(2\sqrt{1-\frac{4 |\omega|\gamma v_F^2}{\Gamma^2}}\sinh^{-1}\left(\sqrt{\frac{\Gamma^2}{4|\omega|\gamma v_F^2}-1}\right)+\ln\left(\frac{|\omega|\gamma v_F^2}{\Gamma^2}\right)\right)\\
  &= \frac{-ig^2 \omega}{2\pi^2 \Gamma}\ln\left(\frac{e \Gamma^2}{|\omega|\gamma v_F^2}\right)+\mathcal{O}(\omega^2)\,.
\end{split}
\end{equation} This logarithmic behavior signatures the break down of Prange-Kadanoff reduction.

\textbf{Alternative calculation of $\Sigma_Q$}:
 In \eqref{eq:SigmaQ0}, we perform the momentum integral over $q$ first:
\begin{equation}\label{}
  \Sigma_Q(i\omega_n)=-\frac{i g^2 T}{2\pi} \sum_{\Omega_n\neq 0}\sgn(\omega_n-\Omega_n)\frac{\cosh^{-1}\left(\frac{|A(\omega_n-\Omega_n)|}{v_F \sqrt{\gamma|\Omega_n|}}\right)}{\sqrt{A^2(\omega_n-\Omega_n)-v_F^2 \gamma|\Omega_n|}}\,.
\end{equation} To evaluate the sum to leading order in $\Gamma$, we replace $A$ by $\Gamma/2$, and we obtain
\begin{equation}\label{eq:SigmaQ_disordered}
\begin{split}
   \Sigma_Q(i\omega_n) &= -\frac{i g^2 T}{\pi}\sgn\omega_n \sum_{0<\Omega_n<|\omega_n|}\frac{2}{\Gamma} \ln\left(\frac{\Gamma}{v_F\sqrt{\gamma|\Omega_n|}}\right) \\
     & =-i\sgn \omega_n \frac{2g^2 T}{\pi\Gamma}\left[\left(\frac{|\omega_n|}{2\pi T}-\frac{1}{2}\right)\ln\frac{\Gamma}{v_F \sqrt{2\pi T\gamma}}-\frac{1}{2}\ln\Gamma_F\left(\frac{|\omega_n|}{2\pi T}+\frac{1}{2}\right)\right]\,,
\end{split}
\end{equation} where the $\Gamma_F$ denotes the gamma function. Taking the $T\to 0$ limit, we recover \eqref{eq:SigmaQ_disordered_T=0}.

Combining \eqref{eq:SigmaT_disordered} and \eqref{eq:SigmaQ_disordered}, we obtain
\begin{equation}\label{eq:SigmaQ+SigmaT}
  \Sigma_Q(i\omega_n)+\Sigma_T(i\omega_n)=-i\sgn \omega_n \frac{2 g^2 T}{\pi\Gamma}\left[\frac{|\omega_n|}{2\pi T}\ln \frac{\Gamma}{v_F\sqrt{2\pi T \gamma}}-\frac{1}{2}\ln \frac{\Delta(T)}{\sqrt{2\pi T \gamma}}-\frac{1}{2}\ln\Gamma_F\left(\frac{|\omega_n|}{2\pi T}+\frac{1}{2}\right)\right]\,.
\end{equation} Including momentum dependence will shift $\Gamma$ to $\Gamma+i\xi_k \sgn\omega_n$, whose primary effect is to introduce a real part to the self energy, which we will ignore.

\subsection{Conductivity in the disordered model}

    Now we calculate the conductivity in the disordered model. The conductivity is given by \eqref{eq:PiAxx13} and \eqref{eq:sgimaxx_subtract}.

    We will have to invert the operator $W_\Sigma^{-1}-W_G$. Since the Prange-Kadanoff method doesn't apply, we will treat disorder scattering exactly and treat fermion-boson scattering perturbatively in $g$. This is justified as $g$ is irrelavent in this $z=2$ theory.

    Let's set up the formalism. Similar to \eqref{eq:partial_inverse}, we integrate out the bosons to write
    \begin{equation}
      \left(\frac{1}{W_\Sigma^{-1}-W_G}\right)_{FF}=\frac{1}{\underbrace{W_{\Sigma,FF}^{-1}}_{W_{\Sigma,0}^{-1}+W_{\Sigma,1}^{-1}}-\underbrace{W_{G,FF}}_{W_{dis}+W_\text{MT}}-\underbrace{W_{G,FB}W_{\Sigma,BB}W_{G,BF}}_{W_\text{AL}}}\,.
    \label{eq:WSWG}
    \end{equation}

    Here $W_{\Sigma,0}$ is a diagonal operator in $k$-space whose expression is
    \begin{equation}\label{}
      W_{\Sigma,0}(k,p)=G_0(k+p/2)G_0(k-p/2)\,,
    \end{equation}where $G_0$ is the Green's function which only includes disorder:
    \begin{equation}\label{}
      G_{0}(i\omega,\vec{k})=\frac{1}{i\omega-\xi_k-\Sigma_{dis}(i\omega)}\,,\quad \Sigma_{dis}(i\omega)=-\frac{i\Gamma}{2}\sgn\omega\,.
    \end{equation} Here $p$ denotes CoM 3-momentum and $k$ denotes relative 3-momentum.

    $W_{\Sigma,1}^{-1}$ is obtained from $W_{\Sigma,0}^{-1}$ by doing first-order expansion in $g^2$:
    \begin{equation}
    \begin{split}
      W_{\Sigma,1}^{-1}(k,p)&=-(\Sigma_T(k+p/2)+\Sigma_Q(k+p/2))G_0^{-1}(k-p/2)\\
      &-(\Sigma_T(k-p/2)+\Sigma_Q(k-p/2))G_0^{-1}(k+p/2)\,.
    \end{split}
    \label{eq:WS}
    \end{equation}

    $W_{dis}$ describes disorder scattering:
    \begin{equation}\label{}
      W_{dis}[F](i\omega,\vec{k})=v^2 \int \frac{\rd^2 \vec{q}}{(2\pi)^2}F(i\omega,\vec{q})\,,
    \end{equation} and in $l$-th angular harmonics, it takes the form
    \begin{equation}\label{}
      W_{dis}^{(l)}[F](i\omega,\xi_k)=\Gamma \delta_{l,0} \int \frac{\rd \xi_q}{2\pi}F(i\omega,\xi_q)\,.
    \end{equation} Here and after the superscript $(l)$ denotes fourier transform in the angular harmonics. The simplicity of $W_{dis}$ allows us to treat it exactly.

    $W_\text{MT}$ describes scattering in Maki-Thompson diagrams:
    \begin{equation}\label{}
      W_\text{MT}[F](i\omega,\vec{k})=g^2 \int \frac{\rd^2\vec{k}'\rd \omega'}{(2\pi)^3} D(\omega-\omega',\vec{k}-\vec{k}')F(i\omega',\vec{k}')
    \end{equation}

     $W_\text{AL}$ describes scattering in Azlamasov-Larkin diagrams:
    \begin{equation}\label{eq:WAL}
    \begin{split}
      W_{\text{AL}}[F](k_1)=&-\frac{g^4}{2} \int \frac{\rd^3 q\rd^3 k_2}{(2\pi)^6}\left(G(k_1-q)+G(k_1+q)\right)\left(G(k_2-q)+G(k_2+q)\right)\\
      &\times D(q+p/2)D(q-p/2)F(k_2)\\%\exp\left(-il\left(\theta_{k_1}-\theta_{k_2}\right)\right)\\
      &=-g^4 \int \frac{\rd^3 q\rd^3 k_2}{(2\pi)^6}G(k_1-q)\left(G(k_2-q)+G(k_2+q)\right)D(q+p/2)D(q-p/2)\\
      &\times F(k_2)\,.
      %\exp\left(-il\left(\theta_{k_1}-\theta_{k_2}\right)\right)\,,
    \end{split}
    \end{equation}

    \subsubsection{Zeroth order}

      The zeroth order polarization is
      \begin{equation}\label{}
        \Pi^{xx}_{0}(i\Omega)=N(\Gamma^x)^T\frac{1}{W_{\Sigma_0}^{-1}-W_{dis}} \Gamma^x\,,
      \end{equation} where the bare vertex function is approximated by
      \begin{equation}\label{}
        \Gamma^x(k)=v_F\cos\theta_k\,.
      \end{equation} Since $\Gamma^x(k)$ only contains first harmonics, $W_{dis}$ vanishes, and we obtain a Drude-like result
      \begin{equation}\label{}
        \Pi^{xx}_{0}(i\Omega)/N=\frac{\calN v_F^2}{2}\frac{\Omega }{\Omega+\Gamma}\,,
      \end{equation}
      \begin{equation}\label{}
        \sigma_{xx,0}(i\omega)/N=\frac{\calN v_F^2}{2} \frac{1}{-i\omega+\Gamma}\,.
      \end{equation}

    \subsubsection{First order: Self-energy and Maki-Thompson diagrams}

        To first order, the polarization is
        \begin{equation}\label{}
          \Pi^{xx}_1=-N (\Gamma^x)^T \frac{1}{W_{\Sigma,0}^{-1}-W_{dis}}\left( W_{\Sigma,1}^{-1}-W_\text{MT}-W_\text{AL}\right)\frac{1}{W_{\Sigma,0}^{-1}-W_{dis}}\Gamma^x\,.
        \end{equation} Using the fact that $\Gamma^{x}$ only contains first harmonics, we have
        \begin{equation}\label{eq:Pixx1}
        \begin{split}
          \Pi^{xx}_1=&N (\Gamma^{x})^T W_{\Sigma,0} \left(W_\text{MT}^{(1)}+W_\text{AL}^{(1)}-W_{\Sigma,1}^{-1}\right)W_{\Sigma,0} \Gamma^{x}\\
          &=N (\Gamma^{x})^T W_{\Sigma,0} \left(\tilde{\Gamma}^{x}_{\text MT}+\tilde{\Gamma}^x_\text{AL}-\tilde{\Gamma}^x_{\Sigma}\right)\,.
        \end{split}
        \end{equation} In the transport limit $p=(\Omega_n,0)$, the kernel $W_{\Sigma,1}$ is rotational invariant so we have dropped the superscript.

        In \eqref{eq:Pixx1} we have defined three types of renormalized vertices $\tilde{\Gamma}^{x}_{\Sigma},~\tilde{\Gamma}^{x}_\text{MT}$ and $\tilde{\Gamma}^x_\text{AL}$. %We will compute the first two and neglect the contributions from AL diagrams.

        \paragraph{$\tilde{\Gamma}^x_\Sigma$}
        The first type $\tilde{\Gamma}^{x}_{\Sigma}$ describes the contribution due to self-energies:
        \begin{equation}\label{}
          \tilde{\Gamma}^x_\Sigma(i\omega_n,\vec{k})=W_{\Sigma,1}^{-1}W_{\Sigma,0}\Gamma^x=-v_F\cos\theta_k\left(\Sigma_{+}G_{+}+\Sigma_{-}G_{-}\right)\,,
        \end{equation}where we have used a shorthand notation
        \begin{equation}\label{eq:Gpm}
          \Sigma_{\pm}=\Sigma_Q(i\omega_n\pm i\Omega_n/2)+\Sigma_T(i\omega_n\pm i\Omega_n/2)\,,\quad G_\pm=G_0(i\omega_n\pm i\Omega_n/2,\vec{k})\,.
        \end{equation}

        \paragraph{$\tilde{\Gamma}^x_\text{MT}$}
        Next we calculate $\tilde{\Gamma}^x_\text{MT}$:
        \begin{equation}\label{eq:tGammaMT}
          \tilde{\Gamma}^x_\text{MT}(i\omega_n,\vec{k})=v_F \cos\theta_k g^2T\sum_{\nu_n}\int \frac{q\rd q}{2\pi}\frac{\rd \theta_q}{2\pi}D(\nu_n,q)e^{-i\theta_{kk'}}\frac{1}{i A'_+-\xi_k+v_F q \cos\theta_q}\frac{1}{iA'_--\xi_k+v_F q\cos\theta_q}\,,
        \end{equation} where
        \begin{equation}\label{}
          A(\omega_n)=\omega_n+\frac{\Gamma}{2}\sgn \omega_n\,,\quad A'_\pm=A(\omega_n-\nu_n\pm \Omega_n/2)\,,
        \end{equation} and $\theta_q$ is the angle between $\vec{k}$ and $\vec{q}$.
        The $\theta_{kk'}$ above is the angle between $\vec{k}$ and $\vec{k'}=\vec{k}-\vec{q}$, and because $\vec{q}$ is small compared to $k_F$, we approximate
        \begin{equation}\label{}
          e^{-i\theta_{kk'}}=1-\frac{q^2}{2 k_F^2}\sin^2 \theta_q\,.
        \end{equation} The boson propagator is given by
        \begin{equation}\label{}
          D(\nu_n,q)=\frac{1}{q^2+M^2(T,\nu_n)}\,,\quad M^2(T,\nu_n)=\begin{cases}
                       \gamma|\nu_n|, & \nu_n\neq 0\,; \\
                       \Delta(T)^2, & \nu_n=0\,.
                     \end{cases}
        \end{equation}

        We can now perform the angular integrals in \eqref{eq:tGammaMT}, which yields
        \begin{equation}\label{eq:tGammaMT1}
        \begin{split}
          \tilde{\Gamma}^x_\text{MT}&=v_F\cos \theta_k g^2 T\sum_{\nu_n}\int_0^\infty \frac{q\rd q}{2\pi}D(\nu_n,q)\frac{1}{A'_+-A'_-}\Bigg\{\left(\frac{\sgn A'_+}{\sqrt{A^{\prime 2}_++v_F^2q^2}}-\frac{\sgn A'_-}{\sqrt{A^{\prime 2}_-+v_F^2q^2}}\right)\\
          &+\frac{1}{v_F^2 k_F^2}\left[\sgn A'_+ \left(\sqrt{A^{\prime 2}_++v_F^2q^2}-|A'_+|\right)-\sgn A'_- \left(\sqrt{A^{\prime 2}_-+v_F^2q^2}-|A'_-|\right) \right]\Bigg\}\,,
        \end{split}
        \end{equation}where we have assumed $\vec{k}$ to be lying on the FS and set $\xi_k=0$.

        The $1/(A'_+-A'_-)$ factor is a piecewise constant function ($\Omega_n>0$):
        \begin{equation}\label{}
          \frac{1}{A'_+-A'_-}=\left\{
                                \begin{array}{ll}
                                  \frac{1}{\Omega_n+\Gamma}, & |\nu_n-\omega_n|<\Omega_n/2; \\
                                  \frac{1}{\Omega_n}, & |\nu_n-\omega_n|>\Omega_n/2.
                                \end{array}
                              \right.
        \end{equation} Plugging the above into the first line of \eqref{eq:tGammaMT1}, we can separate out a part which yields the self energy and a correction term:
        \begin{equation}\label{eq:tGammaMTa}
        \begin{split}
           \tilde{\Gamma}^{x}_{\text{MT},a}&=v_F \cos\theta_k g^2 T \sum_{\nu_n}\int_0^\infty \frac{q\rd q}{2\pi}\frac{1}{\Omega_n}D(q,\nu_n)\left(\frac{\sgn A'_+}{\sqrt{A^{\prime 2}_++v_F^2q^2}}-\frac{\sgn A'_-}{\sqrt{A^{\prime 2}_-+v_F^2q^2}}\right)  \\
             & =\frac{iv_F \cos\theta_k}{\Omega_n}\left(\Sigma_+-\Sigma_-\right)\,,
        \end{split}
        \end{equation}
        \begin{equation}\label{eq:tGammaMTb}
          \begin{split}
             \tilde{\Gamma}^x_{\text{MT},b} & =v_F\cos\theta_k g^2 T \sum_{|\nu_n-\omega_n|<\Omega_n/2} \int_0^\infty \frac{q\rd q}{2\pi}\frac{-\Gamma}{\Omega_n(\Omega_n+\Gamma)}D(q,\nu_n)\left(\frac{\sgn A'_+}{\sqrt{A^{\prime 2}_++v_F^2q^2}}-\frac{\sgn A'_-}{\sqrt{A^{\prime 2}_-+v_F^2q^2}}\right) \\
            &= v_F \cos \theta_k \left(\frac{-\Gamma}{\Omega_n(\Omega_n+\Gamma)}\right)\frac{g^2 T}{2\pi}\sum_{|\nu_n-\omega_n|<\Omega_n/2}\left(\frac{\cosh^{-1}\left(\frac{A_+'}{v_F M(T,\nu_n)}\right)}{\sqrt{A^{'2}_+-v_F^2M^2(T,\nu_n)}}+(+\to -)\right)\\
            &=iv_F \cos \theta_k \left(\frac{-\Gamma}{\Omega_n(\Omega_n+\Gamma)}\right)(\Sigma_+-\Sigma_-)\,.
          \end{split}
        \end{equation}
           \begin{equation}\label{}
             \tilde{\Gamma}^{x}_{\text{MT},a}+\tilde{\Gamma}^{x}_{\text{MT},b}=i\frac{v_F \cos \theta_k}{\Omega_n+\Gamma}(\Sigma_+-\Sigma_-)
           \end{equation}
     To obtain the above results, we evaluated the $q$ integral first and next the $\nu_n$ sum with large $\Gamma$ approximation, and found the result agrees with \eqref{eq:SigmaQ+SigmaT}.

        Finally we compute the second line of \eqref{eq:tGammaMT1}, we again split it into two parts:

        \begin{equation}\label{}
          \begin{split}
             \tilde{\Gamma}^x_{\text{MT},c} & = \frac{v_F \cos\theta_k g^2 T}{v_F^2 k_F^2} \sum_{\nu_n}\int_0^\infty \frac{q\rd q}{2\pi}\frac{1}{\Omega_n}D(q,\nu_n)\left[\sgn A'_+ \left(\sqrt{A^{\prime 2}_++v_F^2q^2}-|A'_+|\right)\right.\\
                                            &\left.-\sgn A'_- \left(\sqrt{A^{\prime 2}_-+v_F^2q^2}-|A'_-|\right) \right] \\
          \end{split}
        \end{equation}
        \begin{equation}\label{}
          \begin{split}
             \tilde{\Gamma}^x_{\text{MT},d} & = \frac{v_F \cos\theta_k g^2 T}{v_F^2 k_F^2} \sum_{|\nu_n-\omega_n|<\Omega_n/2}\int_0^\infty \frac{q\rd q}{2\pi}\frac{-\Gamma}{\Omega_n(\Omega_n+\Gamma)}D(q,\nu_n)\left[\sgn A'_+ \left(\sqrt{A^{\prime 2}_++v_F^2q^2}-|A'_+|\right)\right.\\
                                            &\left.-\sgn A'_- \left(\sqrt{A^{\prime 2}_-+v_F^2q^2}-|A'_-|\right) \right] \\
          \end{split}
        \end{equation}

        The $q$-integral is UV divergent and we cut it off by a Pauli-Vilas regulator $\Lambda\sim k_F$
        \begin{equation}\label{eq:qintegral_disordered}
        \begin{split}
          &\int_0^{\infty} \frac{q\rd q }{2\pi}\left(\frac{1}{q^2+M^2}-\frac{1}{q^2+\Lambda^2}\right)\left(\sqrt{|A|^2+v_F^2 q^2}-|A|\right)\\
            =& \frac{v_F \Lambda}{4}+\frac{1}{2\pi}\left[\sqrt{|A|^2-M^2 v_F^2}\cosh^{-1}\left(\frac{|A|}{M v_F}\right)-|A|\ln\left(\frac{\Lambda e}{M}\right)\right]\\
            \approx & \frac{v_F \Lambda}{4}+\frac{1}{2\pi}|A|\ln\left(\frac{2|A|}{e \Lambda v_F}\right)-\frac{1}{4\pi|A|}M^2 v_F^2 \ln\left(\frac{2\sqrt{e}|A|}{M v_F}\right) \,.
        \end{split}
        \end{equation}

        Computing the frequency sum, we obtain
        \begin{equation}\label{}
          \tilde{\Gamma}^x_{\text{MT},c}+\tilde{\Gamma}^x_{\text{MT},d}=\frac{v_F \cos \theta_k g^2}{\pi v_F^2 k_F^2} \frac{\Omega_n}{\Omega_n+\Gamma}\left[\frac{v_F \Lambda}{4}+\frac{\Gamma}{4\pi}\ln\left(\frac{\Gamma}{e \Lambda v_F}\right)\right]\,.
        \end{equation} Here we have dropped the last term in \eqref{eq:qintegral_disordered} because it scales as $\Omega_n^2/\Gamma$.

        \paragraph{MT+ self energy}

        It's easy to check that
         \begin{equation}\label{}
           (\Gamma^x)^T W_{\Sigma,0}\left(\tilde{\Gamma}^x_{\text{MT},a+b}-\tilde{\Gamma}^x_\Sigma\right)=0\,,
         \end{equation} which can be seen after computing the $\xi_k$ integral.

        The rest from the MT diagrams contribute as
        \begin{equation}\label{}
          (\Gamma^x)^T W_{\Sigma,0}\tilde{\Gamma}^x_{\text{MT},c+d}=\frac{\calN v_F^2}{2} \left(\frac{\Omega_n}{\Omega_n+\Gamma}\right)^2 \frac{2 g^2}{(v_F k_F)^2} \left[\frac{v_F \Lambda}{4}+\frac{\Gamma}{4\pi}\ln\left(\frac{\Gamma}{e \Lambda v_F}\right)\right]\,,
        \end{equation} and its contribution to conductivity is
        \begin{equation}\label{}
          \sigma_{xx,1,\text{MT}}(i\omega)/N=\frac{\calN v_F^2}{2} \frac{-i\omega}{(-i\omega+\Gamma)^2}\frac{2 g^2}{(v_F k_F)^2} \left[\frac{v_F \Lambda}{4}+\frac{\Gamma}{4\pi}\ln\left(\frac{\Gamma}{e \Lambda v_F}\right)\right]\,.
        \end{equation} This result can be interpreted as an additional scattering rate in the Drude formula
        \begin{equation}\label{sigmatauMT}
          \sigma_{xx}=N\frac{\calN v_F^2}{2}\frac{1}{-i\omega+\Gamma+\frac{1}{\tau_\text{MT}(\omega)}}\,,
        \end{equation} where
        \begin{equation}\label{}
          \frac{1}{\tau_\text{MT}(\omega)}=i\omega\frac{2 g^2}{(v_F k_F)^2}\left[\frac{v_F \Lambda}{4}+\frac{\Gamma}{4\pi}\ln\left(\frac{\Gamma}{e \Lambda v_F}\right)\right]\,.
        \end{equation} There is no linear in $T$ resistivity. Higher order corrections in $1/\Gamma$ will start at order $|\omega_n|^2$ or $T^2$, which is Fermi-liquid like. This cancellation is also reminiscent of the U(1) Ward identity introduced in Sec.~\ref{sec:Ward}. However, because the disordered model is less controlled in the sense that Prange-Kadanoff reduction is unavailable, we can't give a rigorous argument as in the clean model.

        \subsubsection{Aslamazov-Larkin Diagrams}
        \label{sec:ALK}

          Now we show that the contributions from AL diagrams are also subdominant.
          The expression to evaluate is
          \begin{equation}\label{eq:PiAL}
          \begin{split}
            \Pi^{xx}_{1,\text{AL}}(i\Omega)/N=&-\frac{g^4}{2}\int\frac{\rd^3 q}{(2\pi)^3}D(q+\Omega/2)D(q-\Omega/2)X(q,\Omega)^2
          \end{split}
          \end{equation} where
          \begin{equation}\label{}
          \begin{split}
             X(q,\Omega) =& \int \frac{\rd^3 k}{(2\pi)^3} v_k \cos\theta_k G_0(k+\Omega/2)G_0(k-\Omega/2)\left[G_0(k+q)+G_0(k-q)\right]\,.
          \end{split}
          \end{equation} Here $q=(\nu,\vec{q})$ and $k=(\omega,\vec{k})$. The notation $q+ \Omega/2$ means adding $\Omega/2$ to the Matsubara component. For conductivity computation we assume $\Omega>0$.

          We first evaluate $X(q,\Omega)$, plugging in the expression for $G_0$ we have
          \begin{equation}\label{eq:Xqall}
          \begin{split}
            X(q,\Omega)&=2\pi\calN \int \frac{\rd \omega}{2\pi}\frac{\rd \xi_k}{2\pi} \frac{\rd \theta_k}{2\pi}v_k\cos\theta_k \frac{1}{iA(\omega+\Omega/2)-\xi_k}\frac{1}{iA(\omega-\Omega/2)-\xi_k}\\
            &\times\left[\frac{1}{iA(\omega+\nu)-\xi_k-v_k q\cos\theta_{kq}-\kappa q^2}+(\nu\to-\nu,\theta_q\to \pi+\theta_q)\right]\,.
          \end{split}
          \end{equation} Here $\theta_{kq}=\theta_k-\theta_q$ measures the angle between $\vec{k}$ and $\vec{q}$, and $A(\omega)=\omega+(\Gamma/2)\sgn \omega$. We have included the Fermi surface curvature $\kappa=1/(2m)$. Noticing that $A(\omega)$ is an odd function of $\omega$, under standard approximations $v_k=v_F$ and $\kappa\to 0$, the integrand is odd under $(\omega,\xi_k)\to-(\omega,\xi_k)$ and we get $X(q,\omega)=0$. Therefore, we need to keep terms that break the $\xi_k\to -\xi_k$ symmetry. There are two sources: Fermi surface curvature and dependence of $v_k$ on $\xi_k$.
          %We expect these two terms contribute at the same order of magnitude, and we consider the Fermi surface curvature only.

          To set up the expansion, we write $v_k=\sqrt{1+2\xi_k/(v_F k_F)}$ and $\kappa=v_F/(2k_F)$, and expand Eq.~\eqref{eq:Xqall} to first order in $1/k_F$, the first nonzero term is
          \begin{equation}\label{}
          \begin{split}
            X(q,\Omega)=\frac{\calN}{2k_F}&\int\rd\omega \frac{\rd \xi_k}{2\pi}\frac{\rd \theta_k}{2\pi}\cos\theta_k\frac{1}{iA(\omega+\Omega/2)-\xi_k}\frac{1}{iA(\omega-\Omega/2)-\xi_k}\\
            &\times\left[\frac{q^2 v_F^2-2\xi_k^2+2i \xi_k A(\nu+\omega)}{\left(iA(\nu+\omega)-\xi_k-q v_F \cos\theta_{kq}\right)^2}+(\nu\to-\nu,\theta_q\to\theta_q+\pi)\right]\,.
          \end{split}
          \end{equation} The integral over $\xi_k$ can be taken to be along the real line, since the finite band width only corrects the result by $\mathcal{O}(1/k_F^2)$. As a result the $\xi_k$ integral can be evaluated by residue method. The angular integral is performed using the formula
          $$
          \int \frac{\rd \theta_k}{2\pi}\frac{\cos \theta_k}{(ia-b \cos\theta_{kq})^2}=\frac{i b\cos\theta_q\sgn a}{(a^2+b^2)^{3/2}}\,,\quad a\in\mathbb{R},b>0.
          $$

          The final result for $X$ contains two analytic branches depending on whether $|\nu|<\Omega/2$ or $|\nu|>\Omega/2$. The branch with $|\nu|<\Omega/2$ will connect to $D_RD_A$ when Eq.~\eqref{eq:PiAL} is continued to real frequency, while the branch with $|\nu|>\Omega/2$ will connect to $D_R D_R$ or $D_A D_A$. It is shown in \cite{KamenevOreg} that only the first branch contributes at the low frequency limit ($|\Omega|<T$). In this limit, we are allowed to expand in small $|\nu|$ and small $|\Omega|$ (both are of the same order), yielding
          \begin{equation}\label{eq:Xnusmall}
            X(|\nu|<\Omega/2,q,\Omega)=\frac{\calN}{k_F}\frac{2iq^3 v_F^3 \nu\cos\theta_q}{(\Omega+\Gamma)(q^2v_F^2+\Gamma^2)^{3/2}}+\mathcal{O}(\nu^2,\Omega^2)\,.
          \end{equation} The numerator of the result has the same scaling as \cite{KamenevOreg}, but the denominator is different because in our large $N$ limit we have dropped vertex correction of Yukawa interaction due to disorders. In obtaining Eq.~\eqref{eq:Xnusmall}, the frequency summation is over a piecewise constant function, and therefore Eq.~\eqref{eq:Xnusmall} should be valid at finite temperature as well.

          Finally, we evaluate the integral \eqref{eq:PiAL} using \eqref{eq:Xnusmall} with $\nu\in[-\Omega/2,\Omega/2]$. To lowest order in $g$ we can set $\gamma=0$ in the boson propagators, and we obtain
          \begin{equation}\label{}
            \Pi^{xx}_{1,\text{AL}}(i\Omega)/N=\frac{\calN v_F^2}{2}\frac{g^4 (\Omega ^3+8\pi T^2\Omega)}{96 \pi ^2 \Gamma ^2 (\Gamma +\Omega )^2 k_F
   v_F}\,.
          \end{equation} Here we have used $\calN=k_F/(2\pi v_F)$.

          Analytically continuing to real frequency $\Omega\to -i\omega+0^+$, we obtain a effective scattering rate
          \begin{equation}\label{eq:tauAL}
            \frac{1}{\tau_{\text{AL}}(\omega)}=\frac{g^4 (\omega^2-8\pi^2 T^2)}{96\pi^2 \Gamma^2 k_Fv_F}\,.
          \end{equation}
          Note that this is a correction to the elastic scattering rate $\Gamma$, as in (\ref{sigmatauMT}). Therefore, the contribution of AL diagrams is less singular than MT + self energy diagrams.

\subsection{Discussion}

        Collecting the above results together, we see that the self energy and Maki-Thompson diagrams only renormalize the $-i\omega$ term in the Drude formula, while the Aslamazov-Larkin diagrams yield a $|\omega|^2$ decay rate. In what follows, we try to interpret the above results in terms of diffusion dynamics of Fermi surface as in the previous model.

        At first glance, because the condition for Prange-Kadanoff reduction \eqref{eq:PKcond} is violated, it seems inappropriate to talk about dynamics using states near the Fermi surface. However, following discussions in Sec.~\ref{sec:PK} we see that the violation of \eqref{eq:PKcond} means that both bosons with momentum normal and transverse to the Fermi surface can be excited (not excluded by Pauli principle). Now we discuss the effect of these two kinds of bosons.

        Let's first discuss the new part, which is the boson with momentum normal to the Fermi surface. Fermions excited by these bosons will have their velocities pointing in the same direction but renormalized a little bit by the bosons. These effects can be captured by renormalizing the $i\omega$ term in the conductivity, i.e. Eq.\eqref{sigmatauMT}.

        Next, for bosons with momentum tranverse to the Fermi surface, its effect can still be described in the context of diffusion on the Fermi surface. Since momentum conservation is no longer present, we won't expect the correlated superdiffusion behavior in the clean model, but instead a conventional diffusion dynamics with $\partial_\theta^2$ diffusion term. The diffusion coefficient can therefore be estimated as
        \begin{equation}
            D\sim \Im\Sigma_{g,R}(\omega)\left(\delta\theta\right)^2\sim |\omega|^2\,.
        \end{equation} Again, the diffusion coefficient is a product of the scattering rate ($\Im\Sigma_{g,R}\sim|\omega|$) with the angular steps ($\delta \theta\sim q/k_F\sim |\omega|^{1/2}$). This result can be matched with AL diagrams \eqref{eq:tauAL} as well as the next order expansion of MT diagrams \eqref{sigmatauMT}.

        In reality, the effect of the above two kinds of bosons are mixed but the qualitative feature should agree with the limiting cases of the above discussions.

        To achieve linear-in-T resistivity, we would need a mechanism which yield diffusion coefficient proportional to $\omega$. Since the thermodynamics experiments favor a marginal Fermi liquid self energy which is linear in $|\omega|$, the only way is to make the angular step $\delta\theta\sim |\omega|^0$. To achieve this, a small momentum boson must cause large momentum change for the fermions, meaning that momentum should not be conserved, either by a disordered interaction or Umklapp process. This also amounts to suppress vertex correction diagrams.

\section{Conclusions}

In this work, we have computed the electrical conductivity of a critical Fermi surface at the leading large $N$ order.

For the translational invariant `clean' model, we found that due to momentum conservation and strong boson drag the DC resistivity is zero. In the optical conductivity the scattering rate scales as $|\omega|^2$ and consequently the correction to Drude optical conductivity scales as $|\omega|^0$, which is a weaker scaling than the $|\omega|^{-2/3}$ correction in previous literature \cite{Kim94}. Our results are in general agreement with those of Refs.~\cite{ChubukovMaslov,Maslov12,Maslov17}, who also argue that the cancellation of the  $|\omega|^{-2/3}$ term is present only for convex Fermi surfaces.

On the experimental side for the clean model, an $|\omega|^{2}$ scattering rate in the optical conductivity cannot be distinguished from Fermi liquid-like corrections from impurities. However, it is known that when momentum-conserving collisions dominate, the system enters viscous (hydrodynamic) regime, and the DC current is determined by the external electric field non-locally through the $k$-dependent conductivity $\sigma(\omega=0,k)$. This non-local conductivity can be measured in transport experiments as proposed in \cite{LedwithGuo2019,Lucas2021}. We leave the computation of $\sigma(k)$ to future study.

For the disordered model, we showed that upon adding disorder potential the critical boson induces a marginal Fermi liquid self energy to the fermions, in addition to the elastic disorder scattering rate. However, the MFL self energy is cancelled by boson vertex corrections, and does not contribute to the transport lifetime. Therefore, to obtain MFL phenomenology in transport coefficients, we need a mechanism which is not cancelled by vertex correction. In the companion paper \cite{Patel:2022gdh}, we achieve this goal by introducing spatially disordered interactions.

\subsection*{Acknowledgements}

We thank A.~Chubukov, D.~Else, M.~Foster, H.~Goldman, D.~Maslov, T.~Senthil, and G. Torroba for valuable discussions. This research was supported by the National Science Foundation under Grant No.~DMR-2002850. I.~E. acknowledges support from the Harvard Quantum Initiative Postdoctoral Fellowship in Science and Engineering. This work was also supported by the Simons Collaboration on Ultra-Quantum Matter, which is a grant from the Simons Foundation (651440, S.S.). This research was supported in part by the Heising-Simons Foundation, the Simons Foundation, and National Science Foundation Grant No. NSF PHY-1748958. The Flatiron Institute is a division of the Simons Foundation.

\noindent
{\it Notes added:} ({\it i\/}) A recent independent work \cite{ShiElse2022,ShiElse2022a} reaches similar conclusions for the clean model without spatial disorder by different methods.
({\it ii}) We learnt of the paper by Wu {\it et al.} \cite{Foster22}. They confirm the cancellation of the linear-$T$ term in the resistivity in the potential disorder model of Section~\ref{sec:potential}. They also considered Altshuler-Aronov corrections, and find a $-1/T$ correction the Drude resistivity.

\appendix
\section{Cancellation of self energy, Maki-Thompson and Aslamazov-Larkin diagrams}
\label{app:Kim}

In this appendix we review the computation of Kim {\it et al.} \cite{Kim94}, and demonstrate that the cancellation of self energy, Maki-Thompson and Aslamazov-Larkin diagrams is already present in their expressions, but was overlooked by them. Our new contribution is the computation of the numerical coefficient $c_1$ in Eq.~(37) of Ref.~\cite{Kim94}, which they did not calculate. A related independent analysis appears in Appendix F of Ref.~\cite{ShiElse2022a}.

We will follow the notation of Ref.~\cite{Kim94} in this appendix. The boson propagator in their convention is
\begin{equation}\label{}
  D(i\nu,\vec{q})=\frac{1}{\gamma \frac{|\nu|}{q}+\chi q^\eta}\,,
\end{equation} and there is a form factor $\vec{k}/m$ in the fermion boson coupling $\psi_{k+q}^\dagger \psi_k \phi_q$.

Following their calculation scheme, we use the above RPA propagator for bosons but treat fermions in flavor large $N$. To leading order, the fermion Green's function is free $G_0(i\omega,\vec{k})=(i\omega-\xi_k)^{-1}$. 

The fermion self energy is
\begin{equation}\label{}
  \Sigma(i\omega,\vec{k})=\int \frac{\rd \nu}{2\pi} \frac{\rd^2 \vec{q}}{(2\pi)^2} \frac{|k\times \hat{q}|^2}{m^2} D(i\nu,\vec{q}) G_0(i\omega+i\nu,\vec{k}+\vec{q})
\end{equation} We evaluate the integral using Prange-Kadanoff reduction, yielding
\begin{equation}\label{}
  \Sigma(i\omega,\vec{k})=2\pi\calN \int \frac{\rd \nu}{2\pi} \int \frac{\rd \theta_{k'}}{2\pi}\frac{k_F^2 \cos^2(\theta_{kk'}/2)}{m^2}D\left(i\nu,k_F\left(\hat{\theta}_{k'}-\hat{\theta}_k\right)\right)\left(-\frac{i}{2}\right)\sgn(\omega+\nu)\,.
\end{equation} The integral can be calculated near the region $|\theta_{kk'}|\ll 1$, yielding
\begin{equation}\label{}
  \Sigma(i\omega,\vec{k})=-i\lambda|\omega|^{\frac{2}{1+\eta}} \sgn\omega\,,
\end{equation} with
\begin{equation}\label{}
  \lambda=\frac{v_F}{4\pi \chi^{\frac{2}{1+\eta}}\gamma^{\frac{\eta-1}{\eta+1}}\sin\left(\frac{2\pi}{1+\eta}\right)}\,.
\end{equation} There are subleading corrections order $|\omega|^{\frac{1}{1+\eta}}/k_F$ due to the form factor and geometry of the Fermi surface, which are not important for the cancellations we demonstrate below.

Next, we calculate the boson polarization in the transport limit of vanishing wavevector $\vec{Q}=0$.

The self energy contribution is
\begin{equation}\label{}
\begin{split}
  \Pi_{xx}^{(1)}(i\Omega)=&-\int \frac{\rd \omega}{2\pi} \frac{\rd^2 \vec{k}}{(2\pi)^2}\Big[G_0(i\omega+i\Omega,\vec{k})^2G_0(i\omega,\vec{k})\Sigma(i\omega+i\Omega,\vec{k})\\
  &+G_0(i\omega+i\Omega,\vec{k})G_0(i\omega,\vec{k})^2 \Sigma(i\omega,\vec{k})\Big]\frac{k^2}{2m^2}\,,
\end{split}
\end{equation} where the last $k^2/(2m^2)$ term is the angle-averaged form factor. The result is
\begin{equation}\label{}
  \Pi^{(1)}_{xx}(i\Omega)=\frac{k_F^2}{2\pi m} \frac{1+\eta}{3+\eta} \lambda \frac{1}{|\Omega|^{\frac{\eta-1}{\eta+1}}}
\end{equation}

The Maki-Thompson diagram contributes
\begin{equation}\label{}
\begin{split}
  \Pi_{xx}^{(2)}(i\Omega)=-&\int \frac{\rd \nu}{2\pi} \frac{\rd^2\vec{q}}{(2\pi)^2} \frac{\rd \omega}{2\pi} \frac{\rd^2\vec{k}}{(2\pi)^2} \frac{k^2-(\vec{k}\cdot \hat{q})^2}{m^2} \frac{\vec{k}\cdot\left(\vec{k}+\vec{q}\right)}{2m^2}D(i\nu,\vec{q})\\
  &\times G_0(i\omega+i\Omega,\vec{k})G_0(i\omega,\vec{k}) G_0(i\omega+i\nu+i\Omega,\vec{k}+\vec{q})G_0(i\omega+i\nu,\vec{k}+\vec{q})\,,
\end{split}
\end{equation} where we have utilized rotation symmetry to write the vertex form factor as $\vec{k}\cdot(\vec{k}+\vec{q})/(2m^2)$. $\Pi_{xx}^{(2)}$ can be evaluated using the Prange-Kadanoff procedure described in the main text. We let $\vec{k}'=\vec{k}+\vec{q}$ and perform integral over $\xi_k$ and $\xi_{k'}$ first, which projects $\vec{k}$ and $\vec{k}'$ onto the fermi surface. The remaining integral is
\begin{equation}\label{}
  \begin{split}
     \Pi_{xx}^{(2)}= & -\calN^2 \int \frac{\rd \omega}{2\pi}\frac{\rd \nu}{2\pi}\int \rd \theta_k\rd \theta_{k'}\frac{1}{(i\Omega)^2}\left(-\frac{1}{4}\right) D\left(i\nu,k_F\left(\hat{\theta}_{k'}-\hat{\theta}_k\right)\right) \\
       & \left(\sgn(\omega)-\sgn(\omega+\Omega)\right)\left(\sgn(\omega+\nu)-\sgn(\omega+\nu+\Omega)\right)\\
      & \frac{k_F^2\cos^2(\theta_{kk'}/2)}{m^2}\frac{k_F^2\cos(\theta_{kk'})}{2m^2}\,.
  \end{split}
\end{equation} Here $\theta_{kk'}=\theta_k-\theta_{k'}$. The above integral is evaluated by expanding the $\cos(\theta_{kk'})=1-\frac{1}{2}\theta_{kk'}^2$, yielding $\Pi_{xx}^{(2)}=\Pi_{xx}^{(2a)}+\Pi_{xx}^{(2b)}$
\begin{equation}\label{}
  \Pi_{xx}^{(2a)}=-\frac{k_F^3}{8m^2 \pi^2\sin\left(\frac{2\pi}{1+\eta}\right)}\frac{1+\eta}{3+\eta}\frac{1}{\chi^{\frac{2}{1+\eta}}\gamma^{\frac{\eta-1}{\eta+1}}|\Omega|^{\frac{\eta-1}{\eta+1}}}\,,
\end{equation} and
\begin{equation}\label{}
  \Pi_{xx}^{(2b)}= \frac{k_F}{32 m^2 \pi^2 \sin\left(\frac{4\pi}{1+\eta}\right)}\frac{1+\eta}{5+\eta}\frac{1}{\gamma^{\frac{\eta-3}{1+\eta}}\chi^{\frac{4}{1+\eta}}|\Omega|^{\frac{\eta-3}{\eta+1}}}\,.
\end{equation} In obtaining the above results, we have only kept higher order terms in $q/k_F$ from the $\cos\theta_{kk'}$ factor but neglected order corrections in coming from the form factor or the Fermi surface geometry. This is justified by directly manipulating the integrands of $\Pi_{xx}^{(1)}+\Pi_{xx}^{(2)}$ using Ward identities discussed in Ref.~\cite{Kim94}, or the main text, which show that the whole integrand is proportional to $(1-\cos\theta_{kk'})$, which is the leading order $q$ dependence.

It is easy to see that $\Pi_{xx}^{(1)}+\Pi_{xx}^{(2a)}=0$ as in Eq.~(29) of Ref.~\cite{Kim94}. However, there appear to be typographical errors in \cite{Kim94}, as our individual results disagree with Ref.~\cite{Kim94} by some powers of 2.

Finally, we look at the Aslamazov-Larkin diagrams, which is 
\begin{equation}\label{}
\begin{split}
   \Pi^{(3)}_{xx}= & \int \frac{\rd \nu}{2\pi}\frac{\rd \omega}{2\pi}\frac{\rd \omega'}{2\pi}\frac{\rd^2\vec{k}}{(2\pi)^2}\frac{\rd^2\vec{q}}{(2\pi)^2}
   D(\vec{q},i\nu+i\Omega/2)D(\vec{q},i\nu-i\Omega/2)G_0(i\omega+i\Omega/2,\vec{k})\\
  & G_0(i\omega-i\Omega/2,\vec{k})  G_0(i\omega'+i\Omega/2,\vec{k}')G_0(i\omega'-i\Omega/2,\vec{k}')G_0(i\omega-i\nu,\vec{k}-\vec{q})\\
     &\left[G_0(i\omega'-i\nu,\vec{k}'-\vec{q})+G_0(i\omega'+i\nu,\vec{k}'+\vec{q})\right]\frac{|\vec{k}\times\hat{q}|^2}{m^2}\frac{|\vec{k}'\times\hat{q}|^2}{m^2}\frac{\vec{k}\cdot \vec{k}'}{2m^2}\,.
\end{split}
\end{equation} This contribution is analyzed using Prange-Kadanoff reduction as described in the main text, with similar steps. The result is 
\begin{equation}\label{}
  \Pi_{xx}^{(3)}=\left(-\frac{k_F}{2\pi\gamma}\right) \frac{k_F}{32 m^2 \pi^2 \sin\left(\frac{4\pi}{1+\eta}\right)}\frac{1+\eta}{5+\eta}\frac{1}{\gamma^{\frac{\eta-3}{1+\eta}}\chi^{\frac{4}{1+\eta}}|\Omega|^{\frac{\eta-3}{\eta+1}}}\,.
\end{equation} Therefore $\Pi_{xx}^{(3)}=\left(-{k_F}/({2\pi\gamma})\right)\Pi_{xx}^{(2b)}$. 
The value of $\gamma$ should be calculated using Landau damping of free fermions \cite{Kim94}, which exactly yields $\gamma=k_F/(2\pi)$, and confirms the cancellation $\Pi_{xx}^{(2b)}+\Pi_{xx}^{(3)}=0$.

\bibliography{refs,fermi}

\begin{thebibliography}{73}%
\makeatletter
\providecommand \@ifxundefined [1]{%
 \@ifx{#1\undefined}
}%
\providecommand \@ifnum [1]{%
 \ifnum #1\expandafter \@firstoftwo
 \else \expandafter \@secondoftwo
 \fi
}%
\providecommand \@ifx [1]{%
 \ifx #1\expandafter \@firstoftwo
 \else \expandafter \@secondoftwo
 \fi
}%
\providecommand \natexlab [1]{#1}%
\providecommand \enquote  [1]{``#1''}%
\providecommand \bibnamefont  [1]{#1}%
\providecommand \bibfnamefont [1]{#1}%
\providecommand \citenamefont [1]{#1}%
\providecommand \href@noop [0]{\@secondoftwo}%
\providecommand \href [0]{\begingroup \@sanitize@url \@href}%
\providecommand \@href[1]{\@@startlink{#1}\@@href}%
\providecommand \@@href[1]{\endgroup#1\@@endlink}%
\providecommand \@sanitize@url [0]{\catcode `\\12\catcode `\$12\catcode
  `\&12\catcode `\#12\catcode `\^12\catcode `\_12\catcode `\%12\relax}%
\providecommand \@@startlink[1]{}%
\providecommand \@@endlink[0]{}%
\providecommand \url  [0]{\begingroup\@sanitize@url \@url }%
\providecommand \@url [1]{\endgroup\@href {#1}{\urlprefix }}%
\providecommand \urlprefix  [0]{URL }%
\providecommand \Eprint [0]{\href }%
\providecommand \doibase [0]{https://doi.org/}%
\providecommand \selectlanguage [0]{\@gobble}%
\providecommand \bibinfo  [0]{\@secondoftwo}%
\providecommand \bibfield  [0]{\@secondoftwo}%
\providecommand \translation [1]{[#1]}%
\providecommand \BibitemOpen [0]{}%
\providecommand \bibitemStop [0]{}%
\providecommand \bibitemNoStop [0]{.\EOS\space}%
\providecommand \EOS [0]{\spacefactor3000\relax}%
\providecommand \BibitemShut  [1]{\csname bibitem#1\endcsname}%
\let\auto@bib@innerbib\@empty
%</preamble>
\bibitem [{\citenamefont {Coleman}(2015)}]{coleman2015}%
  \BibitemOpen
  \bibfield  {author} {\bibinfo {author} {\bibfnamefont {P.}~\bibnamefont
  {Coleman}},\ }\href {https://doi.org/10.1017/CBO9781139020916} {\emph
  {\bibinfo {title} {Introduction to Many-Body Physics}}}\ (\bibinfo
  {publisher} {Cambridge University Press},\ \bibinfo {year}
  {2015})\BibitemShut {NoStop}%
\bibitem [{\citenamefont {Lee}(1989)}]{PALee89}%
  \BibitemOpen
  \bibfield  {author} {\bibinfo {author} {\bibfnamefont {P.~A.}\ \bibnamefont
  {Lee}},\ }\bibfield  {title} {\emph {\bibinfo {title} {{Gauge field,
  Aharonov-Bohm flux, and high-${T}_{c}$ superconductivity}}},\ }\href
  {https://doi.org/10.1103/PhysRevLett.63.680} {\bibfield  {journal} {\bibinfo
  {journal} {Phys. Rev. Lett.}\ }\textbf {\bibinfo {volume} {63}},\ \bibinfo
  {pages} {680} (\bibinfo {year} {1989})}\BibitemShut {NoStop}%
\bibitem [{\citenamefont {Polchinski}(1994)}]{Polchinski:1993ii}%
  \BibitemOpen
  \bibfield  {author} {\bibinfo {author} {\bibfnamefont {J.}~\bibnamefont
  {Polchinski}},\ }\bibfield  {title} {\emph {\bibinfo {title} {{Low-energy
  dynamics of the spinon gauge system}}},\ }\href
  {https://doi.org/10.1016/0550-3213(94)90449-9} {\bibfield  {journal}
  {\bibinfo  {journal} {Nucl. Phys. B}\ }\textbf {\bibinfo {volume} {422}},\
  \bibinfo {pages} {617} (\bibinfo {year} {1994})},\ \Eprint
  {https://arxiv.org/abs/cond-mat/9303037} {arXiv:cond-mat/9303037}
  \BibitemShut {NoStop}%
\bibitem [{\citenamefont {Halperin}\ \emph {et~al.}(1993)\citenamefont
  {Halperin}, \citenamefont {Lee},\ and\ \citenamefont {Read}}]{HLR}%
  \BibitemOpen
  \bibfield  {author} {\bibinfo {author} {\bibfnamefont {B.~I.}\ \bibnamefont
  {Halperin}}, \bibinfo {author} {\bibfnamefont {P.~A.}\ \bibnamefont {Lee}},\
  and\ \bibinfo {author} {\bibfnamefont {N.}~\bibnamefont {Read}},\ }\bibfield
  {title} {\emph {\bibinfo {title} {{Theory of the half-filled Landau
  level}}},\ }\href {https://doi.org/10.1103/PhysRevB.47.7312} {\bibfield
  {journal} {\bibinfo  {journal} {Phys. Rev. B}\ }\textbf {\bibinfo {volume}
  {47}},\ \bibinfo {pages} {7312} (\bibinfo {year} {1993})}\BibitemShut
  {NoStop}%
\bibitem [{\citenamefont {{Kim}}\ \emph {et~al.}(1994)\citenamefont {{Kim}},
  \citenamefont {{Furusaki}}, \citenamefont {{Wen}},\ and\ \citenamefont
  {{Lee}}}]{Kim94}%
  \BibitemOpen
  \bibfield  {author} {\bibinfo {author} {\bibfnamefont {Y.~B.}\ \bibnamefont
  {{Kim}}}, \bibinfo {author} {\bibfnamefont {A.}~\bibnamefont {{Furusaki}}},
  \bibinfo {author} {\bibfnamefont {X.-G.}\ \bibnamefont {{Wen}}},\ and\
  \bibinfo {author} {\bibfnamefont {P.~A.}\ \bibnamefont {{Lee}}},\ }\bibfield
  {title} {\emph {\bibinfo {title} {{Gauge-invariant response functions of
  fermions coupled to a gauge field}}},\ }\href
  {https://doi.org/10.1103/PhysRevB.50.17917} {\bibfield  {journal} {\bibinfo
  {journal} {Phys. Rev. B}\ }\textbf {\bibinfo {volume} {50}},\ \bibinfo
  {pages} {17917} (\bibinfo {year} {1994})},\ \Eprint
  {https://arxiv.org/abs/cond-mat/9405083} {arXiv:cond-mat/9405083 [cond-mat]}
  \BibitemShut {NoStop}%
\bibitem [{\citenamefont {{Altshuler}}\ \emph {et~al.}(1994)\citenamefont
  {{Altshuler}}, \citenamefont {{Ioffe}},\ and\ \citenamefont
  {{Millis}}}]{Altshuler94}%
  \BibitemOpen
  \bibfield  {author} {\bibinfo {author} {\bibfnamefont {B.~L.}\ \bibnamefont
  {{Altshuler}}}, \bibinfo {author} {\bibfnamefont {L.~B.}\ \bibnamefont
  {{Ioffe}}},\ and\ \bibinfo {author} {\bibfnamefont {A.~J.}\ \bibnamefont
  {{Millis}}},\ }\bibfield  {title} {\emph {\bibinfo {title} {{Low-energy
  properties of fermions with singular interactions}}},\ }\href
  {https://doi.org/10.1103/PhysRevB.50.14048} {\bibfield  {journal} {\bibinfo
  {journal} {Phys. Rev. B}\ }\textbf {\bibinfo {volume} {50}},\ \bibinfo
  {pages} {14048} (\bibinfo {year} {1994})},\ \Eprint
  {https://arxiv.org/abs/cond-mat/9406024} {arXiv:cond-mat/9406024 [cond-mat]}
  \BibitemShut {NoStop}%
\bibitem [{\citenamefont {Nayak}\ and\ \citenamefont
  {Wilczek}(1994)}]{Nayak:1994ng}%
  \BibitemOpen
  \bibfield  {author} {\bibinfo {author} {\bibfnamefont {C.}~\bibnamefont
  {Nayak}}\ and\ \bibinfo {author} {\bibfnamefont {F.}~\bibnamefont
  {Wilczek}},\ }\bibfield  {title} {\emph {\bibinfo {title} {{Renormalization
  group approach to low temperature properties of a non-Fermi liquid metal}}},\
  }\href {https://doi.org/10.1016/0550-3213(94)90158-9} {\bibfield  {journal}
  {\bibinfo  {journal} {Nucl. Phys. B}\ }\textbf {\bibinfo {volume} {430}},\
  \bibinfo {pages} {534} (\bibinfo {year} {1994})},\ \Eprint
  {https://arxiv.org/abs/cond-mat/9408016} {arXiv:cond-mat/9408016}
  \BibitemShut {NoStop}%
\bibitem [{\citenamefont {{Lee}}(2009)}]{sungsik1}%
  \BibitemOpen
  \bibfield  {author} {\bibinfo {author} {\bibfnamefont {S.-S.}\ \bibnamefont
  {{Lee}}},\ }\bibfield  {title} {\emph {\bibinfo {title} {{Low-energy
  effective theory of Fermi surface coupled with U(1) gauge field in 2+1
  dimensions}}},\ }\href {https://doi.org/10.1103/PhysRevB.80.165102}
  {\bibfield  {journal} {\bibinfo  {journal} {Phys. Rev. B}\ }\textbf {\bibinfo
  {volume} {80}},\ \bibinfo {eid} {165102} (\bibinfo {year} {2009})},\ \Eprint
  {https://arxiv.org/abs/0905.4532} {arXiv:0905.4532 [cond-mat.str-el]}
  \BibitemShut {NoStop}%
\bibitem [{\citenamefont {{Metlitski}}\ and\ \citenamefont
  {{Sachdev}}(2010)}]{metlitski1}%
  \BibitemOpen
  \bibfield  {author} {\bibinfo {author} {\bibfnamefont {M.~A.}\ \bibnamefont
  {{Metlitski}}}\ and\ \bibinfo {author} {\bibfnamefont {S.}~\bibnamefont
  {{Sachdev}}},\ }\bibfield  {title} {\emph {\bibinfo {title} {{Quantum phase
  transitions of metals in two spatial dimensions. I. Ising-nematic order}}},\
  }\href {https://doi.org/10.1103/PhysRevB.82.075127} {\bibfield  {journal}
  {\bibinfo  {journal} {Phys. Rev. B}\ }\textbf {\bibinfo {volume} {82}},\
  \bibinfo {eid} {075127} (\bibinfo {year} {2010})},\ \Eprint
  {https://arxiv.org/abs/1001.1153} {arXiv:1001.1153 [cond-mat.str-el]}
  \BibitemShut {NoStop}%
\bibitem [{\citenamefont {{Mross}}\ \emph {et~al.}(2010)\citenamefont
  {{Mross}}, \citenamefont {{McGreevy}}, \citenamefont {{Liu}},\ and\
  \citenamefont {{Senthil}}}]{mross}%
  \BibitemOpen
  \bibfield  {author} {\bibinfo {author} {\bibfnamefont {D.~F.}\ \bibnamefont
  {{Mross}}}, \bibinfo {author} {\bibfnamefont {J.}~\bibnamefont {{McGreevy}}},
  \bibinfo {author} {\bibfnamefont {H.}~\bibnamefont {{Liu}}},\ and\ \bibinfo
  {author} {\bibfnamefont {T.}~\bibnamefont {{Senthil}}},\ }\bibfield  {title}
  {\emph {\bibinfo {title} {{Controlled expansion for certain non-Fermi-liquid
  metals}}},\ }\href {https://doi.org/10.1103/PhysRevB.82.045121} {\bibfield
  {journal} {\bibinfo  {journal} {Phys. Rev. B}\ }\textbf {\bibinfo {volume}
  {82}},\ \bibinfo {eid} {045121} (\bibinfo {year} {2010})},\ \Eprint
  {https://arxiv.org/abs/1003.0894} {arXiv:1003.0894 [cond-mat.str-el]}
  \BibitemShut {NoStop}%
\bibitem [{\citenamefont {{Sur}}\ and\ \citenamefont {{Lee}}(2014)}]{sungsik3}%
  \BibitemOpen
  \bibfield  {author} {\bibinfo {author} {\bibfnamefont {S.}~\bibnamefont
  {{Sur}}}\ and\ \bibinfo {author} {\bibfnamefont {S.-S.}\ \bibnamefont
  {{Lee}}},\ }\bibfield  {title} {\emph {\bibinfo {title} {{Chiral non-Fermi
  liquids}}},\ }\href {https://doi.org/10.1103/PhysRevB.90.045121} {\bibfield
  {journal} {\bibinfo  {journal} {Phys. Rev. B}\ }\textbf {\bibinfo {volume}
  {90}},\ \bibinfo {eid} {045121} (\bibinfo {year} {2014})},\ \Eprint
  {https://arxiv.org/abs/1310.7543} {arXiv:1310.7543 [cond-mat.str-el]}
  \BibitemShut {NoStop}%
\bibitem [{\citenamefont {Metlitski}\ \emph {et~al.}(2015)\citenamefont
  {Metlitski}, \citenamefont {Mross}, \citenamefont {Sachdev},\ and\
  \citenamefont {Senthil}}]{metlitski5}%
  \BibitemOpen
  \bibfield  {author} {\bibinfo {author} {\bibfnamefont {M.~A.}\ \bibnamefont
  {Metlitski}}, \bibinfo {author} {\bibfnamefont {D.~F.}\ \bibnamefont
  {Mross}}, \bibinfo {author} {\bibfnamefont {S.}~\bibnamefont {Sachdev}},\
  and\ \bibinfo {author} {\bibfnamefont {T.}~\bibnamefont {Senthil}},\
  }\bibfield  {title} {\emph {\bibinfo {title} {{Cooper pairing in non-Fermi
  liquids}}},\ }\href {https://doi.org/10.1103/PhysRevB.91.115111} {\bibfield
  {journal} {\bibinfo  {journal} {Phys. Rev. B}\ }\textbf {\bibinfo {volume}
  {91}},\ \bibinfo {pages} {115111} (\bibinfo {year} {2015})},\ \Eprint
  {https://arxiv.org/abs/1403.3694} {arXiv:1403.3694 [cond-mat.str-el]}
  \BibitemShut {NoStop}%
\bibitem [{\citenamefont {Hartnoll}\ \emph {et~al.}(2014)\citenamefont
  {Hartnoll}, \citenamefont {Mahajan}, \citenamefont {Punk},\ and\
  \citenamefont {Sachdev}}]{Hartnoll:2014gba}%
  \BibitemOpen
  \bibfield  {author} {\bibinfo {author} {\bibfnamefont {S.~A.}\ \bibnamefont
  {Hartnoll}}, \bibinfo {author} {\bibfnamefont {R.}~\bibnamefont {Mahajan}},
  \bibinfo {author} {\bibfnamefont {M.}~\bibnamefont {Punk}},\ and\ \bibinfo
  {author} {\bibfnamefont {S.}~\bibnamefont {Sachdev}},\ }\bibfield  {title}
  {\emph {\bibinfo {title} {{Transport near the Ising-nematic quantum critical
  point of metals in two dimensions}}},\ }\href
  {https://doi.org/10.1103/PhysRevB.89.155130} {\bibfield  {journal} {\bibinfo
  {journal} {Phys. Rev. B}\ }\textbf {\bibinfo {volume} {89}},\ \bibinfo
  {pages} {155130} (\bibinfo {year} {2014})},\ \Eprint
  {https://arxiv.org/abs/1401.7012} {arXiv:1401.7012 [cond-mat.str-el]}
  \BibitemShut {NoStop}%
\bibitem [{\citenamefont {Eberlein}\ \emph {et~al.}(2017)\citenamefont
  {Eberlein}, \citenamefont {Patel},\ and\ \citenamefont
  {Sachdev}}]{Patel_viscosity}%
  \BibitemOpen
  \bibfield  {author} {\bibinfo {author} {\bibfnamefont {A.}~\bibnamefont
  {Eberlein}}, \bibinfo {author} {\bibfnamefont {A.~A.}\ \bibnamefont
  {Patel}},\ and\ \bibinfo {author} {\bibfnamefont {S.}~\bibnamefont
  {Sachdev}},\ }\bibfield  {title} {\emph {\bibinfo {title} {{Shear viscosity
  at the Ising-nematic quantum critical point in two dimensional metals}}},\
  }\href {https://doi.org/10.1103/PhysRevB.95.075127} {\bibfield  {journal}
  {\bibinfo  {journal} {Phys. Rev. B}\ }\textbf {\bibinfo {volume} {95}},\
  \bibinfo {pages} {075127} (\bibinfo {year} {2017})},\ \Eprint
  {https://arxiv.org/abs/1607.03894} {arXiv:1607.03894 [cond-mat.str-el]}
  \BibitemShut {NoStop}%
\bibitem [{\citenamefont {{Holder}}\ and\ \citenamefont
  {{Metzner}}(2015{\natexlab{a}})}]{HolderMetzner1}%
  \BibitemOpen
  \bibfield  {author} {\bibinfo {author} {\bibfnamefont {T.}~\bibnamefont
  {{Holder}}}\ and\ \bibinfo {author} {\bibfnamefont {W.}~\bibnamefont
  {{Metzner}}},\ }\bibfield  {title} {\emph {\bibinfo {title} {{Anomalous
  dynamical scaling from nematic and U(1) gauge field fluctuations in
  two-dimensional metals}}},\ }\href
  {https://doi.org/10.1103/PhysRevB.92.041112} {\bibfield  {journal} {\bibinfo
  {journal} {Phys. Rev. B}\ }\textbf {\bibinfo {volume} {92}},\ \bibinfo {eid}
  {041112} (\bibinfo {year} {2015}{\natexlab{a}})},\ \Eprint
  {https://arxiv.org/abs/1503.05089} {arXiv:1503.05089 [cond-mat.str-el]}
  \BibitemShut {NoStop}%
\bibitem [{\citenamefont {{Holder}}\ and\ \citenamefont
  {{Metzner}}(2015{\natexlab{b}})}]{HolderMetzner2}%
  \BibitemOpen
  \bibfield  {author} {\bibinfo {author} {\bibfnamefont {T.}~\bibnamefont
  {{Holder}}}\ and\ \bibinfo {author} {\bibfnamefont {W.}~\bibnamefont
  {{Metzner}}},\ }\bibfield  {title} {\emph {\bibinfo {title} {{Fermion loops
  and improved power-counting in two-dimensional critical metals with singular
  forward scattering}}},\ }\href {https://doi.org/10.1103/PhysRevB.92.245128}
  {\bibfield  {journal} {\bibinfo  {journal} {Phys. Rev. B}\ }\textbf {\bibinfo
  {volume} {92}},\ \bibinfo {eid} {245128} (\bibinfo {year}
  {2015}{\natexlab{b}})},\ \Eprint {https://arxiv.org/abs/1509.07783}
  {arXiv:1509.07783 [cond-mat.str-el]} \BibitemShut {NoStop}%
\bibitem [{\citenamefont {Fitzpatrick}\ \emph {et~al.}(2014)\citenamefont
  {Fitzpatrick}, \citenamefont {Kachru}, \citenamefont {Kaplan},\ and\
  \citenamefont {Raghu}}]{Raghu1}%
  \BibitemOpen
  \bibfield  {author} {\bibinfo {author} {\bibfnamefont {A.~L.}\ \bibnamefont
  {Fitzpatrick}}, \bibinfo {author} {\bibfnamefont {S.}~\bibnamefont {Kachru}},
  \bibinfo {author} {\bibfnamefont {J.}~\bibnamefont {Kaplan}},\ and\ \bibinfo
  {author} {\bibfnamefont {S.}~\bibnamefont {Raghu}},\ }\bibfield  {title}
  {\emph {\bibinfo {title} {{Non-Fermi-liquid behavior of large-$N_B$ quantum
  critical metals}}},\ }\href {https://doi.org/10.1103/PhysRevB.89.165114}
  {\bibfield  {journal} {\bibinfo  {journal} {Phys. Rev. B}\ }\textbf {\bibinfo
  {volume} {89}},\ \bibinfo {pages} {165114} (\bibinfo {year} {2014})},\
  \Eprint {https://arxiv.org/abs/1312.3321} {arXiv:1312.3321 [cond-mat.str-el]}
  \BibitemShut {NoStop}%
\bibitem [{\citenamefont {Aguilera~Damia}\ \emph {et~al.}(2019)\citenamefont
  {Aguilera~Damia}, \citenamefont {Kachru}, \citenamefont {Raghu},\ and\
  \citenamefont {Torroba}}]{Raghu2}%
  \BibitemOpen
  \bibfield  {author} {\bibinfo {author} {\bibfnamefont {J.}~\bibnamefont
  {Aguilera~Damia}}, \bibinfo {author} {\bibfnamefont {S.}~\bibnamefont
  {Kachru}}, \bibinfo {author} {\bibfnamefont {S.}~\bibnamefont {Raghu}},\ and\
  \bibinfo {author} {\bibfnamefont {G.}~\bibnamefont {Torroba}},\ }\bibfield
  {title} {\emph {\bibinfo {title} {{Two dimensional non-Fermi liquid metals: a
  solvable large N limit}}},\ }\href
  {https://doi.org/10.1103/PhysRevLett.123.096402} {\bibfield  {journal}
  {\bibinfo  {journal} {Phys. Rev. Lett.}\ }\textbf {\bibinfo {volume} {123}},\
  \bibinfo {pages} {096402} (\bibinfo {year} {2019})},\ \Eprint
  {https://arxiv.org/abs/1905.08256} {arXiv:1905.08256 [cond-mat.str-el]}
  \BibitemShut {NoStop}%
\bibitem [{\citenamefont {Damia}\ \emph
  {et~al.}(2020{\natexlab{a}})\citenamefont {Damia}, \citenamefont
  {Sol\'\i{}s},\ and\ \citenamefont {Torroba}}]{Torroba1}%
  \BibitemOpen
  \bibfield  {author} {\bibinfo {author} {\bibfnamefont {J.~A.}\ \bibnamefont
  {Damia}}, \bibinfo {author} {\bibfnamefont {M.}~\bibnamefont {Sol\'\i{}s}},\
  and\ \bibinfo {author} {\bibfnamefont {G.}~\bibnamefont {Torroba}},\
  }\bibfield  {title} {\emph {\bibinfo {title} {{How non-Fermi liquids cure
  their infrared divergences}}},\ }\href
  {https://doi.org/10.1103/PhysRevB.102.045147} {\bibfield  {journal} {\bibinfo
   {journal} {Phys. Rev. B}\ }\textbf {\bibinfo {volume} {102}},\ \bibinfo
  {pages} {045147} (\bibinfo {year} {2020}{\natexlab{a}})},\ \Eprint
  {https://arxiv.org/abs/2004.05181} {arXiv:2004.05181 [cond-mat.str-el]}
  \BibitemShut {NoStop}%
\bibitem [{\citenamefont {Damia}\ \emph
  {et~al.}(2020{\natexlab{b}})\citenamefont {Damia}, \citenamefont
  {Sol\'\i{}s},\ and\ \citenamefont {Torroba}}]{Torroba2}%
  \BibitemOpen
  \bibfield  {author} {\bibinfo {author} {\bibfnamefont {J.~A.}\ \bibnamefont
  {Damia}}, \bibinfo {author} {\bibfnamefont {M.}~\bibnamefont {Sol\'\i{}s}},\
  and\ \bibinfo {author} {\bibfnamefont {G.}~\bibnamefont {Torroba}},\
  }\bibfield  {title} {\emph {\bibinfo {title} {{Thermal effects in non-Fermi
  liquid superconductivity}}},\ }\href@noop {} {\  (\bibinfo {year}
  {2020}{\natexlab{b}})},\ \Eprint {https://arxiv.org/abs/2009.11887}
  {arXiv:2009.11887 [cond-mat.str-el]} \BibitemShut {NoStop}%
\bibitem [{\citenamefont {{Ridgway}}\ and\ \citenamefont
  {{Hooley}}(2015)}]{Hooley15}%
  \BibitemOpen
  \bibfield  {author} {\bibinfo {author} {\bibfnamefont {S.~P.}\ \bibnamefont
  {{Ridgway}}}\ and\ \bibinfo {author} {\bibfnamefont {C.~A.}\ \bibnamefont
  {{Hooley}}},\ }\bibfield  {title} {\emph {\bibinfo {title} {{Non-Fermi-Liquid
  Behavior and Anomalous Suppression of Landau Damping in Layered Metals Close
  to Ferromagnetism}}},\ }\href
  {https://doi.org/10.1103/PhysRevLett.114.226404} {\bibfield  {journal}
  {\bibinfo  {journal} {\prl}\ }\textbf {\bibinfo {volume} {114}},\ \bibinfo
  {eid} {226404} (\bibinfo {year} {2015})},\ \Eprint
  {https://arxiv.org/abs/1410.2539} {arXiv:1410.2539 [cond-mat.str-el]}
  \BibitemShut {NoStop}%
\bibitem [{\citenamefont {Patel}\ \emph {et~al.}(2018)\citenamefont {Patel},
  \citenamefont {McGreevy}, \citenamefont {Arovas},\ and\ \citenamefont
  {Sachdev}}]{Patel2018mag}%
  \BibitemOpen
  \bibfield  {author} {\bibinfo {author} {\bibfnamefont {A.~A.}\ \bibnamefont
  {Patel}}, \bibinfo {author} {\bibfnamefont {J.}~\bibnamefont {McGreevy}},
  \bibinfo {author} {\bibfnamefont {D.~P.}\ \bibnamefont {Arovas}},\ and\
  \bibinfo {author} {\bibfnamefont {S.}~\bibnamefont {Sachdev}},\ }\bibfield
  {title} {\emph {\bibinfo {title} {Magnetotransport in a model of a disordered
  strange metal}},\ }\href {https://doi.org/10.1103/PhysRevX.8.021049}
  {\bibfield  {journal} {\bibinfo  {journal} {Phys. Rev. X}\ }\textbf {\bibinfo
  {volume} {8}},\ \bibinfo {pages} {021049} (\bibinfo {year}
  {2018})}\BibitemShut {NoStop}%
\bibitem [{\citenamefont {Chowdhury}\ \emph {et~al.}(2018)\citenamefont
  {Chowdhury}, \citenamefont {Werman}, \citenamefont {Berg},\ and\
  \citenamefont {Senthil}}]{Chowdhury:2018sho}%
  \BibitemOpen
  \bibfield  {author} {\bibinfo {author} {\bibfnamefont {D.}~\bibnamefont
  {Chowdhury}}, \bibinfo {author} {\bibfnamefont {Y.}~\bibnamefont {Werman}},
  \bibinfo {author} {\bibfnamefont {E.}~\bibnamefont {Berg}},\ and\ \bibinfo
  {author} {\bibfnamefont {T.}~\bibnamefont {Senthil}},\ }\bibfield  {title}
  {\emph {\bibinfo {title} {{Translationally invariant non-Fermi liquid metals
  with critical Fermi-surfaces: Solvable models}}},\ }\href
  {https://doi.org/10.1103/PhysRevX.8.031024} {\bibfield  {journal} {\bibinfo
  {journal} {Phys. Rev. X}\ }\textbf {\bibinfo {volume} {8}},\ \bibinfo {pages}
  {031024} (\bibinfo {year} {2018})},\ \Eprint
  {https://arxiv.org/abs/1801.06178} {arXiv:1801.06178 [cond-mat.str-el]}
  \BibitemShut {NoStop}%
\bibitem [{\citenamefont {{Moon}}\ and\ \citenamefont
  {{Chubukov}}(2010)}]{Moon2010}%
  \BibitemOpen
  \bibfield  {author} {\bibinfo {author} {\bibfnamefont {E.-G.}\ \bibnamefont
  {{Moon}}}\ and\ \bibinfo {author} {\bibfnamefont {A.}~\bibnamefont
  {{Chubukov}}},\ }\bibfield  {title} {\emph {\bibinfo {title}
  {{Quantum-critical Pairing with Varying Exponents}}},\ }\href
  {https://doi.org/10.1007/s10909-010-0199-y} {\bibfield  {journal} {\bibinfo
  {journal} {Journal of Low Temperature Physics}\ }\textbf {\bibinfo {volume}
  {161}},\ \bibinfo {pages} {263} (\bibinfo {year} {2010})},\ \Eprint
  {https://arxiv.org/abs/1005.0356} {arXiv:1005.0356 [cond-mat.supr-con]}
  \BibitemShut {NoStop}%
\bibitem [{\citenamefont {{Abanov}}\ and\ \citenamefont
  {{Chubukov}}(2020)}]{Chubukov1}%
  \BibitemOpen
  \bibfield  {author} {\bibinfo {author} {\bibfnamefont {A.}~\bibnamefont
  {{Abanov}}}\ and\ \bibinfo {author} {\bibfnamefont {A.~V.}\ \bibnamefont
  {{Chubukov}}},\ }\bibfield  {title} {\emph {\bibinfo {title} {{Interplay
  between superconductivity and non-Fermi liquid at a quantum critical point in
  a metal. I. The $\gamma$ model and its phase diagram at T =0 : The case $0 <
  \gamma <1$}}},\ }\href {https://doi.org/10.1103/PhysRevB.102.024524}
  {\bibfield  {journal} {\bibinfo  {journal} {Phys. Rev. B}\ }\textbf {\bibinfo
  {volume} {102}},\ \bibinfo {eid} {024524} (\bibinfo {year} {2020})},\ \Eprint
  {https://arxiv.org/abs/2004.13220} {arXiv:2004.13220 [cond-mat.str-el]}
  \BibitemShut {NoStop}%
\bibitem [{\citenamefont {{Wu}}\ \emph {et~al.}(2020)\citenamefont {{Wu}},
  \citenamefont {{Abanov}}, \citenamefont {{Wang}},\ and\ \citenamefont
  {{Chubukov}}}]{Chubukov2}%
  \BibitemOpen
  \bibfield  {author} {\bibinfo {author} {\bibfnamefont {Y.-M.}\ \bibnamefont
  {{Wu}}}, \bibinfo {author} {\bibfnamefont {A.}~\bibnamefont {{Abanov}}},
  \bibinfo {author} {\bibfnamefont {Y.}~\bibnamefont {{Wang}}},\ and\ \bibinfo
  {author} {\bibfnamefont {A.~V.}\ \bibnamefont {{Chubukov}}},\ }\bibfield
  {title} {\emph {\bibinfo {title} {{Interplay between superconductivity and
  non-Fermi liquid at a quantum critical point in a metal. II. The $\gamma$
  model at a finite T for $0 < \gamma <1$}}},\ }\href
  {https://doi.org/10.1103/PhysRevB.102.024525} {\bibfield  {journal} {\bibinfo
   {journal} {Phys. Rev. B}\ }\textbf {\bibinfo {volume} {102}},\ \bibinfo
  {eid} {024525} (\bibinfo {year} {2020})},\ \Eprint
  {https://arxiv.org/abs/2006.02968} {arXiv:2006.02968 [cond-mat.supr-con]}
  \BibitemShut {NoStop}%
\bibitem [{\citenamefont {{Chubukov}}\ and\ \citenamefont
  {{Abanov}}(2021)}]{Chubukov3}%
  \BibitemOpen
  \bibfield  {author} {\bibinfo {author} {\bibfnamefont {A.~V.}\ \bibnamefont
  {{Chubukov}}}\ and\ \bibinfo {author} {\bibfnamefont {A.}~\bibnamefont
  {{Abanov}}},\ }\bibfield  {title} {\emph {\bibinfo {title} {{Pairing by a
  Dynamical Interaction in a Metal}}},\ }\href
  {https://doi.org/10.1134/S1063776121040051} {\bibfield  {journal} {\bibinfo
  {journal} {Soviet Journal of Experimental and Theoretical Physics}\ }\textbf
  {\bibinfo {volume} {132}},\ \bibinfo {pages} {606} (\bibinfo {year}
  {2021})},\ \Eprint {https://arxiv.org/abs/2012.11777} {arXiv:2012.11777
  [cond-mat.supr-con]} \BibitemShut {NoStop}%
\bibitem [{\citenamefont {{Wang}}\ and\ \citenamefont {{Berg}}(2019)}]{Berg1}%
  \BibitemOpen
  \bibfield  {author} {\bibinfo {author} {\bibfnamefont {X.}~\bibnamefont
  {{Wang}}}\ and\ \bibinfo {author} {\bibfnamefont {E.}~\bibnamefont
  {{Berg}}},\ }\bibfield  {title} {\emph {\bibinfo {title} {{Scattering
  mechanisms and electrical transport near an Ising nematic quantum critical
  point}}},\ }\href {https://doi.org/10.1103/PhysRevB.99.235136} {\bibfield
  {journal} {\bibinfo  {journal} {\prb}\ }\textbf {\bibinfo {volume} {99}},\
  \bibinfo {eid} {235136} (\bibinfo {year} {2019})},\ \Eprint
  {https://arxiv.org/abs/1902.04590} {arXiv:1902.04590 [cond-mat.str-el]}
  \BibitemShut {NoStop}%
\bibitem [{\citenamefont {{Klein}}\ \emph {et~al.}(2020)\citenamefont
  {{Klein}}, \citenamefont {{Chubukov}}, \citenamefont {{Schattner}},\ and\
  \citenamefont {{Berg}}}]{Berg2}%
  \BibitemOpen
  \bibfield  {author} {\bibinfo {author} {\bibfnamefont {A.}~\bibnamefont
  {{Klein}}}, \bibinfo {author} {\bibfnamefont {A.~V.}\ \bibnamefont
  {{Chubukov}}}, \bibinfo {author} {\bibfnamefont {Y.}~\bibnamefont
  {{Schattner}}},\ and\ \bibinfo {author} {\bibfnamefont {E.}~\bibnamefont
  {{Berg}}},\ }\bibfield  {title} {\emph {\bibinfo {title} {{Normal State
  Properties of Quantum Critical Metals at Finite Temperature}}},\ }\href
  {https://doi.org/10.1103/PhysRevX.10.031053} {\bibfield  {journal} {\bibinfo
  {journal} {Physical Review X}\ }\textbf {\bibinfo {volume} {10}},\ \bibinfo
  {eid} {031053} (\bibinfo {year} {2020})},\ \Eprint
  {https://arxiv.org/abs/2003.09431} {arXiv:2003.09431 [cond-mat.str-el]}
  \BibitemShut {NoStop}%
\bibitem [{\citenamefont {{Grossman}}\ \emph {et~al.}(2021)\citenamefont
  {{Grossman}}, \citenamefont {{Hofmann}}, \citenamefont {{Holder}},\ and\
  \citenamefont {{Berg}}}]{Berg3}%
  \BibitemOpen
  \bibfield  {author} {\bibinfo {author} {\bibfnamefont {O.}~\bibnamefont
  {{Grossman}}}, \bibinfo {author} {\bibfnamefont {J.~S.}\ \bibnamefont
  {{Hofmann}}}, \bibinfo {author} {\bibfnamefont {T.}~\bibnamefont
  {{Holder}}},\ and\ \bibinfo {author} {\bibfnamefont {E.}~\bibnamefont
  {{Berg}}},\ }\bibfield  {title} {\emph {\bibinfo {title} {{Specific Heat of a
  Quantum Critical Metal}}},\ }\href
  {https://doi.org/10.1103/PhysRevLett.127.017601} {\bibfield  {journal}
  {\bibinfo  {journal} {Phys. Rev. Lett.}\ }\textbf {\bibinfo {volume} {127}},\
  \bibinfo {eid} {017601} (\bibinfo {year} {2021})},\ \Eprint
  {https://arxiv.org/abs/2009.11280} {arXiv:2009.11280 [cond-mat.str-el]}
  \BibitemShut {NoStop}%
\bibitem [{\citenamefont {{Chowdhury}}\ and\ \citenamefont
  {{Berg}}(2020)}]{DCBerg}%
  \BibitemOpen
  \bibfield  {author} {\bibinfo {author} {\bibfnamefont {D.}~\bibnamefont
  {{Chowdhury}}}\ and\ \bibinfo {author} {\bibfnamefont {E.}~\bibnamefont
  {{Berg}}},\ }\bibfield  {title} {\emph {\bibinfo {title} {{The unreasonable
  effectiveness of Eliashberg theory for pairing of non-Fermi liquids}}},\
  }\href {https://doi.org/10.1016/j.aop.2020.168125} {\bibfield  {journal}
  {\bibinfo  {journal} {Annals of Physics}\ }\textbf {\bibinfo {volume}
  {417}},\ \bibinfo {eid} {168125} (\bibinfo {year} {2020})},\ \Eprint
  {https://arxiv.org/abs/1912.07646} {arXiv:1912.07646 [cond-mat.supr-con]}
  \BibitemShut {NoStop}%
\bibitem [{\citenamefont {Raether}\ \emph {et~al.}(2020)\citenamefont
  {Raether}, \citenamefont {Blanco},\ and\ \citenamefont
  {Chowdhury}}]{DebanjanAPS}%
  \BibitemOpen
  \bibfield  {author} {\bibinfo {author} {\bibfnamefont {A.~F.~S.}\
  \bibnamefont {Raether}}, \bibinfo {author} {\bibfnamefont {F.~M.}\
  \bibnamefont {Blanco}},\ and\ \bibinfo {author} {\bibfnamefont
  {D.}~\bibnamefont {Chowdhury}},\ }\bibfield  {title} {\emph {\bibinfo {title}
  {{A cascade of non-Fermi liquid crossovers from an interplay of local and
  bosonic quantum criticality}}},\ }\href
  {https://meetings.aps.org/Meeting/MAR20/Session/F50.10} {\bibfield  {journal}
  {\bibinfo  {journal} {Bull. Am. Phys. Soc.}\ }\textbf {\bibinfo {volume}
  {65}},\ \bibinfo {pages} {F50.00010} (\bibinfo {year} {2020})}\BibitemShut
  {NoStop}%
\bibitem [{\citenamefont {Patel}\ and\ \citenamefont
  {Sachdev}(2018)}]{Patel:2018eak}%
  \BibitemOpen
  \bibfield  {author} {\bibinfo {author} {\bibfnamefont {A.~A.}\ \bibnamefont
  {Patel}}\ and\ \bibinfo {author} {\bibfnamefont {S.}~\bibnamefont
  {Sachdev}},\ }\bibfield  {title} {\emph {\bibinfo {title} {Critical strange
  metal from fluctuating gauge fields in a solvable random model}},\ }\href
  {https://doi.org/10.1103/PhysRevB.98.125134} {\bibfield  {journal} {\bibinfo
  {journal} {Phys. Rev. B}\ }\textbf {\bibinfo {volume} {98}},\ \bibinfo
  {pages} {125134} (\bibinfo {year} {2018})}\BibitemShut {NoStop}%
\bibitem [{\citenamefont {Esterlis}\ and\ \citenamefont
  {Schmalian}(2019)}]{Ilya1}%
  \BibitemOpen
  \bibfield  {author} {\bibinfo {author} {\bibfnamefont {I.}~\bibnamefont
  {Esterlis}}\ and\ \bibinfo {author} {\bibfnamefont {J.}~\bibnamefont
  {Schmalian}},\ }\bibfield  {title} {\emph {\bibinfo {title} {{Cooper pairing
  of incoherent electrons: an electron-phonon version of the Sachdev-Ye-Kitaev
  model}}},\ }\href {https://doi.org/10.1103/PhysRevB.100.115132} {\bibfield
  {journal} {\bibinfo  {journal} {Phys. Rev. B}\ }\textbf {\bibinfo {volume}
  {100}},\ \bibinfo {pages} {115132} (\bibinfo {year} {2019})},\ \Eprint
  {https://arxiv.org/abs/1906.04747} {arXiv:1906.04747 [cond-mat.str-el]}
  \BibitemShut {NoStop}%
\bibitem [{\citenamefont {{Hauck}}\ \emph {et~al.}(2020)\citenamefont
  {{Hauck}}, \citenamefont {{Klug}}, \citenamefont {{Esterlis}},\ and\
  \citenamefont {{Schmalian}}}]{Ilya2}%
  \BibitemOpen
  \bibfield  {author} {\bibinfo {author} {\bibfnamefont {D.}~\bibnamefont
  {{Hauck}}}, \bibinfo {author} {\bibfnamefont {M.~J.}\ \bibnamefont {{Klug}}},
  \bibinfo {author} {\bibfnamefont {I.}~\bibnamefont {{Esterlis}}},\ and\
  \bibinfo {author} {\bibfnamefont {J.}~\bibnamefont {{Schmalian}}},\
  }\bibfield  {title} {\emph {\bibinfo {title} {{Eliashberg equations for an
  electron-phonon version of the Sachdev-Ye-Kitaev model: Pair breaking in
  non-Fermi liquid superconductors}}},\ }\href
  {https://doi.org/10.1016/j.aop.2020.168120} {\bibfield  {journal} {\bibinfo
  {journal} {Annals of Physics}\ }\textbf {\bibinfo {volume} {417}},\ \bibinfo
  {eid} {168120} (\bibinfo {year} {2020})},\ \Eprint
  {https://arxiv.org/abs/1911.04328} {arXiv:1911.04328 [cond-mat.str-el]}
  \BibitemShut {NoStop}%
\bibitem [{\citenamefont {Wang}(2020)}]{Wang:2019bpd}%
  \BibitemOpen
  \bibfield  {author} {\bibinfo {author} {\bibfnamefont {Y.}~\bibnamefont
  {Wang}},\ }\bibfield  {title} {\emph {\bibinfo {title} {{Solvable
  Strong-coupling Quantum Dot Model with a Non-Fermi-liquid Pairing
  Transition}}},\ }\href {https://doi.org/10.1103/PhysRevLett.124.017002}
  {\bibfield  {journal} {\bibinfo  {journal} {Phys. Rev. Lett.}\ }\textbf
  {\bibinfo {volume} {124}},\ \bibinfo {pages} {017002} (\bibinfo {year}
  {2020})},\ \Eprint {https://arxiv.org/abs/1904.07240} {arXiv:1904.07240
  [cond-mat.str-el]} \BibitemShut {NoStop}%
\bibitem [{\citenamefont {Wang}\ and\ \citenamefont
  {Chubukov}(2020)}]{Wang:2020dtj}%
  \BibitemOpen
  \bibfield  {author} {\bibinfo {author} {\bibfnamefont {Y.}~\bibnamefont
  {Wang}}\ and\ \bibinfo {author} {\bibfnamefont {A.~V.}\ \bibnamefont
  {Chubukov}},\ }\bibfield  {title} {\emph {\bibinfo {title} {{Quantum Phase
  Transition in the Yukawa-SYK Model}}},\ }\href
  {https://doi.org/10.1103/PhysRevResearch.2.033084} {\bibfield  {journal}
  {\bibinfo  {journal} {Phys. Rev. Res.}\ }\textbf {\bibinfo {volume} {2}},\
  \bibinfo {pages} {033084} (\bibinfo {year} {2020})},\ \Eprint
  {https://arxiv.org/abs/2005.07205} {arXiv:2005.07205 [cond-mat.str-el]}
  \BibitemShut {NoStop}%
\bibitem [{\citenamefont {{Aldape}}\ \emph {et~al.}(2022)\citenamefont
  {{Aldape}}, \citenamefont {{Cookmeyer}}, \citenamefont {{Patel}},\ and\
  \citenamefont {{Altman}}}]{Altman1}%
  \BibitemOpen
  \bibfield  {author} {\bibinfo {author} {\bibfnamefont {E.~E.}\ \bibnamefont
  {{Aldape}}}, \bibinfo {author} {\bibfnamefont {T.}~\bibnamefont
  {{Cookmeyer}}}, \bibinfo {author} {\bibfnamefont {A.~A.}\ \bibnamefont
  {{Patel}}},\ and\ \bibinfo {author} {\bibfnamefont {E.}~\bibnamefont
  {{Altman}}},\ }\bibfield  {title} {\emph {\bibinfo {title} {{Solvable theory
  of a strange metal at the breakdown of a heavy Fermi liquid}}},\ }\href
  {https://doi.org/10.1103/PhysRevB.105.235111} {\bibfield  {journal} {\bibinfo
   {journal} {Phys. Rev. B}\ }\textbf {\bibinfo {volume} {105}},\ \bibinfo
  {eid} {235111} (\bibinfo {year} {2022})},\ \Eprint
  {https://arxiv.org/abs/2012.00763} {arXiv:2012.00763 [cond-mat.str-el]}
  \BibitemShut {NoStop}%
\bibitem [{\citenamefont {Patel}\ and\ \citenamefont
  {Sachdev}(2017)}]{Patel:2016wdy}%
  \BibitemOpen
  \bibfield  {author} {\bibinfo {author} {\bibfnamefont {A.~A.}\ \bibnamefont
  {Patel}}\ and\ \bibinfo {author} {\bibfnamefont {S.}~\bibnamefont
  {Sachdev}},\ }\bibfield  {title} {\emph {\bibinfo {title} {{Quantum chaos on
  a critical Fermi surface}}},\ }\href
  {https://doi.org/10.1073/pnas.1618185114} {\bibfield  {journal} {\bibinfo
  {journal} {Proc. Nat. Acad. Sci.}\ }\textbf {\bibinfo {volume} {114}},\
  \bibinfo {pages} {1844} (\bibinfo {year} {2017})},\ \Eprint
  {https://arxiv.org/abs/1611.00003} {arXiv:1611.00003 [cond-mat.str-el]}
  \BibitemShut {NoStop}%
\bibitem [{\citenamefont {Patel}\ and\ \citenamefont
  {Sachdev}(2019)}]{Patel:2019qce}%
  \BibitemOpen
  \bibfield  {author} {\bibinfo {author} {\bibfnamefont {A.~A.}\ \bibnamefont
  {Patel}}\ and\ \bibinfo {author} {\bibfnamefont {S.}~\bibnamefont
  {Sachdev}},\ }\bibfield  {title} {\emph {\bibinfo {title} {{Theory of a
  Planckian metal}}},\ }\href {https://doi.org/10.1103/PhysRevLett.123.066601}
  {\bibfield  {journal} {\bibinfo  {journal} {Phys. Rev. Lett.}\ }\textbf
  {\bibinfo {volume} {123}},\ \bibinfo {pages} {066601} (\bibinfo {year}
  {2019})},\ \bibinfo {note} {{(The `resonance' condition employed here can be
  viewed as a rationale for interactions mediated by the exchange of a critical
  scalar)}},\ \Eprint {https://arxiv.org/abs/1906.03265} {arXiv:1906.03265
  [cond-mat.str-el]} \BibitemShut {NoStop}%
\bibitem [{\citenamefont {Oganesyan}\ \emph {et~al.}(2001)\citenamefont
  {Oganesyan}, \citenamefont {Kivelson},\ and\ \citenamefont
  {Fradkin}}]{oganesyan2001}%
  \BibitemOpen
  \bibfield  {author} {\bibinfo {author} {\bibfnamefont {V.}~\bibnamefont
  {Oganesyan}}, \bibinfo {author} {\bibfnamefont {S.~A.}\ \bibnamefont
  {Kivelson}},\ and\ \bibinfo {author} {\bibfnamefont {E.}~\bibnamefont
  {Fradkin}},\ }\bibfield  {title} {\emph {\bibinfo {title} {Quantum theory of
  a nematic fermi fluid}},\ }\href {https://doi.org/10.1103/PhysRevB.64.195109}
  {\bibfield  {journal} {\bibinfo  {journal} {Phys. Rev. B}\ }\textbf {\bibinfo
  {volume} {64}},\ \bibinfo {pages} {195109} (\bibinfo {year}
  {2001})}\BibitemShut {NoStop}%
\bibitem [{\citenamefont {{Lee}}(2018)}]{Lee_ARCMP}%
  \BibitemOpen
  \bibfield  {author} {\bibinfo {author} {\bibfnamefont {S.-S.}\ \bibnamefont
  {{Lee}}},\ }\bibfield  {title} {\emph {\bibinfo {title} {{Recent Developments
  in Non-Fermi Liquid Theory}}},\ }\href
  {https://doi.org/10.1146/annurev-conmatphys-031016-025531} {\bibfield
  {journal} {\bibinfo  {journal} {Annual Review of Condensed Matter Physics}\
  }\textbf {\bibinfo {volume} {9}},\ \bibinfo {pages} {227} (\bibinfo {year}
  {2018})},\ \Eprint {https://arxiv.org/abs/1703.08172} {arXiv:1703.08172
  [cond-mat.str-el]} \BibitemShut {NoStop}%
\bibitem [{\citenamefont {{Sachdev}}\ and\ \citenamefont {{Ye}}(1993)}]{SY92}%
  \BibitemOpen
  \bibfield  {author} {\bibinfo {author} {\bibfnamefont {S.}~\bibnamefont
  {{Sachdev}}}\ and\ \bibinfo {author} {\bibfnamefont {J.}~\bibnamefont
  {{Ye}}},\ }\bibfield  {title} {\emph {\bibinfo {title} {{Gapless spin-fluid
  ground state in a random quantum Heisenberg magnet}}},\ }\href
  {https://doi.org/10.1103/PhysRevLett.70.3339} {\bibfield  {journal} {\bibinfo
   {journal} {Phys. Rev. Lett.}\ }\textbf {\bibinfo {volume} {70}},\ \bibinfo
  {pages} {3339} (\bibinfo {year} {1993})},\ \Eprint
  {https://arxiv.org/abs/cond-mat/9212030} {cond-mat/9212030} \BibitemShut
  {NoStop}%
\bibitem [{\citenamefont {{Kitaev}}(2015)}]{kitaev2015talk}%
  \BibitemOpen
  \bibfield  {author} {\bibinfo {author} {\bibfnamefont {A.~Y.}\ \bibnamefont
  {{Kitaev}}},\ }\bibfield  {title} {\emph {\bibinfo {title} {{Talks at KITP,
  University of California, Santa Barbara}}},\ }\href
  {http://online.kitp.ucsb.edu/online/entangled15/} {\bibfield  {journal}
  {\bibinfo  {journal} {Entanglement in Strongly-Correlated Quantum Matter}\ }
  (\bibinfo {year} {2015})}\BibitemShut {NoStop}%
\bibitem [{\citenamefont {{Sachdev}}(2015)}]{SS15}%
  \BibitemOpen
  \bibfield  {author} {\bibinfo {author} {\bibfnamefont {S.}~\bibnamefont
  {{Sachdev}}},\ }\bibfield  {title} {\emph {\bibinfo {title}
  {{Bekenstein-Hawking Entropy and Strange Metals}}},\ }\href
  {https://doi.org/10.1103/PhysRevX.5.041025} {\bibfield  {journal} {\bibinfo
  {journal} {Phys. Rev. X}\ }\textbf {\bibinfo {volume} {5}},\ \bibinfo {eid}
  {041025} (\bibinfo {year} {2015})},\ \Eprint
  {https://arxiv.org/abs/1506.05111} {arXiv:1506.05111 [hep-th]} \BibitemShut
  {NoStop}%
\bibitem [{\citenamefont {Maldacena}\ and\ \citenamefont
  {Stanford}(2016)}]{JMDS16}%
  \BibitemOpen
  \bibfield  {author} {\bibinfo {author} {\bibfnamefont {J.}~\bibnamefont
  {Maldacena}}\ and\ \bibinfo {author} {\bibfnamefont {D.}~\bibnamefont
  {Stanford}},\ }\bibfield  {title} {\emph {\bibinfo {title} {{Remarks on the
  Sachdev-Ye-Kitaev model}}},\ }\href
  {https://doi.org/10.1103/PhysRevD.94.106002} {\bibfield  {journal} {\bibinfo
  {journal} {Phys. Rev. D}\ }\textbf {\bibinfo {volume} {94}},\ \bibinfo
  {pages} {106002} (\bibinfo {year} {2016})},\ \Eprint
  {https://arxiv.org/abs/1604.07818} {arXiv:1604.07818 [hep-th]} \BibitemShut
  {NoStop}%
%%CITATION = ARXIV:1604.07818;%%
\bibitem [{\citenamefont {Esterlis}\ \emph {et~al.}(2021)\citenamefont
  {Esterlis}, \citenamefont {Guo}, \citenamefont {Patel},\ and\ \citenamefont
  {Sachdev}}]{Esterlis:2021eth}%
  \BibitemOpen
  \bibfield  {author} {\bibinfo {author} {\bibfnamefont {I.}~\bibnamefont
  {Esterlis}}, \bibinfo {author} {\bibfnamefont {H.}~\bibnamefont {Guo}},
  \bibinfo {author} {\bibfnamefont {A.~A.}\ \bibnamefont {Patel}},\ and\
  \bibinfo {author} {\bibfnamefont {S.}~\bibnamefont {Sachdev}},\ }\bibfield
  {title} {\emph {\bibinfo {title} {{Large $N$ theory of critical Fermi
  surfaces}}},\ }\href {https://doi.org/10.1103/PhysRevB.103.235129} {\bibfield
   {journal} {\bibinfo  {journal} {Phys. Rev. B}\ }\textbf {\bibinfo {volume}
  {103}},\ \bibinfo {pages} {235129} (\bibinfo {year} {2021})},\ \Eprint
  {https://arxiv.org/abs/2103.08615} {arXiv:2103.08615 [cond-mat.str-el]}
  \BibitemShut {NoStop}%
\bibitem [{\citenamefont {Bagrets}\ \emph {et~al.}(2016)\citenamefont
  {Bagrets}, \citenamefont {Altland},\ and\ \citenamefont
  {Kamenev}}]{Bagrets:2016cdf}%
  \BibitemOpen
  \bibfield  {author} {\bibinfo {author} {\bibfnamefont {D.}~\bibnamefont
  {Bagrets}}, \bibinfo {author} {\bibfnamefont {A.}~\bibnamefont {Altland}},\
  and\ \bibinfo {author} {\bibfnamefont {A.}~\bibnamefont {Kamenev}},\
  }\bibfield  {title} {\emph {\bibinfo {title}
  {{Sachdev\textendash{}Ye\textendash{}Kitaev model as Liouville quantum
  mechanics}}},\ }\href {https://doi.org/10.1016/j.nuclphysb.2016.08.002}
  {\bibfield  {journal} {\bibinfo  {journal} {Nucl. Phys. B}\ }\textbf
  {\bibinfo {volume} {911}},\ \bibinfo {pages} {191} (\bibinfo {year}
  {2016})},\ \Eprint {https://arxiv.org/abs/1607.00694} {arXiv:1607.00694
  [cond-mat.str-el]} \BibitemShut {NoStop}%
\bibitem [{\citenamefont {Kitaev}\ and\ \citenamefont
  {Suh}(2018)}]{Kitaev:2017awl}%
  \BibitemOpen
  \bibfield  {author} {\bibinfo {author} {\bibfnamefont {A.}~\bibnamefont
  {Kitaev}}\ and\ \bibinfo {author} {\bibfnamefont {S.~J.}\ \bibnamefont
  {Suh}},\ }\bibfield  {title} {\emph {\bibinfo {title} {{The soft mode in the
  Sachdev-Ye-Kitaev model and its gravity dual}}},\ }\href
  {https://doi.org/10.1007/JHEP05(2018)183} {\bibfield  {journal} {\bibinfo
  {journal} {JHEP}\ }\textbf {\bibinfo {volume} {05}},\ \bibinfo {pages}
  {183}},\ \Eprint {https://arxiv.org/abs/1711.08467} {arXiv:1711.08467
  [hep-th]} \BibitemShut {NoStop}%
\bibitem [{\citenamefont {Proust}\ and\ \citenamefont
  {Taillefer}(2019)}]{ProustTaillefer2019}%
  \BibitemOpen
  \bibfield  {author} {\bibinfo {author} {\bibfnamefont {C.}~\bibnamefont
  {Proust}}\ and\ \bibinfo {author} {\bibfnamefont {L.}~\bibnamefont
  {Taillefer}},\ }\bibfield  {title} {\emph {\bibinfo {title} {The remarkable
  underlying ground states of cuprate superconductors}},\ }\href
  {https://doi.org/10.1146/annurev-conmatphys-031218-013210} {\bibfield
  {journal} {\bibinfo  {journal} {Annual Review of Condensed Matter Physics}\
  }\textbf {\bibinfo {volume} {10}},\ \bibinfo {pages} {409} (\bibinfo {year}
  {2019})},\ \Eprint
  {https://arxiv.org/abs/https://doi.org/10.1146/annurev-conmatphys-031218-013210}
  {https://doi.org/10.1146/annurev-conmatphys-031218-013210} \BibitemShut
  {NoStop}%
\bibitem [{\citenamefont {Bruin}\ \emph {et~al.}(2013)\citenamefont {Bruin},
  \citenamefont {Sakai}, \citenamefont {Perry},\ and\ \citenamefont
  {Mackenzie}}]{bruin}%
  \BibitemOpen
  \bibfield  {author} {\bibinfo {author} {\bibfnamefont {J.~A.~N.}\
  \bibnamefont {Bruin}}, \bibinfo {author} {\bibfnamefont {H.}~\bibnamefont
  {Sakai}}, \bibinfo {author} {\bibfnamefont {R.~S.}\ \bibnamefont {Perry}},\
  and\ \bibinfo {author} {\bibfnamefont {A.~P.}\ \bibnamefont {Mackenzie}},\
  }\bibfield  {title} {\emph {\bibinfo {title} {{Similarity of Scattering Rates
  in Metals Showing $T$-Linear Resistivity}}},\ }\href
  {https://doi.org/10.1126/science.1227612} {\bibfield  {journal} {\bibinfo
  {journal} {Science}\ }\textbf {\bibinfo {volume} {339}},\ \bibinfo {pages}
  {804} (\bibinfo {year} {2013})}\BibitemShut {NoStop}%
\bibitem [{\citenamefont {Zaanen}(2004)}]{Zaanen}%
  \BibitemOpen
  \bibfield  {author} {\bibinfo {author} {\bibfnamefont {J.}~\bibnamefont
  {Zaanen}},\ }\bibfield  {title} {\emph {\bibinfo {title} {Why the temperature
  is high}},\ }\href {https://doi.org/10.1038/430512a} {\bibfield  {journal}
  {\bibinfo  {journal} {Nature}\ }\textbf {\bibinfo {volume} {430}},\ \bibinfo
  {pages} {512} (\bibinfo {year} {2004})}\BibitemShut {NoStop}%
\bibitem [{\citenamefont {{Grissonnanche}}\ \emph {et~al.}(2021)\citenamefont
  {{Grissonnanche}}, \citenamefont {{Fang}}, \citenamefont {{Legros}},
  \citenamefont {{Verret}}, \citenamefont {{Lalibert{\'e}}}, \citenamefont
  {{Collignon}}, \citenamefont {{Zhou}}, \citenamefont {{Graf}}, \citenamefont
  {{Goddard}}, \citenamefont {{Taillefer}},\ and\ \citenamefont
  {{Ramshaw}}}]{Gael21}%
  \BibitemOpen
  \bibfield  {author} {\bibinfo {author} {\bibfnamefont {G.}~\bibnamefont
  {{Grissonnanche}}}, \bibinfo {author} {\bibfnamefont {Y.}~\bibnamefont
  {{Fang}}}, \bibinfo {author} {\bibfnamefont {A.}~\bibnamefont {{Legros}}},
  \bibinfo {author} {\bibfnamefont {S.}~\bibnamefont {{Verret}}}, \bibinfo
  {author} {\bibfnamefont {F.}~\bibnamefont {{Lalibert{\'e}}}}, \bibinfo
  {author} {\bibfnamefont {C.}~\bibnamefont {{Collignon}}}, \bibinfo {author}
  {\bibfnamefont {J.}~\bibnamefont {{Zhou}}}, \bibinfo {author} {\bibfnamefont
  {D.}~\bibnamefont {{Graf}}}, \bibinfo {author} {\bibfnamefont {P.~A.}\
  \bibnamefont {{Goddard}}}, \bibinfo {author} {\bibfnamefont {L.}~\bibnamefont
  {{Taillefer}}},\ and\ \bibinfo {author} {\bibfnamefont {B.~J.}\ \bibnamefont
  {{Ramshaw}}},\ }\bibfield  {title} {\emph {\bibinfo {title} {{Linear-in
  temperature resistivity from an isotropic Planckian scattering rate}}},\
  }\href {https://doi.org/10.1038/s41586-021-03697-8} {\bibfield  {journal}
  {\bibinfo  {journal} {Nature}\ }\textbf {\bibinfo {volume} {595}},\ \bibinfo
  {pages} {667} (\bibinfo {year} {2021})},\ \Eprint
  {https://arxiv.org/abs/2011.13054} {arXiv:2011.13054 [cond-mat.str-el]}
  \BibitemShut {NoStop}%
\bibitem [{\citenamefont {Taupin}\ and\ \citenamefont
  {Paschen}(2022)}]{Paschen22}%
  \BibitemOpen
  \bibfield  {author} {\bibinfo {author} {\bibfnamefont {M.}~\bibnamefont
  {Taupin}}\ and\ \bibinfo {author} {\bibfnamefont {S.}~\bibnamefont
  {Paschen}},\ }\bibfield  {title} {\emph {\bibinfo {title} {Are heavy fermion
  strange metals planckian?}},\ }\href {https://doi.org/10.3390/cryst12020251}
  {\bibfield  {journal} {\bibinfo  {journal} {Crystals}\ }\textbf {\bibinfo
  {volume} {12}},\ \bibinfo {pages} {251} (\bibinfo {year} {2022})},\ \Eprint
  {https://arxiv.org/abs/2201.02820} {arXiv:2201.02820 [cond-mat.str-el]}
  \BibitemShut {NoStop}%
\bibitem [{\citenamefont {{Ahn}}\ and\ \citenamefont {{Das
  Sarma}}(2022)}]{Sankar2022}%
  \BibitemOpen
  \bibfield  {author} {\bibinfo {author} {\bibfnamefont {S.}~\bibnamefont
  {{Ahn}}}\ and\ \bibinfo {author} {\bibfnamefont {S.}~\bibnamefont {{Das
  Sarma}}},\ }\bibfield  {title} {\emph {\bibinfo {title} {{Planckian
  properties of two-dimensional semiconductor systems}}},\ }\href
  {https://doi.org/10.1103/PhysRevB.106.155427} {\bibfield  {journal} {\bibinfo
   {journal} {Phys. Rev. B}\ }\textbf {\bibinfo {volume} {106}},\ \bibinfo
  {eid} {155427} (\bibinfo {year} {2022})},\ \Eprint
  {https://arxiv.org/abs/2204.02982} {arXiv:2204.02982 [cond-mat.mes-hall]}
  \BibitemShut {NoStop}%
\bibitem [{\citenamefont {Varma}\ \emph {et~al.}(1989)\citenamefont {Varma},
  \citenamefont {Littlewood}, \citenamefont {Schmitt-Rink}, \citenamefont
  {Abrahams},\ and\ \citenamefont {Ruckenstein}}]{MFL89}%
  \BibitemOpen
  \bibfield  {author} {\bibinfo {author} {\bibfnamefont {C.~M.}\ \bibnamefont
  {Varma}}, \bibinfo {author} {\bibfnamefont {P.~B.}\ \bibnamefont
  {Littlewood}}, \bibinfo {author} {\bibfnamefont {S.}~\bibnamefont
  {Schmitt-Rink}}, \bibinfo {author} {\bibfnamefont {E.}~\bibnamefont
  {Abrahams}},\ and\ \bibinfo {author} {\bibfnamefont {A.~E.}\ \bibnamefont
  {Ruckenstein}},\ }\bibfield  {title} {\emph {\bibinfo {title} {Phenomenology
  of the normal state of {C}u-{O} high-temperature superconductors}},\ }\href
  {https://doi.org/10.1103/PhysRevLett.63.1996} {\bibfield  {journal} {\bibinfo
   {journal} {Phys. Rev. Lett.}\ }\textbf {\bibinfo {volume} {63}},\ \bibinfo
  {pages} {1996} (\bibinfo {year} {1989})}\BibitemShut {NoStop}%
\bibitem [{\citenamefont {Patel}\ \emph {et~al.}(2022)\citenamefont {Patel},
  \citenamefont {Guo}, \citenamefont {Esterlis},\ and\ \citenamefont
  {Sachdev}}]{Patel:2022gdh}%
  \BibitemOpen
  \bibfield  {author} {\bibinfo {author} {\bibfnamefont {A.~A.}\ \bibnamefont
  {Patel}}, \bibinfo {author} {\bibfnamefont {H.}~\bibnamefont {Guo}}, \bibinfo
  {author} {\bibfnamefont {I.}~\bibnamefont {Esterlis}},\ and\ \bibinfo
  {author} {\bibfnamefont {S.}~\bibnamefont {Sachdev}},\ }\bibfield  {title}
  {\emph {\bibinfo {title} {{Universal theory of strange metals from spatially
  random interactions}}},\ }\href@noop {} {\bibfield  {journal} {\bibinfo
  {journal} {Science, to appear}\ } (\bibinfo {year} {2022})},\ \Eprint
  {https://arxiv.org/abs/2203.04990} {arXiv:2203.04990 [cond-mat.str-el]}
  \BibitemShut {NoStop}%
\bibitem [{\citenamefont {{Rech}}\ \emph {et~al.}(2006)\citenamefont {{Rech}},
  \citenamefont {{P{\'e}pin}},\ and\ \citenamefont {{Chubukov}}}]{Rech06}%
  \BibitemOpen
  \bibfield  {author} {\bibinfo {author} {\bibfnamefont {J.}~\bibnamefont
  {{Rech}}}, \bibinfo {author} {\bibfnamefont {C.}~\bibnamefont
  {{P{\'e}pin}}},\ and\ \bibinfo {author} {\bibfnamefont {A.~V.}\ \bibnamefont
  {{Chubukov}}},\ }\bibfield  {title} {\emph {\bibinfo {title} {{Quantum
  critical behavior in itinerant electron systems: Eliashberg theory and
  instability of a ferromagnetic quantum critical point}}},\ }\href
  {https://doi.org/10.1103/PhysRevB.74.195126} {\bibfield  {journal} {\bibinfo
  {journal} {Phys. Rev. B}\ }\textbf {\bibinfo {volume} {74}},\ \bibinfo {eid}
  {195126} (\bibinfo {year} {2006})},\ \Eprint
  {https://arxiv.org/abs/cond-mat/0605306} {arXiv:cond-mat/0605306
  [cond-mat.str-el]} \BibitemShut {NoStop}%
\bibitem [{\citenamefont {{Maslov}}\ \emph {et~al.}(2011)\citenamefont
  {{Maslov}}, \citenamefont {{Yudson}},\ and\ \citenamefont
  {{Chubukov}}}]{ChubukovMaslov}%
  \BibitemOpen
  \bibfield  {author} {\bibinfo {author} {\bibfnamefont {D.~L.}\ \bibnamefont
  {{Maslov}}}, \bibinfo {author} {\bibfnamefont {V.~I.}\ \bibnamefont
  {{Yudson}}},\ and\ \bibinfo {author} {\bibfnamefont {A.~V.}\ \bibnamefont
  {{Chubukov}}},\ }\bibfield  {title} {\emph {\bibinfo {title} {{Resistivity of
  a Non-Galilean-Invariant Fermi Liquid near Pomeranchuk Quantum
  Criticality}}},\ }\href {https://doi.org/10.1103/PhysRevLett.106.106403}
  {\bibfield  {journal} {\bibinfo  {journal} {Phys. Rev. Lett.}\ }\textbf
  {\bibinfo {volume} {106}},\ \bibinfo {eid} {106403} (\bibinfo {year}
  {2011})},\ \Eprint {https://arxiv.org/abs/1012.0069} {arXiv:1012.0069
  [cond-mat.str-el]} \BibitemShut {NoStop}%
\bibitem [{\citenamefont {{Pal}}\ \emph {et~al.}(2012)\citenamefont {{Pal}},
  \citenamefont {{Yudson}},\ and\ \citenamefont {{Maslov}}}]{Maslov12}%
  \BibitemOpen
  \bibfield  {author} {\bibinfo {author} {\bibfnamefont {H.~K.}\ \bibnamefont
  {{Pal}}}, \bibinfo {author} {\bibfnamefont {V.~I.}\ \bibnamefont
  {{Yudson}}},\ and\ \bibinfo {author} {\bibfnamefont {D.~L.}\ \bibnamefont
  {{Maslov}}},\ }\bibfield  {title} {\emph {\bibinfo {title} {{Resistivity of
  non-Galilean-invariant Fermi- and non-Fermi liquids}}},\ }\href
  {https://doi.org/10.3952/lithjphys.52207} {\bibfield  {journal} {\bibinfo
  {journal} {Lithuanian Journal of Physics and Technical Sciences}\ }\textbf
  {\bibinfo {volume} {52}},\ \bibinfo {pages} {142} (\bibinfo {year} {2012})},\
  \Eprint {https://arxiv.org/abs/1204.3591} {arXiv:1204.3591 [cond-mat.str-el]}
  \BibitemShut {NoStop}%
\bibitem [{\citenamefont {{Maslov}}\ and\ \citenamefont
  {{Chubukov}}(2017)}]{Maslov17}%
  \BibitemOpen
  \bibfield  {author} {\bibinfo {author} {\bibfnamefont {D.~L.}\ \bibnamefont
  {{Maslov}}}\ and\ \bibinfo {author} {\bibfnamefont {A.~V.}\ \bibnamefont
  {{Chubukov}}},\ }\bibfield  {title} {\emph {\bibinfo {title} {{Optical
  response of correlated electron systems}}},\ }\href
  {https://doi.org/10.1088/1361-6633/80/2/026503} {\bibfield  {journal}
  {\bibinfo  {journal} {Reports on Progress in Physics}\ }\textbf {\bibinfo
  {volume} {80}},\ \bibinfo {eid} {026503} (\bibinfo {year} {2017})},\ \Eprint
  {https://arxiv.org/abs/1608.02514} {arXiv:1608.02514 [cond-mat.str-el]}
  \BibitemShut {NoStop}%
\bibitem [{\citenamefont {{Ledwith}}\ \emph {et~al.}(2019)\citenamefont
  {{Ledwith}}, \citenamefont {{Guo}},\ and\ \citenamefont
  {{Levitov}}}]{LedwithAoP}%
  \BibitemOpen
  \bibfield  {author} {\bibinfo {author} {\bibfnamefont {P.~J.}\ \bibnamefont
  {{Ledwith}}}, \bibinfo {author} {\bibfnamefont {H.}~\bibnamefont {{Guo}}},\
  and\ \bibinfo {author} {\bibfnamefont {L.}~\bibnamefont {{Levitov}}},\
  }\bibfield  {title} {\emph {\bibinfo {title} {{The hierarchy of excitation
  lifetimes in two-dimensional Fermi gases}}},\ }\href
  {https://doi.org/10.1016/j.aop.2019.167913} {\bibfield  {journal} {\bibinfo
  {journal} {Annals of Physics}\ }\textbf {\bibinfo {volume} {411}},\ \bibinfo
  {pages} {167913} (\bibinfo {year} {2019})},\ \Eprint
  {https://arxiv.org/abs/1905.03751} {arXiv:1905.03751 [cond-mat.mes-hall]}
  \BibitemShut {NoStop}%
\bibitem [{\citenamefont {{Ledwith}}\ \emph {et~al.}(2017)\citenamefont
  {{Ledwith}}, \citenamefont {{Guo}},\ and\ \citenamefont
  {{Levitov}}}]{LedwithArxiv}%
  \BibitemOpen
  \bibfield  {author} {\bibinfo {author} {\bibfnamefont {P.}~\bibnamefont
  {{Ledwith}}}, \bibinfo {author} {\bibfnamefont {H.}~\bibnamefont {{Guo}}},\
  and\ \bibinfo {author} {\bibfnamefont {L.}~\bibnamefont {{Levitov}}},\
  }\bibfield  {title} {\emph {\bibinfo {title} {{Angular Superdiffusion and
  Directional Memory in Two-Dimensional Electron Fluids}}},\ }\href@noop {} {\
  (\bibinfo {year} {2017})},\ \Eprint {https://arxiv.org/abs/1708.01915}
  {arXiv:1708.01915 [cond-mat.mes-hall]} \BibitemShut {NoStop}%
\bibitem [{\citenamefont {Shi}\ \emph {et~al.}(2022)\citenamefont {Shi},
  \citenamefont {Goldman}, \citenamefont {Else},\ and\ \citenamefont
  {Senthil}}]{ShiElse2022}%
  \BibitemOpen
  \bibfield  {author} {\bibinfo {author} {\bibfnamefont {Z.~D.}\ \bibnamefont
  {Shi}}, \bibinfo {author} {\bibfnamefont {H.}~\bibnamefont {Goldman}},
  \bibinfo {author} {\bibfnamefont {D.~V.}\ \bibnamefont {Else}},\ and\
  \bibinfo {author} {\bibfnamefont {T.}~\bibnamefont {Senthil}},\ }\bibfield
  {title} {\emph {\bibinfo {title} {{Gifts from anomalies: Exact results for
  Landau phase transitions in metals}}},\ }\href
  {https://doi.org/10.21468/SciPostPhys.13.5.102} {\bibfield  {journal}
  {\bibinfo  {journal} {SciPost Phys.}\ }\textbf {\bibinfo {volume} {13}},\
  \bibinfo {pages} {102} (\bibinfo {year} {2022})},\ \Eprint
  {https://arxiv.org/abs/2204.07585} {arXiv:2204.07585 [cond-mat.str-el]}
  \BibitemShut {NoStop}%
\bibitem [{\citenamefont {{Wu}}\ \emph {et~al.}(2022)\citenamefont {{Wu}},
  \citenamefont {{Liao}},\ and\ \citenamefont {{Foster}}}]{Foster22}%
  \BibitemOpen
  \bibfield  {author} {\bibinfo {author} {\bibfnamefont {T.~C.}\ \bibnamefont
  {{Wu}}}, \bibinfo {author} {\bibfnamefont {Y.}~\bibnamefont {{Liao}}},\ and\
  \bibinfo {author} {\bibfnamefont {M.~S.}\ \bibnamefont {{Foster}}},\
  }\bibfield  {title} {\emph {\bibinfo {title} {{Quantum Interference of
  Hydrodynamic Modes in a Dirty Marginal Fermi Liquid}}},\ }\href@noop {} {\
  (\bibinfo {year} {2022})},\ \Eprint {https://arxiv.org/abs/2206.01762}
  {arXiv:2206.01762 [cond-mat.str-el]} \BibitemShut {NoStop}%
\bibitem [{\citenamefont {Gu}\ \emph {et~al.}(2020)\citenamefont {Gu},
  \citenamefont {Kitaev}, \citenamefont {Sachdev},\ and\ \citenamefont
  {Tarnopolsky}}]{Gu:2019jub}%
  \BibitemOpen
  \bibfield  {author} {\bibinfo {author} {\bibfnamefont {Y.}~\bibnamefont
  {Gu}}, \bibinfo {author} {\bibfnamefont {A.}~\bibnamefont {Kitaev}}, \bibinfo
  {author} {\bibfnamefont {S.}~\bibnamefont {Sachdev}},\ and\ \bibinfo {author}
  {\bibfnamefont {G.}~\bibnamefont {Tarnopolsky}},\ }\bibfield  {title} {\emph
  {\bibinfo {title} {{Notes on the complex Sachdev-Ye-Kitaev model}}},\ }\href
  {https://doi.org/10.1007/JHEP02(2020)157} {\bibfield  {journal} {\bibinfo
  {journal} {JHEP}\ }\textbf {\bibinfo {volume} {02}},\ \bibinfo {pages}
  {157}},\ \Eprint {https://arxiv.org/abs/1910.14099} {arXiv:1910.14099
  [hep-th]} \BibitemShut {NoStop}%
\bibitem [{\citenamefont {Prange}\ and\ \citenamefont
  {Kadanoff}(1964)}]{PrangeKadanoff}%
  \BibitemOpen
  \bibfield  {author} {\bibinfo {author} {\bibfnamefont {R.~E.}\ \bibnamefont
  {Prange}}\ and\ \bibinfo {author} {\bibfnamefont {L.~P.}\ \bibnamefont
  {Kadanoff}},\ }\bibfield  {title} {\emph {\bibinfo {title} {{Transport Theory
  for Electron-Phonon Interactions in Metals}}},\ }\href
  {https://doi.org/10.1103/PhysRev.134.A566} {\bibfield  {journal} {\bibinfo
  {journal} {Phys. Rev.}\ }\textbf {\bibinfo {volume} {134}},\ \bibinfo {pages}
  {A566} (\bibinfo {year} {1964})}\BibitemShut {NoStop}%
\bibitem [{\citenamefont {{Chubukov}}\ and\ \citenamefont
  {{Maslov}}(2017)}]{ChubukovMaslov0}%
  \BibitemOpen
  \bibfield  {author} {\bibinfo {author} {\bibfnamefont {A.~V.}\ \bibnamefont
  {{Chubukov}}}\ and\ \bibinfo {author} {\bibfnamefont {D.~L.}\ \bibnamefont
  {{Maslov}}},\ }\bibfield  {title} {\emph {\bibinfo {title} {{Optical
  conductivity of a two-dimensional metal near a quantum critical point: The
  status of the extended Drude formula}}},\ }\href
  {https://doi.org/10.1103/PhysRevB.96.205136} {\bibfield  {journal} {\bibinfo
  {journal} {Phys. Rev. B}\ }\textbf {\bibinfo {volume} {96}},\ \bibinfo {eid}
  {205136} (\bibinfo {year} {2017})},\ \Eprint
  {https://arxiv.org/abs/1707.07352} {arXiv:1707.07352 [cond-mat.str-el]}
  \BibitemShut {NoStop}%
\bibitem [{\citenamefont {Millis}(1993)}]{millis}%
  \BibitemOpen
  \bibfield  {author} {\bibinfo {author} {\bibfnamefont {A.~J.}\ \bibnamefont
  {Millis}},\ }\bibfield  {title} {\emph {\bibinfo {title} {Effect of a nonzero
  temperature on quantum critical points in itinerant fermion systems}},\
  }\href {https://doi.org/10.1103/PhysRevB.48.7183} {\bibfield  {journal}
  {\bibinfo  {journal} {Phys. Rev. B}\ }\textbf {\bibinfo {volume} {48}},\
  \bibinfo {pages} {7183} (\bibinfo {year} {1993})}\BibitemShut {NoStop}%
\bibitem [{\citenamefont {Kamenev}\ and\ \citenamefont
  {Oreg}(1995)}]{KamenevOreg}%
  \BibitemOpen
  \bibfield  {author} {\bibinfo {author} {\bibfnamefont {A.}~\bibnamefont
  {Kamenev}}\ and\ \bibinfo {author} {\bibfnamefont {Y.}~\bibnamefont {Oreg}},\
  }\bibfield  {title} {\emph {\bibinfo {title} {Coulomb drag in normal metals
  and superconductors: Diagrammatic approach}},\ }\href
  {https://doi.org/10.1103/PhysRevB.52.7516} {\bibfield  {journal} {\bibinfo
  {journal} {Phys. Rev. B}\ }\textbf {\bibinfo {volume} {52}},\ \bibinfo
  {pages} {7516} (\bibinfo {year} {1995})}\BibitemShut {NoStop}%
\bibitem [{\citenamefont {Ledwith}\ \emph {et~al.}(2019)\citenamefont
  {Ledwith}, \citenamefont {Guo}, \citenamefont {Shytov},\ and\ \citenamefont
  {Levitov}}]{LedwithGuo2019}%
  \BibitemOpen
  \bibfield  {author} {\bibinfo {author} {\bibfnamefont {P.}~\bibnamefont
  {Ledwith}}, \bibinfo {author} {\bibfnamefont {H.}~\bibnamefont {Guo}},
  \bibinfo {author} {\bibfnamefont {A.}~\bibnamefont {Shytov}},\ and\ \bibinfo
  {author} {\bibfnamefont {L.}~\bibnamefont {Levitov}},\ }\bibfield  {title}
  {\emph {\bibinfo {title} {Tomographic dynamics and scale-dependent viscosity
  in 2d electron systems}},\ }\href
  {https://doi.org/10.1103/PhysRevLett.123.116601} {\bibfield  {journal}
  {\bibinfo  {journal} {Phys. Rev. Lett.}\ }\textbf {\bibinfo {volume} {123}},\
  \bibinfo {pages} {116601} (\bibinfo {year} {2019})}\BibitemShut {NoStop}%
\bibitem [{\citenamefont {Huang}\ and\ \citenamefont
  {Lucas}(2021)}]{Lucas2021}%
  \BibitemOpen
  \bibfield  {author} {\bibinfo {author} {\bibfnamefont {X.}~\bibnamefont
  {Huang}}\ and\ \bibinfo {author} {\bibfnamefont {A.}~\bibnamefont {Lucas}},\
  }\bibfield  {title} {\emph {\bibinfo {title} {{Fingerprints of quantum
  criticality in locally resolved transport}}},\ }\href@noop {} {\  (\bibinfo
  {year} {2021})},\ \Eprint {https://arxiv.org/abs/2105.01075}
  {arXiv:2105.01075 [cond-mat.str-el]} \BibitemShut {NoStop}%
\bibitem [{\citenamefont {{Darius Shi}}\ \emph {et~al.}(2022)\citenamefont
  {{Darius Shi}}, \citenamefont {{Else}},\ and\ \citenamefont
  {{Goldman}}}]{ShiElse2022a}%
  \BibitemOpen
  \bibfield  {author} {\bibinfo {author} {\bibfnamefont {Z.}~\bibnamefont
  {{Darius Shi}}}, \bibinfo {author} {\bibfnamefont {D.~V.}\ \bibnamefont
  {{Else}}},\ and\ \bibinfo {author} {\bibfnamefont {H.}~\bibnamefont
  {{Goldman}}},\ }\bibfield  {title} {\emph {\bibinfo {title} {{Loop current
  fluctuations and quantum critical transport}}},\ }\href@noop {} {\bibfield
  {journal} {\bibinfo  {journal} {arXiv e-prints}\ } (\bibinfo {year}
  {2022})},\ \Eprint {https://arxiv.org/abs/2208.04328} {arXiv:2208.04328
  [cond-mat.str-el]} \BibitemShut {NoStop}%
\end{thebibliography}%
\end{document}